\documentclass[prb,aps,floats,amssymb,showkeys,showpacs,
 bibnotes,tightenlines,floatfix]{revtex4}
\usepackage{longtable}
\usepackage{graphicx}
\usepackage{graphics}
\usepackage{epsfig,bm}
\usepackage{pdfpages}
\usepackage{rotating}
\begin{document}

\title{The Blume-Capel model  for spins $S=1$ and $3/2$ \\ 
in  dimensions $d=2$ and $3$}

\author{P. Butera}

\email{paolo.butera@mib.infn.it}

\affiliation
{Dipartimento di Fisica Universita' di Milano-Bicocca\\
and\\
Istituto Nazionale di Fisica Nucleare \\
Sezione di Milano-Bicocca\\
 3 Piazza della Scienza, 20126 Milano, Italy}

\author{M. Pernici} 

\email{mario.pernici@mi.infn.it}

\affiliation
{Istituto Nazionale di Fisica Nucleare \\
Sezione di Milano\\
 16 Via Celoria, 20133 Milano, Italy}

\date{\today}

\date{\today}
\begin{abstract}
 Expansions through the 24th order at high-temperature and up to 11th
order at low-temperature are derived for the main observables of the
Blume-Capel model on bipartite lattices ($sq$,$sc$ and $bcc$) in $2d$
and $3d$ with various values of the spin and in presence of a magnetic
field.  All expansion coefficients are computed exactly as functions
of the crystal and magnetic fields. Several critical properties of the
model are analyzed in the two most studied cases of spin $S=1$ and
$S=3/2$.
\end{abstract}
\maketitle

\section{Introduction}
The Ising system with spin $S=1$, subject to a uniform
``single-ion-splitting crystal-field'' (also called ``anisotropy
field'')\cite{Blume,Capel}, in absence of magnetic field, is the
simplest lattice spin-model exhibiting a
tricritical\cite{griff,griff2} point (TCP), namely a point of the
phase-boundary at which a smooth line of second-order
phase-transitions  undergoes a sudden change into a line of
first-order transitions.  It was originally studied by M. Blume and
H.W. Capel\cite{Blume,Capel} and thereby it is known as the
Blume-Capel (BC) model.  Extensions of the model featuring
interactions of higher spins\cite{taggart,plascak}, in
particular of spin $S=3/2$, were also considered in the literature.

The BC model has been explored in a variety of analytical
 approximations (mean field (MF), effective field, Renormalization
 Group, etc.)\cite{landau,Blume,Capel,taggart,BEG,yuk,
 strecka,plascak,berkwor,adler,oliv,ilko,tensor,grollau,
 grollau32,lara,xavier,costa}, by transfer-matrix
 methods\cite{beale,quian,jung} and by MonteCarlo (MC) simulation
 methods\cite{wilding,yuk,care,min1,min2,deserno,deng,Oz,ziere2d,ziere3d,
 arora,jain, xavier,plascak_landau,silva,berker,martins,kwak}, but
 only in a handful of
 papers\cite{Oitmaa1,Oitmaa2,SWS,wang,brankov,jose} extrapolations of
  series-expansions were employed, in spite of the potential
 reliability and accuracy of this technique.  The high-temperature
 (HT) and the low-temperature (LT) expansions have been jointly
 used\cite{SWS} to map out the phase diagram, (the former being
 generally sufficient to locate the second-order part of the
 phase-boundary and to determine its universal parameters, the latter
 being helpful for the first-order part).

This work fills a gap in the literature by contributing {\it exact} HT
 expansions (based on the computations of
 Ref. [\onlinecite{bp1,bp2,bp3}]), that reach the 24th order   for the square
 ($sq$), simple cubic ($sc$) and body-centered cubic ($bcc$) lattices,
 {\it in presence of a magnetic field}, in the two most
 representative cases of the BC model with spin $S=1$ and $S=3/2$.  In
 the spin $S=1$ case, we have also computed {\it exact} LT expansions
 for the $sq$ and the $sc$ lattices through the 11th order, while our
 LT results reach only the 10th order for the $bcc$ lattice. In the
 spin $S=3/2$ case, the LT expansions extend through the 9th
 order. All series coefficients, both at HT and LT, are expressed in a
 closed form as {\it explicit} functions of the crystal and magnetic
 fields.  The analysis of these expansions leads to results 
 directly comparable with those of the modern simulation studies,
 generally carried out on the simple cubic lattices in $2d$ and $3d$.
 Moreover this work makes now possible to achieve high accuracy in the
 determination of the phase-boundaries, in the tests of universality
 of the critical exponents as well as of appropriate combinations of
 critical amplitudes, and to describe the dependence of the observables
 on the magnetic field, in regions of the model's parameter-space
 that are wider than those covered by other numerical approaches.

The longest series expansions so far available, that date back to long
 ago, were limited to systems with spin $S=1$ on the (close-packed)
 face-centered-cubic ($fcc$) lattice\cite{Oitmaa1,Oitmaa2,SWS}, in
 {\it absence} of magnetic field. At best\cite{SWS}, the HT expansions
 extended only through the 12th order while the LT expansions reached
 the 7th.  The results of their analysis are not comparable with those
 of the modern MC studies for the non-universal features such as the 
 phase-contours, but are relevant only for the universal critical
 parameters.  The $fcc$ lattice was chosen because its non-bipartite
 nature entails the absence of a nearby antiferromagnetic singularity
 in the complex plane of the inverse temperature and together with its
 large coordination-number $q=12$, ensures smooth
 expansion-coefficient sequences. It was therefore hoped that
 reasonable series extrapolations might be performed in spite of the
 short order of the expansions.

 Also for the bipartite lattices, at the time of
Ref. [\onlinecite{SWS}], by employing essentially the same code,
expansions might be derived with the same extension as for the $fcc$
lattice and with the same limitations, but they were not. Only shorter
expansions\cite{jose} through the 8th order were obtained in $2d$ and
$3d$ for these lattices, but were not analyzed.  In
Ref.[\onlinecite{brankov}], HT expansions up to the 5th-order, valid
for general spin on the $sc$ and the $bcc$ lattices, were briefly
discussed, yielding qualitative indications.  In conclusion, the state
of the expansions for the $sq$, the $sc$ and the $bcc$ lattices has
been so far inadequate, while substantial extensions and updates of
the $fcc$ lattice study remain desirable, but not as urgent.

 The paper is organized as follows.  Section II introduces the BC
 model with general spin $S$ and sketches its main features. The 
  following subsections recall the established results of the MF
 approximation,  and contain a few comments on recent numerical studies
 to be later compared with ours.  The following subsections
 describe
 some details of the algebraic structure of the HT and LT expansion
 coefficients and state the expected singular asymptotic behaviors of
 the thermodynamical quantities under scrutiny, to which reference is
 made in discussing the estimates of the critical parameters.  This
 Section is supplemented by the Appendices A and B, the first one
 being a very detailed pedagogical reminder of the known methods and
 results of the MF approximation for each value of the spin considered
 in our study while the second one recalls known arguments about the
 tricritical scaling.

  Section III contains a few general remarks on the analysis of the
  expansions, valid independently of the structure and dimension of
  the lattices considered, while the main properties of the numerical
  tools such as ``modified ratio approximants''(MRA), ``Pad\'e
  approximants''(PA) and ``differential approximants''(DA) employed
  to estimate the critical parameters of the model, are recalled in
  Appendix C.

 In the Sections IV-VII, we
  discuss the results of our series analyses in dimension $d=2$, and
  $d=3$ for spin $S=1$ and $S=3/2$ and compare them with the known
  closed-form expressions for the phase-boundaries in the MF
  approximation and with numerical results obtained by
  transfer-matrix in $2d$ and/or by MC methods both in $2d$ and in
  $3d$.  In the Section VIII, we summarize our work and draw some
  conclusions.

\section{The BC model with spin $S$ }
An extensive discussion of tricriticality, covering a wide range of
the physical situations in which it occurs and a thorough account of
the numerous experimental and theoretical investigations of the BC
model before 1984 can be found in the review
paper[\onlinecite{lawrie}], from which we have largely drawn
throughout the paper.

 The BC model is a simple representative\cite{lawrie} of a class of
pure or disordered spin models capable of describing the phenomenology
of a variety of physical systems, that includes liquid mixtures of
$^3He - ^4He$, many-component classical solutions\cite{griff,griff2},
antiferromagnets and metamagnets\cite{stanley}, alloys of magnetic and
non-magnetic materials\cite{aharony}, ferroelectrics, liquid crystals
and several others.

The  BC model with spin $S$  is defined by the reduced Hamiltonian 
\begin{equation}
\frac{1} {k_BT} {\cal H}_N[s]=
-\frac{K}{S^2}\sum_{\langle ij \rangle} s_is_j+
\frac{D}{S^2}\sum_i s^2_i - \frac{h}{S}\sum_i s_i
\label{hamBC}
\end{equation}
 The Hamiltonian ${\cal H}_N[s]$ is evaluated in a finite lattice of 
 volume $N$. The first sum on the right hand side, describing the
 nearest-neighbor spin-pair interaction, runs over all ordered links,
 while the second and third extend to all sites.  If $S=1$, the spin
 $s_i$ at the site $i$ can take the values $s_i=\pm 1, 0$.  In the
 $S=3/2$ case also discussed here, the spins can take the values
 $s_i=\pm 1/2, \pm 3/2$.

The natural variables appearing in Eq. (\ref{hamBC}) are defined as
 follows:  $K=J/k_BT$, with $T$ the temperature, $k_B$ the Boltzmann
 constant (set to 1 in what follows) and $J > 0$ the
 exchange-interaction energy, is an adimensional quantity such that
 $K/S^2$ can be taken as HT expansion parameter; the quantity $h$ 
 represents the reduced magnetic field $h=-K H/J$ with $H$ the
 external uniform magnetic field, and finally $D=K\Delta/J$ called
 ``reduced crystal-field'' (also ``reduced anisotropy field''), with
 $\Delta$ the ``crystal field'' (also ``anisotropy field'').  A
 further generalization of the BC model, called the BEG
 model\cite{Rys,BEG} includes an additional biquadratic
 exchange-coupling interaction between nearest-neighbor spins.

The thermodynamics of the model is described in terms of three
conjugate pairs of field-density variables: the temperature and the
specific entropy; the magnetic field $H$ and the specific
magnetization $M$ (``order parameter''); the crystal field $\Delta$
and its conjugate density $X$ called ``concentration'' (or
``non-ordering parameter''). For systems with spin $S=1$, the
concentration (of the spins with $s_i=0$) is usually defined as
$X=1-\langle s_0^2 \rangle$. 
In the case of spin $S=3/2$, the
concentration (of the spins with $s_i= \pm 1/2$) defined as
$X=1/2(9/4-\langle s_0^2 \rangle)$, retains the same range $[0,1]$ of
variation.
The main thermodynamic observables of the model are the response
functions, e.g. the specific-heat, the derivative of $X$ with respect
to $\Delta$, denoted by $Y$ and called ``concentration
susceptibility'' (or ``quadrupolar susceptibility''), and the derivatives
of $M$ with respect to $H$, for example the ordinary magnetic
susceptibility, as well as the other higher-order derivatives.

We can intuitively figure out what the critical behavior of the BC
 system  will be in zero magnetic field, as follows.
 For $D = 0$, the model reduces to the Ising model with spin $S$,
 that, in a vanishing magnetic field, displays a power-law critical
 transition from a disordered HT phase to an ordered LT phase at a
 temperature $T_c(0)$. When $D$ is switched on to a small enough
 positive (negative) value, it raises (lowers) slightly the energy of
 the spins with $s_i=\pm S$ above (below) that of the spins with
 $s_i< S$. The field $D$ does not break the symmetry of the disordered
 phase and the exchange interaction remains dominant, so that a
 continuous transition still occurs, though at a slightly smaller
 (higher) temperature $T_c(D)$, since by increasing (decreasing) $D$,
 the onset of order is disfavored (favored). Thus, over some range of
 nonzero values of $D$, a line of second-order transitions $T=T_c(D)$
 will be observed.  Moreover the universal features of the asymptotic
 critical behavior will not be affected by $D$ and thus a single set
 of  exponents with the Ising values\cite{griff,griff2} associated with 
 $D=0$, will characterize this  line of critical points.

For large enough values of the ratio $D/K$, this picture might change
 drastically, as indicated by the emergence of a first-order
 transition line rooted at $T=0$, which is discussed below.

\subsection{The main predictions of the MF approximation}

A spin-dependent description of the essential features of the phase
 diagram of the BC model, in particular an answer to the question
 whether and how the above mentioned line of continuous Ising-like
 critical points is connected with the first-order line, can be
 obtained\cite{Blume,Capel,plascak,BEG} in the conventional MF
 approximation of the Hamiltonian of Eq. (\ref{hamBC}).  In what
 follows, we shall summarize the main indications that emerge, leaving
 a formalization of this approach to the Appendix A that rephrases in
 greater detail the general solutions of Ref. [\onlinecite{plascak}]
 for the two values of the spin examined in this paper.
First, let us set 
 \begin{equation}
 \tilde  \Delta = \Delta/Jq  \qquad  \qquad   and \qquad \qquad \tilde T=T/Jq
 \label{defvar}
 \end{equation}
 and sketch the main features of the BC models with {\it
integer} spin in the MF approximation. For $H=0$ and $-\infty <\Delta
\le \Delta_{tr}$, a continuous transition line separates a
ferromagnetically ordered LT phase from a paramagnetic HT phase.  In
the case of spin $S=1$, this line terminates with a TCP of coordinates
$(\tilde \Delta_{tr}= \frac{2}{3} {\rm ln}2$, $\tilde T_{tr}=\frac {1}{3})$.

The ground-state specific energy for $S=1$ can be computed exactly and
it is observed that its behavior changes at $\tilde \Delta=1/2$. All
spins are in the same state $s_i= S$ for $\tilde \Delta <1/2$, and the
point ($\tilde \Delta=1/2, T=0$), at which the energy becomes
independent of the spin value, is the tip of a first-order transition line
that joins the second-order line at the TCP. For $\tilde \Delta > 1/2$,
the BC system remains paramagnetic at all temperatures.

\begin{figure}[p]
\begin{center}
\leavevmode
\includegraphics[width=5.00 in]{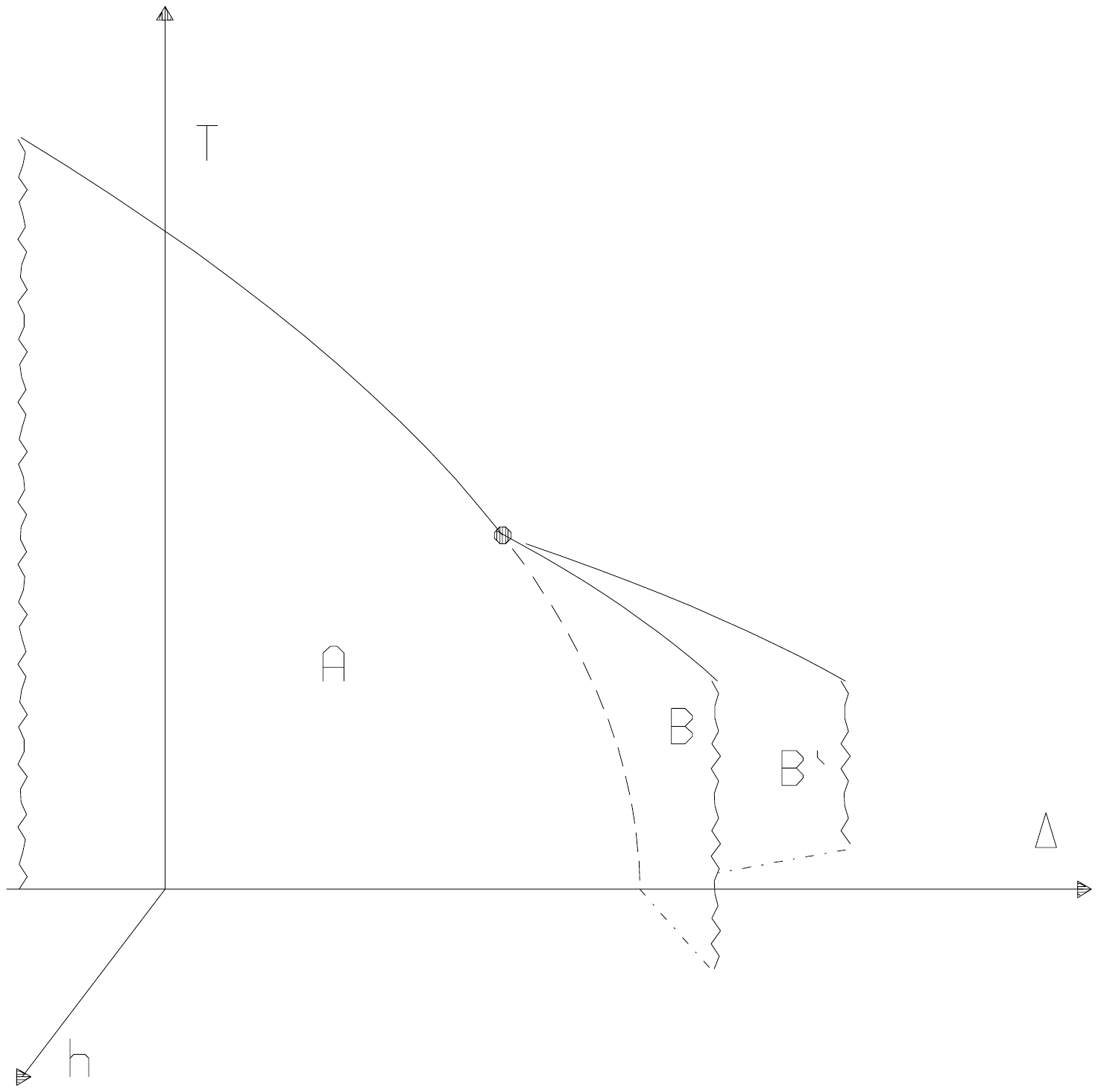}
\caption{ \label{fig_diag_tricr} Schematic\cite{griff} phase diagram
in the $\Delta, T, h$ space for a BC model with spin $S=1$.
First-order transitions occur across the surfaces $A$, $B$ and $B'$.  The
long-dashed line is a first-order line in the $h=0$ plane.  The solid lines that meet at the TCP(full circle) are
critical lines  bounding the surfaces. The
surface $A$ extends to $\Delta= -\infty$, but in the figure is cut along
the wavy line.  Also the surfaces $B$ and $B'$, called
``wings'', extend to large $\Delta$ and $|h|$, but are cut along the
wavy lines. The dot-dashed lines are the wing intersections with the
$T=0$ plane.  Phases with opposite values of $\langle s \rangle$
coexist in the $A$-plane while phases with different values of
$\langle s^2 \rangle$ coexist in the planes $B'$ and $B$, so that
three phases coexist along the  first-order line.}
\end{center}
\end{figure}

For all integer spins $S>1$, the overall phase structure remains
essentially the same: in the $H=0$ plane, there is a second order transition
line that terminates with a TCP. Various first-order transition lines,
all of them rooted at $(\tilde \Delta=1/2, T=0)$, appear in the LT phase
and only the rightmost first-order line joins the second-order line at
the TCP, while the others display a branched structure and terminate
with double-critical endpoints\cite{plascak_landau}. The structure of
this phase diagram for spin $S=1$ is schematized in
Fig. \ref{fig_diag_tricr}.

 If viewed in the extended space of the fields $T,\Delta, H$, the TCP
in the case of spin $S=1$ reveals a complex structure.  As indicated
schematically in Fig. \ref{fig_diag_tricr}, at this point three
distinct critical lines (only one of which lies in the $H=0$ plane)
and a line of first-order transitions meet together. It was this confluence
that suggested\cite{griff,griff2} the name ``tricritical point''.
Thus we can as well say that the TCP is the end-point of a first-order
line of three-phase coexistence (a line of triple-points), at which
these phases become simultaneously critical.  The critical lines bound
three surfaces of first-order transitions, one of them in the $H=0$
plane and two surfaces usually called ``wings'', extending
symmetrically for $H\ne 0$.

On the other hand, the models with {\it half-odd} values of the spins
 $S > 1 $, show the distinctive feature that the second-order line
 does not terminate with a TCP and does not join there with a
 first-order line, but extends indefinitely for $-\infty \leq \Delta
 \leq \infty$. As to the ground-state specific energy, one observes
 that all spins are in the same state $s_i= S$ for $\tilde \Delta
 <1/2$, while they are in the state $s_i= 1/2$ for $\tilde
 \Delta>1/2$.  For $S>3/2$, several first-order transition lines occur
 within the LT phase, all of them rooted at ( $\tilde \Delta=1/2,
 \tilde T=0$). As the spin gets large, they display an increasingly
 branched structure and terminate with double-critical endpoints.

\begin{figure}[p]
\begin{center}
\leavevmode
\includegraphics[width=5.00 in]{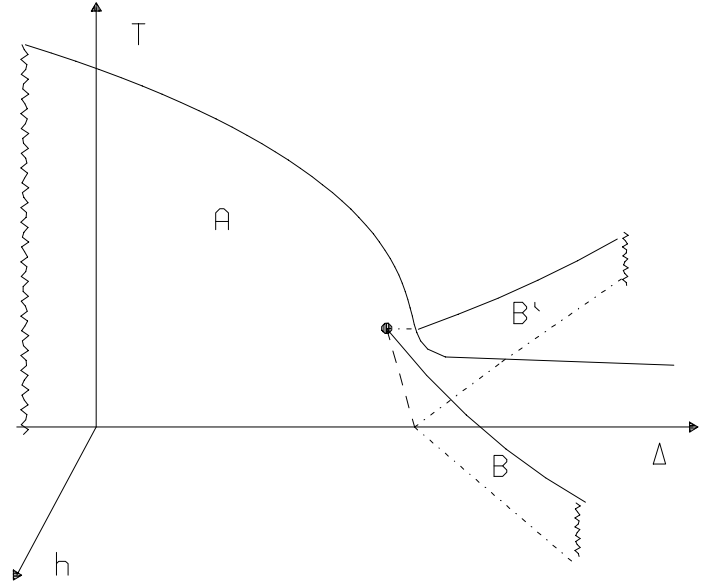}
\caption{ \label{fig_diag_32} Schematic\cite{plascak_landau} phase
diagram in the $ \Delta ,T, h$ space for a BC model with spin $S=3/2$.
First-order transitions occur across the surfaces $A$, $B$ and $B'$.
Phases with opposite values of $\langle s \rangle$ coexist in the
$A$-plane and phases with different values of $\langle s^2 \rangle$
coexist in the planes $B'$ and $B$. Four phases coexist on a
first-order line in the $h=0$ plane, that is the long-dashed line
terminating at a ``double critical endpoint''(full circle).  Solid
lines bounding the surfaces $B$ and $B'$ are critical transitions
that meet at the ``double critical endpoint''.  The surface $A$
extends to $\Delta=-\infty$, but is cut along the wavy line.  Also the
surfaces $B$ and $B'$, called ``wings'', extend to large $\Delta$ and
$|h|$, but are cut along the wavy lines. Short dashed lines are the wing
intersections with the $T=0$ plane. }
\end{center}
\end{figure}

In the spin $S=3/2$ case, a single first-order line appears in the LT
phase, but no TCP occurs on the phase-boundary in the $H=0$ plane and
again two wings depart, as for spin $S=1$, from the
first-order surface.  The structure of this phase diagram is
schematized\cite{plascak_landau} in Fig. \ref{fig_diag_32}.  In this
case, one can also say that a line of four-phase coexistence
terminates at the double-critical endpoint at which these phases
become simultaneously critical.

 It will be interesting to compare the qualitative indications
 obtained from the MF closed-form expressions\cite{plascak} with the
  results of a numerical analysis of the series expansions.

Also for the TCP, there exists an upper tricritical lattice dimension
 $d_{tr}^*$ (the subscript $tr$ will be henceforth attached to the
 quantities associated to the TCP). For $d \leq d_{tr}^*$ the critical
 fluctuations are sufficiently strong that the TCP occurs with non-MF
 exponents.  Thus,  the MF approximation can only
 be a first guide to the critical behavior and its predictions must be
 qualitatively validated and quantitatively refined.  Determining the
 upper critical dimension of the TCP is then a necessary step to get a
 complete understanding of the model.  Both by
 renormalization-group\cite{riede1,riede2,wegne} (RG) methods, as
 mentioned below, and by applying\cite{baus} the Ginzburg criterion,
 it was argued that $d_{tr}^* =2$. At the borderline dimension $d=3$,
 the MF critical behaviors, possibly modified by logarithmic
 correction factors, are expected to set in at the TCP, just as it is
 observed at $d=4$ for the ordinary critical phenomena. For $d > 3$,
 the tricritical exponents should retain $d$-independent values since
 the MF approximation becomes consistent and therefore also
 quantitatively reliable.  In $d=2$ dimensions, the field theory
 defined by Eq. (\ref{Hlandginz}) can be solved within the conformal
 field-theory(CFT) approach \cite{cardy,cardy2,henkel} yielding the
 exact tricritical values reported in Tab. \ref{tab01}. The remaining
 exponents can be calculated by the usual scaling relations.  The $2d$
 exponents are markedly different from the corresponding MF values.

Also for $d>4$, the MF values of the critical exponents should change
  discontinuously at the TCP, into the tricritical ones indicated in
  Tabs. \ref{tab0} and \ref{tab01}.

\begin{table}[ht]
  \caption{The exponents of the Ising
 universality-class\cite{bcisiesse} for lattices of dimension $d=2$
 and $d=3$.  We have also indicated the values of these exponents in
 the MF approximation, valid for $d>3$. }
\begin{tabular}{|c| c c c c| }
\hline
  &${ \alpha}$& ${\beta}$&${ \gamma}$&${ \nu}$\\
 \hline
 \hline
 $d=2$&0&1/8&7/4&1 \\
\hline
 $d=3$&0.11(1)  &0.326(5)&1.2371(2)&0.6299(4) \\
\hline
$MF$&0&1/2&1&1/2\\
\hline
 \end{tabular} 
 \label{tab0}
\end{table}

\begin{table}[ht]
  \caption{The exponents of the BC model at the TCP. 
 They are distinguished from the  exponents on the critical phase-contour 
 by the subscript $tr$.  For lattices of
    dimension $d=3$, MF values are expected, possibly modified by 
logarithmic correction factors. }
\begin{tabular}{|c| c c c c  c| }
\hline
  &$\alpha_{tr}$& $\beta_{tr}$&$\gamma_{tr}$&$\nu_{tr}$&$\phi_{tr}$\\
 \hline
 \hline
 $d=2$ (CFT)&8/9&1/24&37/36&5/9&4/9 \\
\hline
 $d=3$ (MF)&1/2&1/4&1&1/2&1/2 \\
\hline
 \end{tabular} 
 \label{tab01}
\end{table}

Along the second-order lines that border the wings, the BC model
is expected to display exponents {\it independent} of $D$, retaining
by universality the Ising-like values taken at $H=D=0$ and appropriate
to the lattice dimension.

We shall return later on the emergence at the TCP, of the additional
{\it crossover} exponent $\phi$ reported in Tab. \ref{tab01}.

\subsection{Further comments on the recent numerical studies}

 No exact results are known for the BC model (or even for the Ising
 models with spin $S > 1/2$) on lattices of dimension $d>1$.  A number
 of analytic approximations of uncontrolled convergence rate were
 proposed to improve the MF approximation\cite{strecka}, but so far
 the transfer-matrix method and the MC simulations remain the only
 safe alternatives to series methods.

Many MC simulations of accuracy increasing with the speed and memory
 of the available computers, were carried out on the
 $sq$\cite{arora,wilding,yuk,ziere2d,ziere3d}, on the
 $sc$\cite{deserno,Oz,deng}, and the $fcc$ lattices\cite{jain} for the
 BC model with various values of the spin.  In particular a very
 recent multicanonical simulation\cite{ziere2d} for the $sq$ lattice
 reaches a very high accuracy.  In $d=2$ dimensions, for $S=1$, one
 can take also advantage of the feasibility\cite{jung,beale,quian} of very
 accurate transfer-matrix computations.

  The qualitative indications of the MF approximation were supported
 by a transfer-matrix calculation\cite{xavier} also in the case of the
 spin $S=3/2$ BC model in $2d$ and by a MC simulation\cite{lara} in
 $3d$.  For the $sq$ lattice, an effective-field theory with
 correlation was  studied\cite{costa} for all values of the
 spin. Additionally, we can cite a MC simulation of the BC model on
 the $sq$ lattice with quenched disorder\cite{berker} in which 
 accurate data,  comparable with the
 results of this paper, are reported also for the pure case.

  So far no simulations have been carried out for the $bcc$
 lattice. An estimate for the location of the TCP came however from a
 calculation\cite{grollau} in a self-consistent Ornstein-Zernike
 approximation.

The estimates of the critical parameters obtained in some of the cited
studies will be later confronted with those from our analysis of the
LT and HT expansions.

The Euclidean field theory with the potential $V(\phi)$ of
Eq. (\ref{Hlandginz}) can be used as a basis for the RG based
approximations, if we assume that its tricritical behavior belongs to
the same universality class as the BC model.  RG ideas
applied\cite{griff,griff2,riede1,riede2,wegne} in $d=3$ suggested MF
behavior with logarithmic correction factors at the TCP.  The above
mentioned identification of the borderline-critical-dimension was thus
supported. Quantitative results came also from studies of the
Landau-Ginzburg model of Eq. (\ref{Hlandginz}) in $d=3-\epsilon$ by
forming $\epsilon-$ expansion approximations of the exponents at the
TCP and from position-space RG studies of the $S=1$ BC model
\cite{berkwor,adler} on the $sq$ lattice.

In the case of spin $S=3/2$, the first approximate RG
computations\cite{oliv} failed to confirm the MF prediction concerning
the absence of a TCP, but these difficulties were overcome in a more
recent\cite{tensor} approach.

\subsection{Structure of the Series Expansions}

For all values of the spin $S$, the HT expansion of the
dimensionless free-energy per site of the BC model on a lattice of $N$ sites
 can be written in the thermodynamical limit as 
\begin{equation}
 -(K/J) f_{HT}(K,D,h;S)=\lim_{N \to \infty} (1/N)  {\rm ln}Z_N(K,D,h;S)=
V_0(D,h; S)+\sum_{n=1}^{\infty} g_n(D,h;S) K^n
\label{HTexpansion}
\end{equation}
with $Z_N(K,D,h;S)$ the partition function for the Hamiltonian of
Eq. (\ref{hamBC})
 \begin{equation}
 V_0(D,h; S) = -D + h + {\rm ln}(\sum_{s=-S}^S x^{S^2-s^2} \mu^{S-s})
 \label{V0}
 \end{equation}
 and 
\begin{eqnarray}
\mu = exp(-\frac{h}{S})  \qquad \qquad x = exp(\frac{D}{S^2})
 \label{ws}
 \end{eqnarray}

 Using the HT linked-cluster
expansion\cite{hardsquares1,wortis,bp1,bp2,bp3} method, the
coefficients $g_n(D,h;S)$ of the expansion in powers of $K$ can be
computed {\it exactly in $D/S^2$ and $h/S$} as multivariate
polynomials in appropriately defined ``vertex functions''
$V_i(D,h;S)$, with rational coefficients depending on the lattice
structure and dimension.

  If the dependence of the HT series on the magnetic field is
 analytically known, the LT expansions can be derived by a
 transformation of variables and  a direct
 graphical analysis is not necessary.

The structure of the LT expansion for the free-energy density is 
\begin{equation}
 -(K/J)f_{LT}(u,D,h;S)=\frac {1}{2} qK + h -D  +\sum_{n=1}^{\infty} 
L_n(u,x;S)\mu^n.
\label{LTexpansion}
\end{equation} 

Here $u=\exp(-K/S^2)$, while $\mu$ plays a role of high-field
  expansion variable.  For large magnetic fields, i.e. around $\mu=0$,
  the quantity $-\frac{K} {J} f_{LT} - h$ is a convergent series in
  $\mu$, with expansion coefficients $L_k(u,x;S)$ that are polynomials
  in $u$ and $x$.  Below the critical line, this expansion is
  convergent also for $\mu=1$, hence the name LT expansion.

From our series, either at HT or at LT, we can obtain all ``mixed
susceptibilities'', i.e. the derivatives of the free-energy density
with respect to the ordering field $h$ and/or to the ``non-ordering''
field $D$.
\begin{eqnarray}
\chi_{(r;p)}(K,D,h;S)= (-1)^{p+1} \frac {K}
{J}S^{r+2p}\frac{\partial^{r+p}f(K,D,h;S)} {\partial h^r \partial D^p}
= \sum_{i_2,...,i_{r+p}} \langle s_0s_{i_2}...s_{i_r}
s^2_{i_{r+1}}...s^2_{i_{r+p}} \rangle_c
\label{highsus}
\end{eqnarray}

In particular for $S=1$, the density $X(K,D;1)$ conjugated
to the reduced  crystal-field, namely the concentration, is
\begin{equation}
X(K,D;1)=1 - \frac {K} {J} \frac{\partial
f(K,D;1)}{\partial D}= 1-\chi_{(0;1)}(K,D;1)=1-\langle s_0^2 \rangle
\label{xdens}
\end{equation}

The response function with respect to $\Delta$, namely the concentration
susceptibility, is the $D$-derivative of $X(K,D;1)$
\begin{equation}
  Y(K,D;1) = \frac{\partial X(K,D;1)}{\partial D}=
-\frac {K} {J} \frac{\partial^2 f(K,D;1)}{\partial D^2}=  
\chi_{(0;2)}(K,D;1) = \sum_i \langle s^2_0 s^2_i \rangle_c 
\label{yrf}
\end{equation}

For $H=0$, only the susceptibilities $\chi_{(r;p)}(K,D,h;S)$ of even
order $r$ are non-trivial at HT, whereas all field-derivatives are
non-trivial at LT, in particular the specific magnetization $
M(K,D,h;S)=\chi_{(1;0)}(K,D,h;S)$.  For brevity, we shall often adopt
the notations $ f(K,D;S) \equiv f(K,D,h;S)|_{h=0} $ and $
\chi_{(r;p)}(K,D;S) \equiv \chi_{(r;p)}(K,D,h;S)|_{h=0} $
($r=1,2,...$), simply dropping the $h$-dependence.

\subsection{The computation of the LT expansions} 

The MF approximation indicates that for spin $S \ge 1$, 
  $[\frac{2S+1}{2}]$ ordered phases appear in the LT region.  The
 phase in which at $T=0$ all spins take the value $s_i = S$, can be
 described by a LT expansion.

The HT and LT expansions of the free energy must match in the region 
in which $h \to \infty$ and
$K \to 0$, so that the following equation is valid in a vicinity of
$\mu = 0$, $u=1$
\begin{equation}
K \frac{q}{2} + \sum_{k \ge 1} L_k(u,x;S) \mu^k = 
{\rm ln}(\sum_{s=-S}^{S} x^{S^2-s^2} \mu^{S-s}) + 
\sum_{i\ge 1} g_i(h(\mu), D(x))K(u)^i
\label{htlt1}
\end{equation}

Considering $K$ as a function of $z=1-u$,
from Eq. (\ref{htlt1}) it follows that, for $n \ge 1$,
\begin{equation}
L_n(u, x;S) = [\mu^n]\Big ({\rm ln}(\sum_{s=-S}^{S} x^{S^2-s^2} \mu^{S-s}) + 
  \sum_{i \ge 1} g_i(h(\mu), D(x)) K(z)^i |_{z=1-u}\Big )
\label{htlt2}
\end{equation}
here $[x^n]v$ is defined as the coefficient of the $x^n$ term in the
expansion of the function $v(x)$ in powers of $x$.

Since $K(z) = O(z)$ and $L_n$ has order $Sqn$ in $u$, 
only the HT coefficients $g_i$ with $i \le Sqn$
contribute to $L_n$, in Eq. (\ref{htlt2}) so that the polynomials $L_n$ can be
computed from the HT expansion through the order $Sqn$.

A few more $L_n$ can be computed from the same HT expansion, 
using the fact that $L_n$ has a zero of order $b_n$ in $u$.
For any integer exponent $0 \le b \le b_n$, $u^{-b}L_n(u,x;S)$
is a polynomial in
$u$, and hence in $z$, of degree $Sqn-b$, so that 
\begin{equation}
u^{-b}L_n(u,x;S) = {\cal T}_{Sqn - b}\Big \{(1-z)^{-b}[\mu^n]\Big(
 {\rm ln}(\sum_{s=-S}^{S} x^{S^2-s^2} \mu^{S-s}) + \sum_{i \ge 1} g_i(h(\mu),
 D(x)) K(z)^i|_{z=1-u}\Big ) \Big \}
\label{htlt3a}
\end{equation}

Here the series truncation operator is defined by 
${\cal T}_m F = \sum_{k=0}^m z^k \frac{d^k F}{d z^k}|_{z=0} $ with 
 $F$ is a series in the variable $z$.

Choosing $b=b_n$ is the most efficient way to get $L_n$ from Eq. (\ref{htlt3a}).
If $b_n$ is not known,  the RHS of Eq. (\ref{htlt3a}) has to  be computed 
for some $b \ge b_{n-1}$,
using the heuristic observation that $b_{n} \ge b_{n-1}$.
Since the lowest power in $u$ of $u^{-b}L_n$
 must be $b_n-b$, the presence of a multiple zero in $u=0$ in the RHS of
 Eq. (\ref{htlt3a}) indicates that the result is correct.

In the case of spin $S=1$, the $b_n$ are
known\cite{hardsquares2,foxgutt, foxgaunt} for $D=0$, namely for the
Ising model with spin $S=1$. Thus, we can take $b=b_n$ and derive the
$L_n$ through $n=11$ for the $sq$ and $sc$ lattices, and through
$n=10$ in the $bcc$ case.  For spin $S=3/2$, in the case of the $sq$,
$sc$ and $bcc$ lattices, we can compare our results with the
polynomials $L_n$ obtained in Ref. [\onlinecite{saulferer}] through
$n= 7$.  For $n = 8, 9$, we have used in Eq. (\ref{htlt3a}) values of
$b$ such that there are at least zeros of multiplicity $4$ in the
above procedure, so that we are confident that the expressions
obtained for $L_8$ and $L_9$ are correct.

  The arrangement of the LT expansion as a series of powers of $\mu$
with coefficients  polynomial  in $u$ and $x$, is called ``field
grouping''.  However the LT series can be as well thought of as an
expansion in powers of $u$, with coefficients that are polynomials in
$x$ and $\mu$ (``temperature grouping'').  In numerical use at
moderate orders, the approximations obtained in the two cases may show
some difference.

 As an example of the results of this procedure, for any lattice and
 spin the two lowest-order LT polynomials $L_n(u, x;S)$ are
\begin{equation}
L_1(u, x;S) = u^{qS} x^{2S-1}
\end{equation}
\begin{equation}
L_2(u, x;S) = u^{2qS} x^{4S-4} + (\frac{q}{2}u^{2qS-1} - \frac{q+1}{2} u^{2qS})
x^{4S-2}
\end{equation}

Some properties of the LT polynomials thus obtained provide a useful
 partial check of the computation. For example, in the case of spin
 $S=1$, the polynomials have the structure
 $L_n(u,x;1)=\Sigma^n_{m=0}L_{n,m}(u)x^m$ where the quantities
 $L_{n,m}(u)$ are nonvanishing  only if $n$ and $m$
 are both even or both odd\cite{SWS}.  Moreover, as $D \to -\infty$ the
 polynomials  $L_{2n+1}(u,x;1) \to 0$, while the $L_{2n}(u,x;1)$ reduce
 to the  polynomials $L_{n}$ of the Ising model with $S=1/2$ on the
 same lattice i.e.  $L_{2n}(u,x;1) \to L_{n}(u^{1/4};1/2) $.
 Similar properties are valid in the case of spin $S=3/2$. 

 In
 addition, in the $S=1$ case, dividing out the highest power of $x^n$
 in $L_n(u,x;1)$ and taking the $D \to \infty $ limit, a polynomial
 in $u$ is obtained containing the powers $u^{6n}$, $u^{6n-1}$,
 $u^{6n-2}$, ..., whose coefficients reproduce orderly the
 coefficients of $u^{3n}$, $u^{3n-1}$ ... of the polynomials $
 L_{n}(u;1/2)$ of the Ising model with $S=1/2$ on the same lattice. An
 analogous property is observed for $S=3/2$.

 As an example of our results for the LT expansions for the BC model,
in the Table \ref{tab5a} we have reported the first 11 LT polynomials
for the $sc$ lattice in the case of spin $S=1$.

\subsection{The computation of
 the HT expansions} The HT expansion coefficients $g_n(D,h;S)$
 Eq. (\ref{HTexpansion}), are polynomials with rational
 coefficients\cite{wortis} in the bare vertices $V_j(D,h;S) = \frac
 {d^j} {dh^j} V_0(D,h;S)$ (for $j \ge 1 $),  that are the successive
 $h$-derivatives of the vertex generating-function $V_0(D,h;S)$
 defined by Eq. (\ref{V0}).

They  are evidently independent of the lattice structure and
dimensionality, but depend  on the value $S$ of the spin.
The coefficients $g_n$ can be computed directly by
the ``unrenormalized linked-cluster
expansion''\cite{wortis}. 
More efficiently, the quantities $\frac{d g_i}{d h}$ can be derived
using the ``renormalized linked-cluster expansion''. By integration
with respect to $h$, the polynomials $g_n$ in the bare vertices
are recovered and it can be proved that the integration constant vanishes.
The bare vertices
are regular series in $\mu$ and this implies that the same property is valid
 for the coefficients $g_n(h(\mu), D(x))$.

\subsection{ The vertex functions for spin $S=1$}

For spin $S=1$, all vertex functions with $n>0$
 can also be expressed\cite{wang}
as polynomials in  {\it two} auxiliary functions
\begin{eqnarray}
{\rm A}(D,h;1) = \frac{1-\mu^2} {1+x\mu+\mu^2} \qquad \qquad
{\rm B}(D,h;1)= \frac{1 + \mu^2} {1+x\mu+\mu^2} 
\label{gs}
\end{eqnarray}

For  $g_0(D,h;1) $, we have
\begin{equation} 
g_0(D,h;1)=V_0(D,h;1) ={\rm  ln}[1+2\exp(-D)\cosh(h)]=
-{\rm ln}(1 - {\rm B}(D,h;1))
\label{gen1}
\end{equation}

The HT expansion coefficients of the free-energy
density $g_n(D,h;1)$ with $n>0$ can be  rewritten more simply as
bivariate polynomials (with rational coefficients) in ${\rm A}(D,h;1)$ and $
B(D,h;1)$.
 This property follows from the equations

\begin{eqnarray}
\frac{\partial {\rm A}(D,h;1)}{\partial h} 
= {\rm B} - {\rm A}^2   \qquad \qquad
\frac{\partial {\rm B}(D,h;1)}{\partial h} = {\rm A} - {\rm A} {\rm B}
\end{eqnarray}

\begin{eqnarray}
\frac{\partial {\rm A}(D,h;1)}{\partial D} 
= {\rm A} {\rm B} -{\rm A} \qquad \qquad
 \frac{\partial {\rm B}(D,h;1)}{\partial D}= {\rm B}^2 -{\rm B}
\end{eqnarray}
It is also useful to remember that
\begin{eqnarray}
\frac{\partial V_0(D,h;1)}{\partial h} = {\rm A}(D,h;1), \qquad
\frac{\partial V_0(D,h;1)}{\partial D} = -{\rm B}(D,h;1)
\end{eqnarray}

{\rm A} similar remark applies to the expansion coefficients of  the
higher (mixed) susceptibilities $\chi_{(r;p)}(K,D,h;1)$.

From Eq. (\ref{gen1}), it is clear that, for $D=0$ (i.e.  for $x=1$)
the coefficients $g_n(D,h;1)$ reduce to those of an Ising system with
$S=1$ in a magnetic field. On the other hand, in the limit $D \to
-\infty$, the $S=0$ state is suppressed and we have ${\rm B} \to 1$,
${\rm A} \to {\rm tanh}(h) $, so that the expansion coefficients
reduce essentially to those of an $S=1/2$ Ising system in a field. For
$h=0$, we have $T_c(D \to -\infty;1)=T_c(1/2)$, where $T_c(1/2)$
denotes the critical temperature of the spin $S=1/2$ system.

For $h=0$, the auxiliary function ${\rm A}(D,0;1)$ vanishes and the
coefficients $g_n(D,0;1)$ reduce to polynomials in the single
variable\cite{SWS}
\begin{equation}
    \tau \equiv {\rm B}(D,0;1)=1/(1+x/2)
    \label{tau}
\end{equation}
that  coincides with the variable $\tau$ defined in
Refs.[\onlinecite{SWS,Oitmaa1,Oitmaa2}].
Notice finally that  in the limit $D \to \infty$,
 the spin $S=\pm 1$ states are suppressed.

\subsection{ The vertex functions for spin  $S=3/2$}

The field-derivatives of the vertex-generating function $V_0(D,h;3/2)$
can be expressed as polynomials in the {\it three} auxiliary functions
\begin{equation}
{\rm A}(D,h;3/2)=\frac{1-\mu^3}{1 + x^2(\mu + \mu^2) + \mu^3}
\label{abc1}
\end{equation}
\begin{equation}
{\rm B}(D,h;3/2)=\frac{1 + x^2(\mu-\mu^2) - \mu^3}{1 + x^2(\mu + \mu^2) + \mu^3}
\label{abc2}
\end{equation}
\begin{equation}
{\rm C}(D,h;3/2)=\frac{1+\mu^3}{1 + x^2(\mu + \mu^2) + \mu^3}
\label{abc3}
\end{equation}

The expansion coefficients 
$g_n(D,h;3/2)$ (with $n>0$) are polynomials 
in ${\rm A} , {\rm B}$ and ${\rm C}$.
 For $g_0(D,h,3/2)$, we have
\begin{equation} 
g_0(D,h;3/2)=V_0(D,h;3/2) ={\rm ln}2-D-h -{\rm ln} \Big ({\rm C}(D,h;3/2)-
{\rm A}(D,h;3/2) \Big )
\label{gen32}
\end{equation}
Similar remarks as for the spin $S=1$ case are valid also for the
 derivatives of ${\rm A}$, ${\rm B}$ and ${\rm C}$ with respect to $D$
 and $h$, so that the expansions of the mixed susceptibilities are
 polynomials in these variables.  This property follows from the
 equations
\begin{equation}
\frac{\partial {\rm A}(D,h;3/2)}{\partial
  h}=\frac{2}{3}(\frac{3}{2}{\rm C}-{\rm A}^2 -\frac{{\rm A}{\rm B}}{2})
\end{equation}
\begin{equation} 
 \frac{\partial {\rm B}(D,h;3/2)}{\partial h}=\frac{2}{3}(\frac{1}{2}+{\rm C}
-{\rm A}{\rm B}-\frac{{\rm B}^2}{2})
\end{equation}
\begin{equation} 
  \frac{\partial {\rm C}(D,h;3/2)}{\partial h}
=\frac{2}{3}(\frac{{3\rm A}}{2}-{\rm A}{\rm C}-\frac{{\rm B}{\rm C}}{2})
\end{equation}

\begin{equation}
\frac{\partial {\rm A}(D,h;3/2)}{\partial D}=\frac{8}{9}(-{\rm A}
+{\rm A}{\rm C})
\end{equation}
\begin{equation} 
 \frac{\partial {\rm B}(D,h;3/2)}{\partial D}=\frac{8}{9}(-{\rm A}
+{\rm B}{\rm C})
\end{equation}
\begin{equation} 
  \frac{\partial {\rm C}(D,h;3/2)}{\partial D}
=\frac{8}{9}(-{\rm C}+{\rm C}^2)
\end{equation}

and

\begin{equation}
\frac{\partial V_0(D,h;3/2)}{\partial h} = \frac{1}{3}(2{\rm A} + {\rm
B}) \qquad \frac{\partial V_0(D,h;3/2)}{\partial D} = -\frac{1}{9}(1
+ 8{\rm C})
\end{equation}

The polynomial in the variables ${\rm A}$, ${\rm B}$ and ${\rm C}$
representing the $n$th coefficient of the expansion of
$\chi_{(r;p)}(K,D,h;3/2)$ in powers of $K$, has order $ 2n+r+p$,
irrespectively of the lattice dimensionality and structure.

 From Eqs. (\ref{abc1}-\ref{abc3}), it follows that in the limit $D
\to -\infty$, the set of the auxiliary functions simplifies (${\rm
A}={\rm B}$ and ${\rm C} \to 1$) and the $ s= \pm 1/2$ states are
suppressed.  Analogously the $ s= \pm 3/2$ states are suppressed in
the $D \to +\infty$ limit. Thus in both limits, the series expansions
essentially reduce to those of the spin $S=1/2$ Ising model in a
field.

For $h=0$, we have
$T_c(D=-\infty;3/2)=T_c(1/2)=9T_c(D=+\infty;3/2)$. For $D=0$, the
series reduce to those of an Ising system with spin $S=3/2$ in a
field.  For $h=0$, the auxiliary functions $A(D,0;3/2)$ and
${\rm B}(D,0;3/2)$ vanish and the coefficients $g_n(D,h;3/2)$ become
polynomials in the single variable
\begin{equation}
\tilde \tau \equiv {\rm C}(D,0;3/2)=1/(1+x^2). 
\label{tildetau}
\end{equation}

Let us add that for higher spin values, the HT coefficients can be
expressed polynomially in terms of larger sets of auxiliary functions
and simplifications analogous to those indicated above  still
occur in the same limits of $D$ and $h$.  These properties can be used
for a partial (but non-trivial) check of the correctness of the series
derivation.  In particular, for all half-odd spin systems, the series
expansions essentially reduce to those of the spin $S=1/2$ Ising model
in a field, in both limits $D \to \pm \infty$.

To give an idea of the structure of the HT expansions that we have
computed, we have shown in Table \ref{tab1} the first nine
coefficients $g_n(D,h;1)$ of the free-energy expansion for the $sc$
lattice in the spin $S=1$ case. When $h \ne 0$, the complexity of the
polynomial structure of $g_n(D,h;1)$ in $\rm A$ and ${\rm B}$ increases so
rapidly with the order of expansion, that the set of series
coefficients becomes very cumbersome beyond the 9th order. The
expression of the 10th-order HT coefficient is as long as the whole
set of the preceding ones shown in the table. Therefore an extensive
tabulation of our series data does not fit the format of this paper
and requires a separate report\cite{arXiv}.

Our derivation\cite{bp1,bp2,bp3} of the HT expansions for any value of
the spin, was made possible by an efficient coding of the
linked-cluster\cite{hardsquares1,hardsquares2,wortis} graphical
computation algorithms and by an extensive use of the symbolic
manipulation softwares $Sagemath$\cite{sage} and $Python$. At the
order of expansion we have reached, the performance of the current
personal computers is still adequate to face the exponential growth of
the computational complexity with the order of series expansion.

No software for symbolic manipulation and multiprecision arithmetic
was available at the time of the pioneering series-study of
Ref.[\onlinecite{SWS}], so that an exact calculation of the expansion
coefficients (always rational numbers) as functions of $D$ and $h$ was
not easy.  Thus in these studies, approximate numerical
procedures giving rise to sizable rounding errors, even at
relatively low orders, were used\cite{SWS}  to evaluate the series
coefficients. On the contrary, in our work this source of error is
eliminated by an extensive use of the softwares for exact symbolic
computation.

\subsection{The critical behaviors of the main observables}

If we define $t(K,D;S) \equiv (K_c(D;S)/K -1)$, the critical behaviors
  as $K \rightarrow K_c(D;S)$ at fixed $D$, of the mixed
  susceptibilities in zero magnetic field, expected from the scaling
  laws, are
\begin{equation}
\chi_{(r;p)}(K,D;S) \approx A_{(r;p)}(D;S)
|t(K,D;S)|^{-\gamma^{(r;p)}(D;S)}\Big[1+a_{(r;p)}(D;S) t^{\theta}...\Big]
\label{chi2nas}
\end{equation}
  Here $\gamma^{(r;p)}(D;S)$ denotes the leading critical exponent of
the mixed susceptibility of order $(r;p)$, while $\theta$ is the
exponent of the leading correction-to-scaling. A priori,
$\gamma^{(r;p)}(D;S)$ might depend on $D$ and $S$, but to keep
notation simpler we shall often drop this dependence.  The values of
the critical amplitudes $A_{(r;p)}(D;S)$ and $a_{(r;p)}(D;S)$ obtained
approaching the critical line from the HT side will differ from those
obtained on the LT side.  We might distinguish them by appropriate
superscripts, which however we shall omit to avoid overloading the
notation, whenever it is clear from the context which is the relevant
limit.

For $d\le 4$
(or $d \le 3$ in the case of the TCP),  the hyperscaling relation is
\begin{equation}
\gamma^{(r;p)} =p+\gamma^{(2;0)}+(r-2)\hat \Delta
\label{hyperscal}
\end{equation}
with $\hat \Delta=\beta+\gamma^{(2;0)}$.

The asymptotic form\cite{RW,riede1,riede2,riedel2}
 on the tricritical path, i.e. as $T \to T_{tr}$ at fixed $D = D_{tr}$,  
of the density $X(K,D;S)$ conjugated
to the crystal field, is
\begin{equation}
X(K,D;S) \approx X_c(D_{tr};S) +
A_X (D_{tr};S)|t(K,D_{tr};S)|^{\omega}
\label{xas}
\end{equation}
Eq. (\ref{scalingp})
 implies
 $\omega=1-\alpha_{tr}$. 
In the same limit, the
 concentration susceptibility, namely the derivative of $X(K,D;S)$
 with respect to $D$, behaves\cite{riede1,riede2,riedel2} as
\begin{equation}
  Y(K,D;S) 
\approx A_Y(D_{tr};S) |t(K,D_{tr};S)|^{-\lambda}
\label{yas}
\end{equation}

with $\lambda=\alpha_{tr}$.

  The critical parameters defined by the asymptotic form
Eq. (\ref{chi2nas})) are calculable by series extrapolations and
depend on the lattice dimension $d$ and (as far as the amplitudes are
concerned) on the lattice structure, although we have not explicitly
indicated this fact.  The critical amplitudes $A_{(r;p)}(D;S)$ of the
susceptibilities $\chi_{(r;p)}(K,D;S)$ with $r=1,2,..$, can be
obtained by forming the expansions of the {\it effective amplitudes}
biased with the estimated values of $K_c(D;S)$ and the expected values
of $\gamma^{(r;p)}$
\begin{equation}
A^{eff}_{(r;p)}(K,D;S)= |t(K,D;S)|^{\gamma^{(r;p)}} \chi_{(r;p)}(K,D;S)
\label{ampef}
\end{equation}
 and extrapolating them to $K=K_c(D;S)$ from the appropriate side of
 the critical line, namely $A_{(r;p)}(D;S)=A^{eff}_{(r;p)}(K_c;D;S)$.

 The validity of the universality property along all the critical line
for the main and the correction exponents, (while the single
amplitudes are non-universal), can be confirmed by simply checking
that they keep the Ising model values expected for the given lattice
dimensionality.

All over the  critical line, we can also evaluate a few
 appropriate ratios of higher susceptibilities, expected to be
 universal, such as the lowest order terms in the sequences ${\cal
 I}^+_{2r+4}(D;S)$ and ${\cal J}^+_{2r+4}(D;S)$
defined\cite{watson} by
\begin{equation}
{\cal I}^+_{2r+4}(D;S) =\lim_{K \to K^-_c}\frac{\chi_{(2;0)}(K,D;S)^r
\chi_{(2r+4;0)}(K,D;S)} {\chi_{(4;0)}(K,D;S)^{r+1}}=\frac{A_{(2;0)}(D;S)^r
A_{(2r+4;0)}(D;S)} {A_{(4;0)}(D;S)^{r+1}}
\label{Ii} 
\end{equation}
\begin{equation} 
{\cal J}^+_{2r+4}(D;S)=\lim_{K \to K^-_c}\frac{\chi_{(2r;0)}(K,D;S) 
\chi_{(2r+4;0)}(K,D;S)}{\chi_{(2r+2;0)}(K,D;S)^2}
=\frac{A_{(2r;0)}(D;S)A_{(2r+4;0)}(D;S)} {A_{(2r+2;0)}(D;S)^2} 
\label{Ai} 
\end{equation}
 for $r > 0$.  ( Observe that for $r=1$, we have ${\cal I}^+_{6}
 \equiv {\cal J}^+_{6}$).  These amplitude ratios, along with
 additional ones involving also the susceptibilities
 $\chi_{(r;0)}(K;D;S)$ with odd indices, can also be studied for $H=0$
 on the LT side of the critical point, i.e. in the limit $K
 \rightarrow K_c(D;S)^+$.  Analogous ratios can be defined in terms of
 the susceptibilities with $p>0$. Of course, they do not reduce to
 known Ising quantities for $D=0$.

\section{Numerical analysis of the  expansions}

In the following subsections, we shall make a general discussion of
 our analyses of the HT and LT series, valid for all lattices under
 study. 

\subsection{The variables}

 As already observed, in the $H=0$ plane, the HT series analyses are
 performed in terms of the ``natural'' variables $\tau$ (or
 equivalently $D$) and $K$, namely along lines of
 constant $\tau$ in the $(\tau, \tilde T)$ plane.  When comparing our
 estimates of the critical phase-boundaries with those from simulation
 methods, it should be remarked that the simulations are carried out
 at fixed $\Delta$ (or $T$) so that the uncertainties (reported in
 Tabs. \ref{tab2}, \ref{tab2c} and \ref{tab32}) affect only the corresponding $T_c$
 (or $\Delta_c$), while our results are obtained at fixed $D$ (or
 $\tau$) so that uncertainties should affect both $T_c$ and 
 $\Delta_c=DT_c$. However, for convenience in the comparison with the
 estimates from other sources, we can give either results at fixed
 $\Delta$ and shift the uncertainty in this variable into sufficiently
 enlarged error bars for $T_c$ or results at fixed $T$.  Analogous
 remarks apply for the LT series.

\begin{figure}[p]
\begin{center}
\leavevmode
\includegraphics[width=5.00 in]{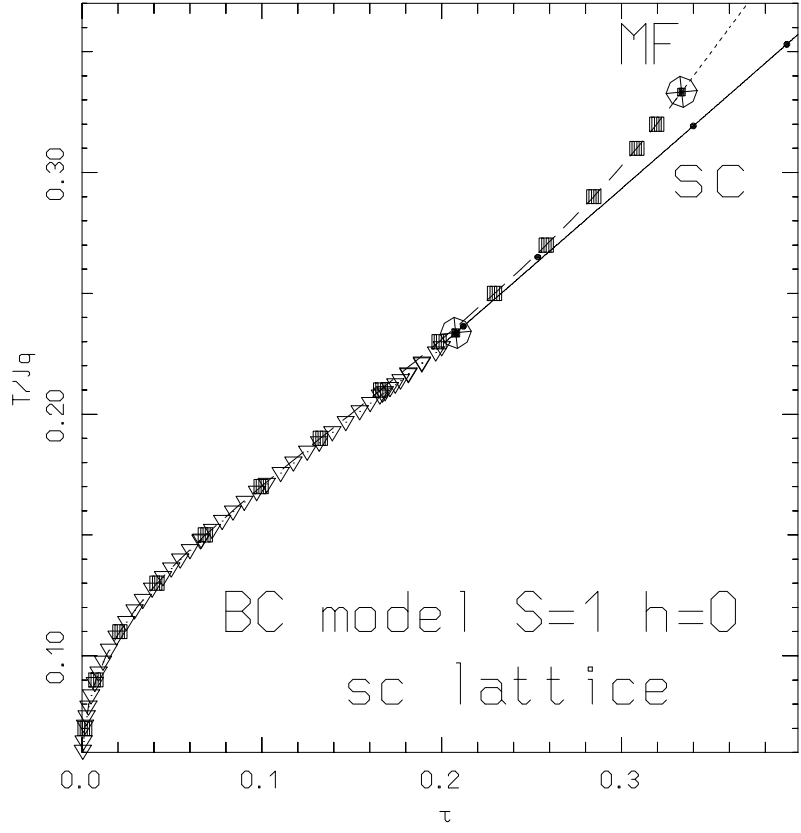}
\caption{ \label{fig_TF_sc_tau} BC model with spin $S=1$, on the $sc$
 lattice for $h=0$.  Phase diagram in the ($\tau$, $ T/Jq$) plane by
 series and in the MF approximation. Big crossed open circles are the
 TCPs from MF and a simulation\cite{deng}.  The upper short-dashed
 line is the MF critical phase-contour and full squares connected by a
 long-dashed line are the MF first-order boundary.  The solid curve
 shows the critical phase-boundary from series.  Open triangles are
 the first-order part of the phase-boundary from intersections of the
 LT and HT expansions of the free energy.}
\end{center}
\end{figure}

\begin{figure}[p]
\begin{center}
\leavevmode
\includegraphics[width=5.00 in]{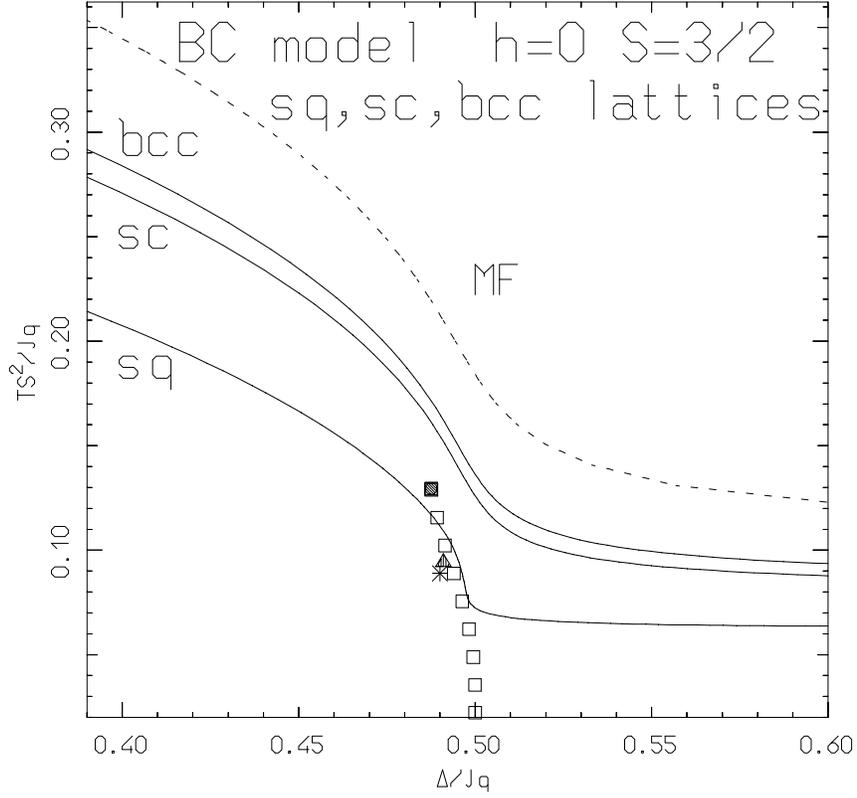}
\caption{ \label{figura_fase_32_Ing} BC model with spin $S=3/2$ in the
 $h=0$ plane. Phase diagrams in the anisotropy-temperature plane for
 the $sq$,$sc$ and $bcc$ lattices.  
    The MF critical border is the
 upper dashed-line.  The MF first-order contour is a sequence of open
 squares ending with a full square and not joining the critical
 border. The lowermost solid line is the critical phase-boundary for the
 $sq$ lattice, the intermediate one for the $sc$ lattice and the
 uppermost one for the $bcc$ lattice.  There is no evidence of a TCP
 for these lattices.  A star is the end-point\cite{xavier} of a 
 first-order line rooted at the point $(\tilde \Delta=1/2, \tilde T=0)$ for the
 $sq$ lattice, and similarly a black triangle for the $sc$
 lattice\cite{grollau32}.  }
\end{center}
\end{figure}

For the systems under scrutiny, we shall map out the phase-diagrams in
 the $(\tilde \Delta, \tilde T)$ plane (or in the $(\tau, \tilde T)$
 plane) and in most cases shall obtain the critical phase-contour from the
 HT expansion of the ordinary susceptibility. 
Then we shall be able also to estimate the exponents and
 the critical amplitudes of a few other susceptibilities (including
 those of higher (mixed) orders  $\chi_{(2r;p)}(K,D;S)$.  Our {\it
 unbiased} HT series estimates for the $sq$, the $sc$ and the $bcc$
 lattices in the case of spin $S=1$, produce the phase diagrams of
 Figs. \ref{Graf_fig3_first_sq}, \ref{Graf_fig3_first_sc}, and
 \ref{Graf_fig3_first_bcc} respectively.  For comparison, in all these
 figures we have also drawn the phase-boundary in the MF
 approximation, which is independent of the lattice dimension and
 structure, due to the choice of the plotting variables. Generally the MF
 approximation is qualitatively correct but, for a given $D$, it
 leads to a phase-contour systematically higher in temperature than
 those from the series or other approximation methods.  The phase
 diagrams in the $(\tau, \tilde T)$ plane, are particularly
 suggestive, not only because for spin $S=1$, the MF critical
 phase-contour reduces exactly (see the Appendix A) to the straight
 line between the points $(1/3,1/3)$ and $(1,1)$, but also because 
  the critical boundaries computed by series remain nearly
 straight lines although shifted to lower temperatures with respect to the
 MF approximation and with smaller slopes, except very near the TCP.
 As an example, we shall show only Fig. \ref{fig_TF_sc_tau} for the
 $sc$ lattice.  The phase-contours in the $(\tau,\tilde T)$ plane
 obtained for the $sq$ and the $bcc$ lattices are completely similar
 and are not reported for brevity.  For the spin $S=3/2$ systems the
 MF critical boundary in the $(\tilde \tau,\tilde T)$ plane is the
 straight line $\tilde T= 8\tilde \tau/9 +1/9$ with $0 \leq \tilde \tau
 \leq 1$  and also the phase-contours obtained from the series for the
 $sq$, $sc$ and $bcc$ lattices do not show a pronounced curvature
 except in the range $ 0 \le \tilde \tau \lesssim 0.1$ in which the
 infinite region $\tilde \Delta \gtrsim 0.4$ is mapped. For brevity,
 we shall generally show only the phase diagram in the $(\tilde
 \Delta,\tilde T)$ plane (see Fig. \ref{figura_fase_32_Ing}).

 For each value of $D$, our unbiased estimates of the
 points of the critical boundaries are obtained fitting the asymptotic
 form Eq. (\ref{ansatz}) to the last few terms of the MRA
 estimator-sequences introduced in Appendix C and formed in most cases
  with the HT
 expansions of the  susceptibility.  As a rule, we have
 extrapolated only the estimator-sequences that appear to have settled
 down in their expected asymptotic forms (see the Appendix C), always
 after making sure that the final estimates are consistent with those
 from unbiased second-order DAs.

\subsection{The LT expansions}

In the analysis of the LT expansions, both
simple ratio-methods and their extensions\cite{guttda,zinnmra}, such as the
MRAs introduced in the Appendix C, are unfit to locate the critical
points and estimate the exponents, because  generally the large-order
behavior of the LT expansion coefficients is dominated by nonphysical
complex singularities closer to the origin than the critical
singularity. For these analyses, only the PA and DA techniques are
useful.

  All methods described in the Appendix C fail to map the first-order
 lines which are associated with singularities too weak to be detected
 by the MRAs, PAs or DAs.  In the study of the spin $S=1$ BC model on the $fcc$
 lattice, the LT series have been employed\cite{SWS} jointly with the
 HT series to map the first-order part of the phase-boundary. It is
 expected that the free-energy is continuous across the
 phase-boundary and that also its temperature- (or field-)
 derivatives are continuous across the second-order line, while they
 are discontinuous across the first-order line. 
Thus, at fixed $D$, the values of the free-energy computed by HT and
LT expansions should intersect, {\it if} these expansions share a
common region of approximate numerical validity. This makes it possible to
 locate the first-order phase-boundary. 
 This
intersection-method is controversial\cite{baxterenting1,baxterenting2}
to some extent since the features of the singularity associated to the
first-order transition are not known in detail and anyway one is
evaluating the two series near the borders of their convergence regions.
Using the intersection method, whenever we have observed that the
curve of the LT free-energy shows two nearby intersections with the HT
curve, as a rule we have chosen the lowest temperature one.

 We should finally add that in our case this procedure unfortunately
 does not lead to reliable results in some small range of temperatures
 $T \lesssim T_{tr}$, but otherwise it appears to work reasonably
 well.  The method can be applied also for $T > T_{tr}$, by looking
 for tangency points, (since the free energy is continuous across the
 second-order line) instead of intersections.  The resulting critical
 phase-boundary generally lies at a temperature slightly smaller than
 that determined by the MRAs of the HT susceptibility, as it was
 already observed\cite{SWS} in the $fcc$ lattice analysis.

\begin{figure}[p]
\begin{center}
\leavevmode
\includegraphics[width=5.00 in]{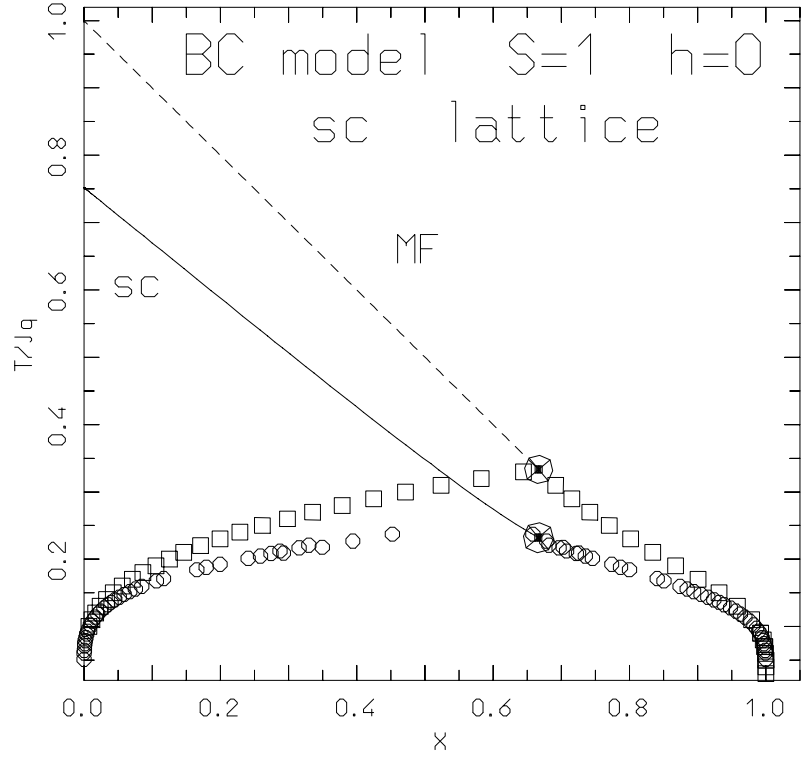}
\caption{ \label{TcscvsX} BC model with spin $S=1$ for $h=0$ on the
  $sc$ lattice.  Phase diagram in the concentration-temperature plane.
  The dashed line is the MF approximation of
  the critical phase-contour. Sequences of open squares are the two
  branches of the phase-contour associated with the HT and the LT
  sides of the MF first-order line.  The open circles are the
  corresponding branches from the series. Big crossed circles are
  TCPs.  }
\end{center}
\end{figure}

\begin{figure}[p]
\begin{center}
\leavevmode
\includegraphics[width=5.00 in]{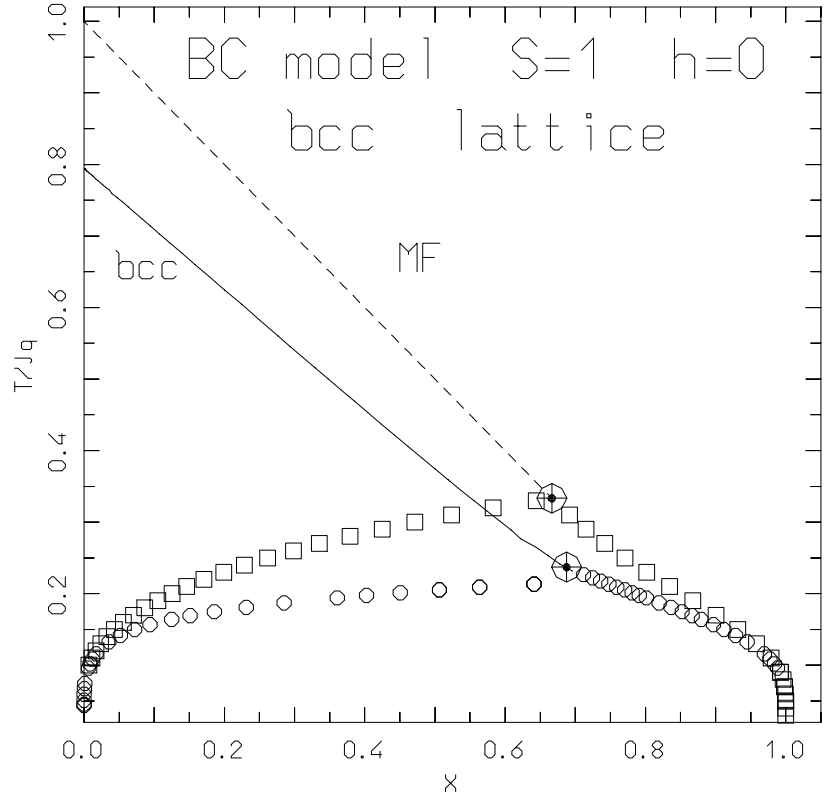}
\caption{ \label{TcbccvsX} BC model with spin $S=1$ on the $bcc$
 lattice for $h=0$.  The phase diagram in the concentration-temperature plane,
 computed by series and in the MF approximation.  The dashed line is
 the MF approximation of the critical border, while sequences of
 open squares are  the HT and the LT branches of the first-order
 contour. The open circles are the two branches of this line
 computed by series. Big crossed circles are TCPs.  }
\end{center}
\end{figure}

In Figs. \ref{TcsqvsX}, \ref{TcscvsX}, \ref{TcbccvsX}, the phase
  diagram is shown at $h=0$, in the concentration-temperature plane
  $(X,\tilde T)$ for the $sq$, the $sc$ and the $bcc$ lattices, in the
  case of spin $S=1$. These critical phase-contours are obtained
  evaluating the expansions of $X(K,D;S) $ along the curves
  $T_c=T_c(D;S)$ that represent the critical phase-boundary in the
  anisotropy-temperature plane as obtained from the HT susceptibility
  expansions, while the first order branches are determined using the
  intersections of the HT-LT free-energy expansions (and/or the
  simulation data when they exist, as in $2d$).

\subsection{Determination of the exponents along the critical boundaries }

  It is expected that by universality, along the second-order part of
the phase-boundary the exponents of the ordinary susceptibility
$\chi_{(2;0)}(K,D;1)$, of the fourth-order susceptibility
$\chi_{(4;0)}(K,D;1)$, of the correlation-length, etc.  should retain
the known Ising values reported in Table \ref{tab0}.  As the TCP is
reached, these exponents must change discontinuously into the sharply
different tricritical values of Table \ref{tab01}.  Series of finite
length cannot possibly reproduce such a step-like change of the
exponents, but only approximate it by a smooth transition extending
over some interval of values of $D < D_{tr}$, whose width depends on
the extension of the available series.  From a numerical point of
view, the smooth exponent change and the crossover phenomenology, that
occur in a left-hand neighborhood of $D_{tr}$ may be related to an
increasingly complex pattern of corrections to scaling near the
TCP. This makes the (unbiased) MRA estimates of the exponents by
Eq. (\ref{zinnesp}) a delicate issue, particularly so in the analysis of
spin $S=1$ on the $sq$ lattice.

  So far, the crossover behavior of the exponent approximations was
 not described systematically in the series context.  In a somewhat
 simplified RG model\cite{riede1,riede2}, that made possible to study
 the competition between a tricritical and a critical fixed point, the
 crossover was illustrated by forming ``effective
 exponents''\cite{kouv}, that measure the exponents as they are
 ``locally felt'' by the singular observables at the point $(D,
 K)$. These local quantities are studied as functions of $K$ for
 $K<K_c(D)$, at fixed values of $D$ close to $D_{tr}$. As $K \to
 K_c(D)$, they tend to the asymptotic critical exponents.  For
 example, for the ordinary susceptibility, an effective
 exponent can be defined by
\begin{equation}
\gamma^{(2;0)}_{eff}(K,D)=(K_c(D)-K)\frac{d {\rm ln}
\chi_{(2;0)}(K,D;S)}{dK}
\label{espeff}
\end{equation} 
It is observed\cite{riede1,riede2} that , for small $(K_c(D)-K)$ in the
crossover region, the quantity $\gamma_{eff}^{(2;0)}(K;D)$ initially
appears to tend to $\gamma^{(2;0)}(D_{tr};1)$ and then approaches
$\gamma^{(2;0)}(D;S)$ only as $(K_c(D)-K) \to 0$. In other words,
initially the renormalized Hamiltonian is attracted by the tricritical
fixed point and only eventually it feels the attraction of the
ordinary critical fixed point. A calculation of this kind can be
repeated in the series context using PAs to resum the HT expansions of
the effective exponents.  The procedure is biased, and might be
sensitive to the accuracy of the estimate of $K_c(D)$.

 Alternatively, the crossover can also be studied, for example by
 plotting vs $1/n^{\theta}$ the MRA estimator-sequences for the
 exponent $\gamma^{(2;0)}(D;1)$ formed with the HT coefficients of the
 susceptibility, at fixed values of $D$.  This procedure is not biased
 by $K_c(D)$, unlike that for the calculation of the effective
 exponent.

\subsection{Accuracy problems near the TCP }

\begin{figure}[p]
\begin{center}
\leavevmode
\includegraphics[width=5.00 in]{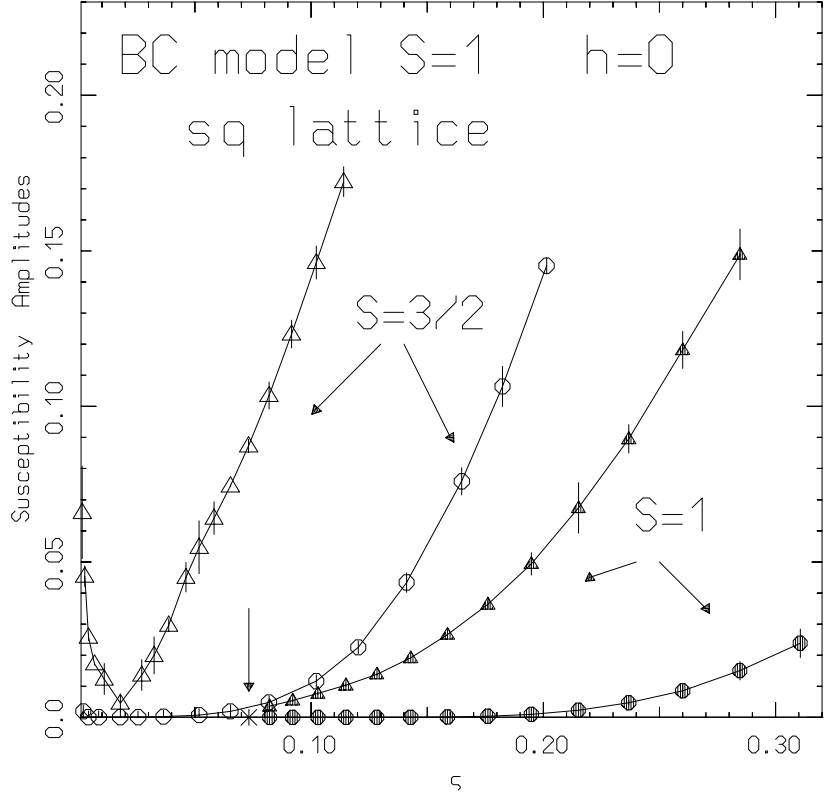}
\caption{\label{graf_ampl_sus_sq_S1} BC model with spin $S=1$ and
$S=3/2$ on the $sq$ lattice, for $h=0$.  The critical amplitudes
$A_{(2,0)}$ of $\chi_{(2;0)}(K,D;S)$ and $-A_{(4,0)}$ of
$\chi_{(4;0)}(K,D;S)$ vs $\zeta$. The variable $\zeta$ is
 $\tau$ in the $S=1$ case, while for graphical convenience 
 $\zeta=(1+exp(2d))^{-1}$  in the $S=3/2$
case. For $S=1$ the amplitudes are  full triangles and
full circles, respectively and  for $S=3/2$, they are  the
analogous open symbols.   A vertical arrow
points to the expected location (star)\cite{quian} of the TCP.
   The minima in the curves of the $S=3/2$
susceptibility amplitudes occur at $\Delta/Jq \approx 1/2$.}
\end{center}
\end{figure}

\begin{figure}[p]
\begin{center}
\leavevmode
\includegraphics[width=5.00 in]{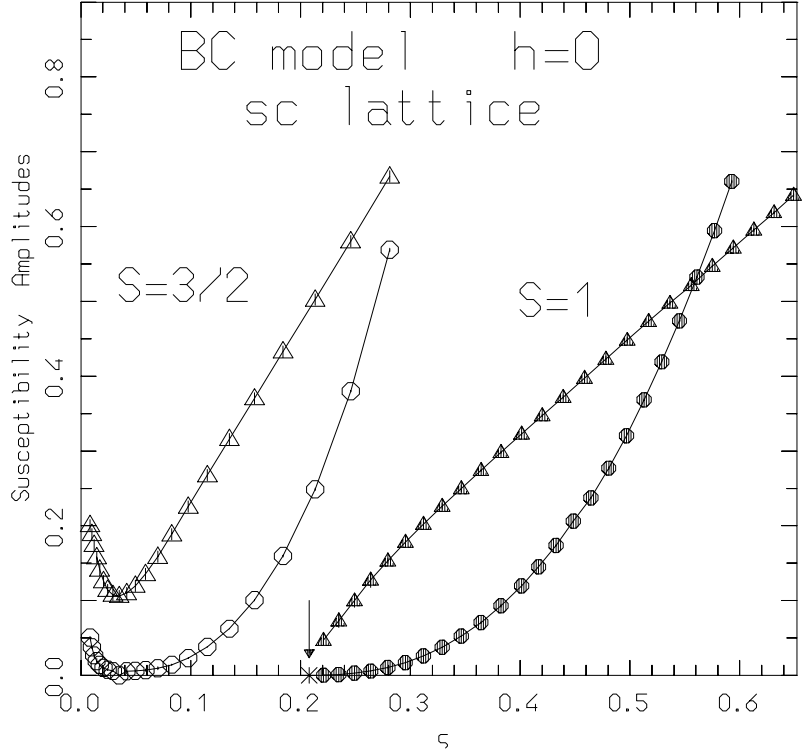}
\caption{ \label{graf_ampl_sus_sc_S1} BC model with spin $S=1$ and
$S=3/2$ on the $sc$ lattice, for $h=0$. The critical amplitudes
$A_{2,0}$ of $\chi_{(2;0)}(K,D;S)$ and $-A_{4,0}$ of
$\chi_{(4;0)}(K,D;S)$ vs $\zeta$. The variable $\zeta$ stands for
$\tau$ in the $S=1$ case, while  for graphical convenience is
$(1+exp(2d))^{-1}$ in the $S=3/2$ case. For $S=1$ these quantities are
full triangles and full circles respectively, and for $S=3/2$ are the
analogous open symbols.  In the $S=1$ case, a vertical arrow points to
the expected location (star) of the tricritical point.  The minima in
the curves of $A_{2,0}$ and $-A_{4,0}$ for $S=3/2$ occur at $\Delta/Jq
\approx 1/2$.}

\end{center}
\end{figure}

\begin{figure}[p]
\begin{center}
\leavevmode
\includegraphics[width=5.00 in]{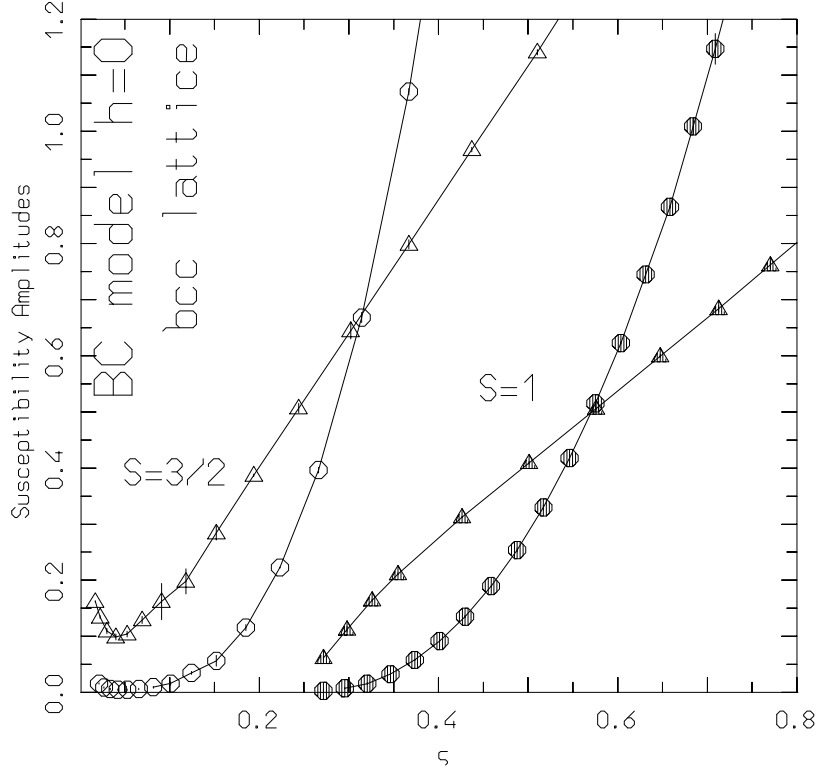}
\caption{ \label{graf_ampl_sus_bcc_S1} BC model with spin $S=1$ and
$S=3/2$ on the $bcc$ lattice, for $h=0$.  The critical amplitudes
$A_{2,0}$ of $\chi_{(2;0)}(K,D;S)$, and $-A_{(4,0)}$ of
$\chi_{(4;0)}(K,D;S)$ plotted vs $\zeta$ that stands for $\tau$ in the
$S=1$ case, while for graphical convenience is $(1+exp(2d))^{-1}$ in the
$S=3/2$ case. For $S=1$ these
quantities are full triangles and full circles, respectively.  For
$S=3/2$, they are the analogous open symbols.  The minima in the
curves of $A_{2,0}$ and $-A_{(4,0)}$ for $S=3/2$ occur at $\Delta/Jq
\approx 1/2$.}
\end{center}
\end{figure}

The numerical accuracy of the MRAs and the PAs or DAs, formed with the
 ordinary susceptibilities to determine either the critical contour in
 the anisotropy-temperature plane or the critical exponents, varies
 greatly along the critical phase-boundary. A marked worsening of the
 accuracy is observed when entering in the crossover region. This
 shortcoming of the series approach had been pointed out already in
 the pioneering analyses for the $fcc$ lattice\cite{SWS}, but remained
 unexplained.  We conjecture that it might be related to a clustering
 of unphysical singularities nearby the border of the convergence disk
 of the series, accompanied by the vanishing of the strength of the
 physical singularities as $D \to D_{tr}^-$, that gradually enhances
 the relative strength of the unphysical singularities as the TCP is
 approached. Then, the pattern of corrections to scaling, that rules
 the rate of convergence of all approximants, becomes increasingly
 complex and makes the analysis of limited-order expansions
 problematic. Such a mechanism might explain why, in these conditions,
 pronounced oscillations appear in the MRA estimator-sequences and at
 the same time both the PAs and the DAs loose precision or even miss
 altogether the physical singularities. (The weakening of the
 amplitudes of the physical singularities is prescribed by the
 tricritical-scaling property discussed in Appendix B and the behavior
 of the amplitudes of $\chi_{(2,0)}(K,D;S)$ and of
 $-\chi_{(4,0)}(K,D;S)$ as $D \to D_{tr}$ is outlined in the
 Figs. \ref{graf_ampl_sus_sq_S1}, \ref{graf_ampl_sus_sc_S1} and
 \ref{graf_ampl_sus_bcc_S1}. The clustering of the unphysical
 singularities is indicated by inspection of PA maps of the
 singularities in the complex $K$ plane.)

This effect, especially evident in the analysis of the $sq$ lattice,
 is less important in $3d$.  A similar phenomenon is observed also in
 the case of the BC system with spin $S=3/2$, in a neighborhood of
 $\tilde \Delta=1/2$, in spite of the absence of a TCP, for reasons
 probably similar to those explained for the spin $S=1$. We must add
 that by itself, the weakening of the critical amplitudes does not
 hinder the efficiency of the MRAs and we cannot remedy to the problem
 considering quantities that should be explicitly independent of these
 amplitudes, such as for example the log-derivative of the
 susceptibilites, because the effect of the unphysical singularities
 clustering remains unmodified.  However, one can extend the range pf
 values of $D$ in which the estimates of the parameters of the
 critical singularity are still reasonably accurate, by analyzing the
 expansions of a $D$- or $K$-derivative of the susceptibilities
 instead of the ordinary susceptibilities, because generally these
 quantities have different unphysical singularities and their critical
 singularities are sharper. Sometimes, also smoothing out the
 oscillations of the MRA sequences, for example by computing
 ``weighted moving-averages'' over several terms, might be
 helpful. Anyway, we have to conclude that in some small left-hand
 vicinity of the TCP, our simple extrapolations of the HT (and LT)
 expansions, might be unable to improve the precision of the best
 present MC simulations or transfer-matrix methods in the
 determination of the critical boundary.

\section{The  BC model with spin $S=1$    on the $sq$ lattice}

The earliest MC and transfer-matrix calculations on the $sq$ lattice
determined\cite{arora,wilding,yuk,beale,martins} the tricritical value
of the crystal field $\tilde \Delta_{tr}$ between 0.4912 and 0.4915,
with uncertainties of a few units in the last figure.  The values of
the tricritical temperature $\tilde T_{tr}$ ranged between 0.1520 and
0.1525, with similar uncertainties. More recently, a Wang-Landau
simulation\cite{kwak} determined $\tilde \Delta_{tr} =0.49151(1) $ at
$\tilde T_{tr}=0.152$ and a sparse transfer-matrix technique of higher
estimated accuracy\cite{quian} yielded $ \tilde \Delta_{tr}
=0.49145373(6)$ with $\tilde T_{tr}=0.15214439(1)$, (equivalently
$D_{tr}=3.2301797(2)$ and $K_{tr}=1.6431759(1)$). In the same
Ref. [\onlinecite{quian}], the value $X_{tr}=0.4549506(2)$ was
proposed for the tricritical concentration.  Except for the value of
$X_{tr}$, we shall assume the validity of the cited estimates,
whenever convenient.
  
\begin{figure}[p]
\begin{center}
\leavevmode
\includegraphics[width=5.00 in]{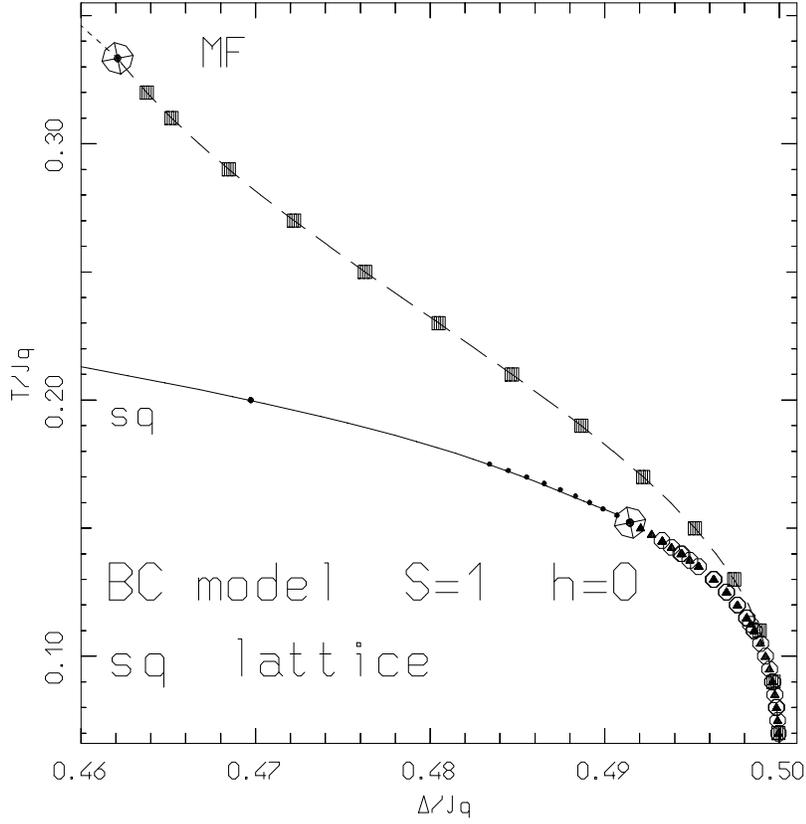}
\caption{ \label{Graf_fig3_first_sq} BC model with spin $S=1$ for
$h=0$ on the $sq$ lattice.  Phase diagram in the $( \Delta/Jq,T/Jq)$
plane, in the MF approximation and from series analysis.  Big crossed
open circles are TCPs. The short-dashed line is the MF critical
phase-contour. Full squares connected by a long-dashed line are the MF
first-order line.  The solid curve is the critical phase-contour from
series, ending with a TCP computed\cite{quian} by transfer-matrix. Full
dots on this line are from MC or transfer-matrix. Small black triangles
are MC points\cite{jung,kwak,ziere2d} on the first-order line.  The
small open circles around the triangles are obtained by intersecting
the LT and HT expansions of the lattice free-energy.}
\end{center}
\end{figure}

For  $D \lesssim 0.3 D_{tr} $, the procedure of employing MRA
estimator-sequences of the ordinary susceptibility to determine the
 critical contour, yields results that agree
well with and probably are more accurate than other determinations.
It was anticipated that as $D$ gets closer to $D_{tr}$, the MRA
sequences begin to show oscillations and at the same time also the
accuracy of the PAs and of the DAs deteriorates.  However, forming MRA
estimator-sequences of the mixed susceptibility 
 $\chi_{(2;2)}(K,D;1)$ (i.e. taking two
$D$-derivatives of the ordinary susceptibility to modify the pattern
of the unphysical singularities) and in some cases, performing a ``weighted
moving-average'' over six terms of the sequences, smoother estimate sequences 
are obtained, so that the determination of the $sq$ critical phase-boundary can
be extended up to $ D \lesssim 0.98 D_{tr}$, keeping within $\approx
0.01\%$ the relative deviations from the best recent
estimates\cite{kwak,jung,ziere2d} obtained by transfer
matrix\cite{jung} or MC simulations\cite{kwak,ziere2d}. This is  shown in
Fig. \ref{Graf_fig3_first_sq}. For $D > 0.98 D_{tr}$, the $sq$
critical boundary obtained from the HT expansion and represented as a solid
line in Fig. \ref{Graf_fig3_first_sq} is only slightly higher in
temperature than the cited recent estimates that for comparison are
reported as full dots on the curve. For graphical clarity, this figure
blows up only a small vicinity of the TCP. A more extensive comparison
of the data in the literature with the series estimates obtained in
our study, appears in the Tables \ref{tab2} and \ref{tab2c}.  
The MRA value of the
critical temperature at $D=0$ ( i.e. for the Ising model with spin
$S=1$) to which we have attached a very generous uncertainty, also appears
in this Table  to give an idea of the precision of the various
simulations. Similarly for higher values of $D$, the spread among the
simulation data from the various sources suggests that, in some
of the older studies, the uncertainties were underestimated.  A set of
very accurate recent estimates of points in the first-order part of
the phase-boundary\cite{jung,ziere2d,kwak} is also reported in
Fig. \ref{Graf_fig3_first_sq}.  The series estimates of these points,
indicated as small open circles in the figure, are obtained by
determining the intersections of the values of the
free-energies computed from  LT and HT expansions.  They are completely consistent with the cited results
except perhaps just beneath the tricritical temperature,
 in the range $0.14 \lesssim \tilde T \lesssim 0.15$, in
which the intersection method produces less accurate results. For this
calculation, it is convenient to use the temperature-grouped $sq$
lattice LT expansion of the free-energy, that extends to order
$u^{20}$.  Applying the intersection method  of the LT and HT
expansions of the free-energy for $T> T_{tr}$, but looking for
tangency points instead of crossings, yields a critical 
phase-contour slightly lower in temperature than that obtained by the
MRAs of the (mixed) susceptibility HT expansion, an effect which 
probably reflects only the need of longer LT series as T increases. A similar
defect was observed in the earliest series study\cite{SWS}.

\begin{figure}[p]
\begin{center}
\leavevmode
\includegraphics[width=5.00 in]{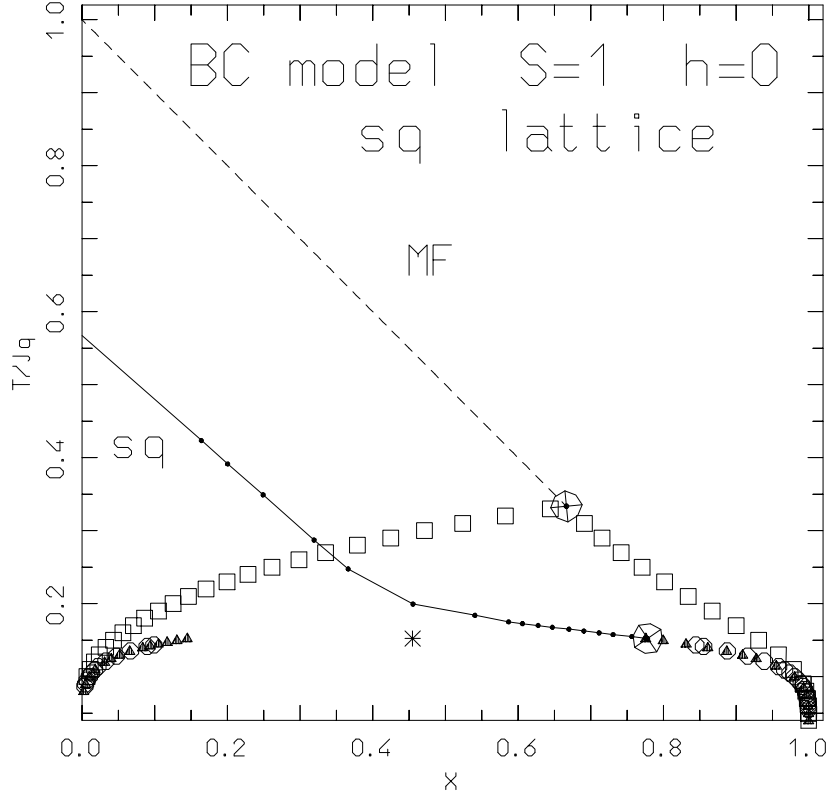}
\caption{ \label{TcsqvsX} BC model with spin $S=1$ on the $sq$ lattice
  for $h=0$.  Phase diagram in the concentration-temperature plane.
  The dashed line is the MF approximation of the critical phase
  boundary. Sequences of open squares are the two branches of the MF
  first-order contour.  The solid line is the critical contour from series,
  the black dots on it and the full triangles are the branches of the
  first-order contour  from Refs. [\onlinecite{kwak,jung,ziere2d}].
  The small open circles surrounding the full triangles are obtained
  by intersecting the LT and HT expansions of the  free-energy.
  Big crossed circles are tricritical concentrations.  The star is a
  transfer-matrix estimate\cite{quian} of $X_{tr}$.  }
\end{center}
\end{figure}

Fig. \ref{TcsqvsX} shows the phase diagram in the
concentration-temperature plane $(X, \tilde T)$. The concentration is
defined by Eq. (\ref{xdens}). The second-order part of this
phase-contour is obtained forming the highest-order available PAs of
the HT expansion of $X(K,D;1)$ in powers of $K$, evaluating them on
the critical contour in the $(\tilde \Delta, \tilde T)$ plane,
represented in Fig. \ref{Graf_fig3_first_sq} and averaging these
estimates.  The same prescription can be directly used also starting
with with the recent simulation and transfer-matrix
data\cite{jung,ziere2d,kwak} for the critical phase-contour of the
$sq$ lattice in the $(\tilde \Delta, \tilde T)$ plane, thus obtaining
the points indicated by black dots in the Fig. \ref{TcsqvsX}. The
results agree with those from the series within less than $0.1\%$.
The prescription adopted for the PA resummation of the concentration
series generally converges well except for $0.35 \lesssim X \lesssim
0.45$. In this interval, we might have to allow for uncertainties in
$\tilde T$ of the order of $ 1\%$.  It must be stressed that the shape
of this critical phase-contour, differs markedly from its MF
counterpart, unlike what is observed in the $3d$ analysis. We are
finally led to the estimate $X_{tr}=X(K(D_{tr}),D_{tr};1)=0.779(1)$,
so that the value $X_{tr}=0.4549506(2) $ of the critical concentration
proposed in Ref. [\onlinecite{quian}] and represented by a star in the
Fig. \ref{TcsqvsX}, or a similar value suggested by an earlier
simulation\cite{arora} should be corrected by nearly a factor $2$.
The right-hand branch of the first-order phase-contour, is obtained
evaluating the HT expansion of $X(K,D;1)$ on the upper rim of the
first-order phase-contour in the $(\tilde \Delta, \tilde T)$ plane,
shown in Fig. \ref{Graf_fig3_first_sq} (these points are thus
approached from the right-hand side of the $T = T(D;1)$ curve).  The
LT expansion of $X$ should be evaluated at the same points, to
represent the lower rim of the first-order line in the $(\tilde
\Delta, \tilde T)$ plane and thus also the left-hand branch of the
phase-boundary in the $(X,\tilde T)$ plane is obtained.  Unfortunately
this branch fails to reach the value $X_{tr}$ and therefore to
corroborate our estimate of this quantity.
It is useful to add that for this last
computation, a PA resummed ``field grouped'' LT expansion turns out to
be more convenient than a PA resummed ``temperature grouped''
expansion.

\begin{figure}[p]
\begin{center}
\leavevmode
\includegraphics[width=5.00 in]{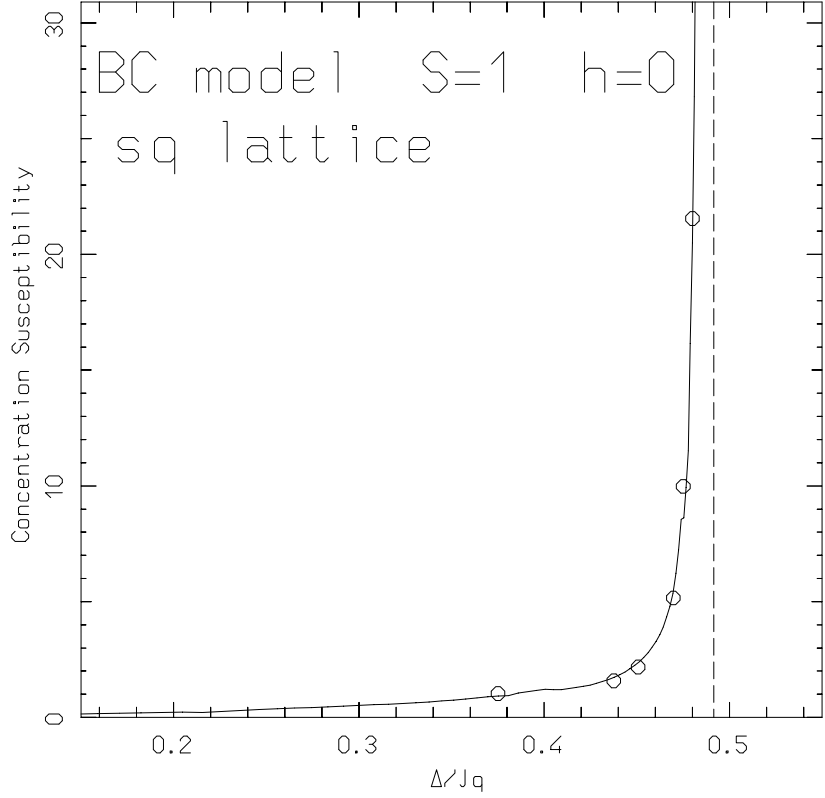}
\caption{ \label{conc_susc_sc_Delta} BC model with spin $S=1$ for
$h=0$ on the $sq$ lattice.  The HT expansion of the concentration
susceptibility $Y(K,D;1)$ evaluated along the critical phase-boundary
(solid line) vs $ \Delta/Jq$.  Open circles are obtained evaluating $Y(K,D;1)$
on high-precision estimates of critical points\cite{kwak,jung,ziere2d}. The
vertical dashed line is the value\cite{quian} of $\Delta_{tr}/Jq$.}
\end{center}
\end{figure}

The fluctuation of the concentration, namely the concentration
susceptibility $ Y(K,D;1)$ defined by Eq. (\ref{yrf}), is expected
from Eq. (\ref{yas}) to show only a mild specific-heat-like
singularity along the critical phase-contour and to diverge strongly
at the TCP.  However, our series analysis in $2d$ suggests in
Fig. \ref{conc_susc_sc_Delta} that $ Y(K,D;1)$ remains finite as $T$
approaches the critical phase-contour from above, for all fixed
$D<D_{tr}$. No theoretical explanation of this behavior is
known\cite{riedel2,riedel3}. The same figure shows that along the
critical phase-contour, $Y(K,D;1)$ grows steeply as $D \to D_{tr}$.

\scriptsize
\begin{table}[ht]
  \caption{ BC model with $S=1$ on the $sq$ lattice in zero magnetic
    field.  Phase-contour $T_c=T_c(\Delta;1)$ from MC simulations,
    transfer-matrix and  the analysis of the expansions of the
    ordinary susceptibility (or of its $D$-derivatives).
     Temperatures and crystal fields are
    not normalized to the coordination number $q$, unlike elsewhere in
    the text.}
\begin{tabular}{|c| c c c c c c  c c c |}
 \hline
 $\Delta/J$ &&&& $T/J$&&&&&Trans.Ord.\\
\hline
&This paper& Ref.[\onlinecite{jung}] & Ref.[\onlinecite{ziere2d}]& 
 Ref.[\onlinecite{kwak}] & Ref.[\onlinecite{beale}]
&Ref.[\onlinecite{silva}]& Ref.[\onlinecite{berker}]& 
 Ref.[\onlinecite{quian}]&\\
 \hline
0.&1.69378(4) &&&&1.695 &1.714(2) &1.693(3)  &&Second \\
0.5& 1.5664(1)&&&& 1.567&1.584(4) &1.564(3)  &&Second\\
1.&1.3986(1)  &&&& 1.398&1.413(1) &1.398(2)  &&Second\\
1.5&1.1467(1) &&&& 1.150&1.155(1) &1.151(1)  && Second\\
1.7027(1)&0.994(5)&1.&1.&&&&&& Second\\
1.75 &0.950(1)&&&&&&0.958(1)&&Second\\
 1.80280(6)&&&0.80&&&&&&Second\\
1.87&0.812(1)&&&&0.800&0.800(3)&&&Second\\
1.87879(3)&&0.80&&&&&&&Second\\
1.9&0.766(1)&&&&&0.755(3)&0.769(1)&&Second\\
1.92&0.7289(2)&&&&0.700&0.713(2)&&&Second\\
1.9336(4)&&0.70&&&&&&&Second\\
1.93296(2)&&0.70&&&&&&&Second\\
1.9379(5)&&0.69&&&&&&&Second\\
1.9421(5)&&0.68&&&&&&&Second\\
1.9461(5)&&0.67&&&&&&&Second\\
1.9501(2)&0.656(4)&0.66&&0.66&0.650&0.651(2)&0.659(2)&&Second\\
1.95273(1)&&&0.65&&&&&&Second\\
1.9533(1)&&0.65&&&&&&&Second\\
1.9534(1)&&&&0.65&&&&&Second\\
1.9565(1)&&0.64&&0.64&&&&&Second\\
1.9596(2)&&0.63&&&&&&&Second\\
1.95980(5)&&&&0.63&&&&&Second\\
1.96270(1)&&&&0.62&&&&&Second\\
1.96539(1)&&0.61&&&&&&&Second\\
1.96550(1)&&&&0.61&0.61&&&&Second\\
1.9658149(2)&&&&&&&&0.60857756(4) &TCP\\
1.96582(1)&&0.60858(5)&&&&&&&TCP\\
1.96604(1)&&&&0.608&&&&&TCP\\
 \hline
 \end{tabular} 
 \label{tab2}
\end{table}

\begin{table}{l}
  \caption {{\it (Continued from the preceding Table)} 
 BC model with $S=1$ on the $sq$ lattice in zero magnetic
    field. Phase-contour  from MC simulations,
    transfer-matrix and  the analysis of the expansions of the
    ordinary susceptibility (or of its $D$-derivatives). Our estimates
    of the first-order part of the phase-contour are obtained by the
    LT-HT intersection method.    Temperatures
    and  crystal fields are not normalized to the
    coordination number $q$, unlike elsewhere  in the
    text. }
\begin{tabular}{|c| c c c c c c    c|}
 \hline
 $\Delta/J$ &&&& $T/J$&&&Trans.Ord.\\
\hline
&This paper& Ref.[\onlinecite{jung}] & Ref.[\onlinecite{ziere2d}]& 
 Ref.[\onlinecite{kwak}] & Ref.[\onlinecite{beale}]& &\\
 \hline

1.968174(3)&&&0.60&&&&First\\
1.96820(3)&&0.60&&&&&First\\
1.96825(1)&&&&0.60&&&First\\
1.97080(5)&&&&0.59&&&First\\
1.97072(5)&&0.59&&&&&First\\
1.97308(4)&0.581(2)&0.58&&&&&First\\
1.97323(1)&&&&0.58&&&First\\
1.97528(4)&&0.57&&&&&First\\
1.9777(1)&0.565&&&&&&First\\
1.97744(3)&0.5602(4)&0.56&&&&&First\\
1.97766(1)&&&&0.56&&&First\\
1.97950(3)&&0.55&&&&&First\\
1.98142(2)&0.543(4)&0.54&&&&&First\\
1.98490(2)&0.519(2)&0.52&&&&&First\\
1.98786(1)&0.502(2)&0.50&&&&&First\\
1.99036(1)&0.481(2)&0.48&0.48&&&&First\\
1.992479(1)&0.460(1)&0.46&0.46&&&&First\\
1.994232(5)&0.440((1)&0.44&0.44&&&&First\\
1.99681357&0.399(2)&0.40&0.40&0.40&0.40&&First\\
1.99842103&0.36&0.36&&0.36&&&First\\
1.99932488&0.32&0.32&&0.32&&&First\\
1.968174(3)&&&0.60&&&&First\\
1.96820(3)&&0.60&&&&&First\\
1.96825(1)&&&&0.60&&&First\\
1.97080(5)&&&&0.59&&&First\\
1.97072(5)&&0.59&&&&&First\\
1.97308(4)&0.581(2)&0.58&&&&&First\\
1.97323(1)&&&&0.58&&&First\\
 \hline
 \end{tabular} 
 \label{tab2c}
\end{table}

\normalsize

\begin{figure}[p]
\begin{center}
\leavevmode
\includegraphics[width=5.00 in]{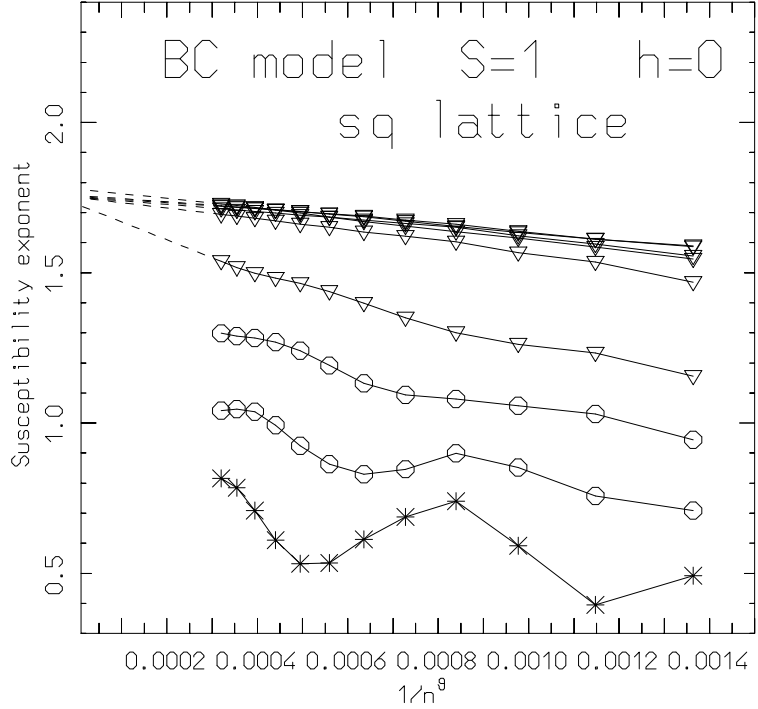}
\caption{ \label{graf_chi2d2_sq_Zinn_S1} BC model with spin $S=1$ on
 the $sq$ lattice, for $h=0$. The twelve highest-order terms of
 several MRA estimator-sequences for the critical exponent
 $\gamma^{(2;0)}(D;1)$, obtained from the mixed susceptibility
 $\chi_{(2;2)}(K,D;1)$, vs $1/n^{\theta}$, with $n$ the number of
 terms included in the series and $\theta= 2.5$.  The lowermost MRA
 sequence (stars) is computed for $D \approx D_{tr}$, the successive
 two (open circles) for $ D = 0.9D_{tr}$ and $ 0.8D_{tr}$, the
 remaining ones (open triangles) for $0.7 D_{tr}$, $ D = 0.4D_{tr}$,
 $0.25D_{tr}$ , $ 0.15D_{tr}$, etc.  The symbols of the successive
 terms are connected by segments to profile the behavior of each
 sequence.  Extrapolations to large $n$, indicated only for $D<
 0.8D_{tr}$, are dashed lines. }
\end{center}
\end{figure}

For several values of $D \ll D_{tr}$ and a few values closer to
$D_{tr}$, Fig. \ref{graf_chi2d2_sq_Zinn_S1} shows the twelve
highest-order terms (out of the 24 available) in the MRA
estimator-sequences of the susceptibility exponent
$\gamma^{(2;0)}(D;1)$.  The terms in the sequences are formed with the
expansions of the mixed susceptibility $\chi_{(2;2)}(K,D;1)$ and are
plotted vs the power $1/n^{\theta}$ of the number $n$ of HT
coefficients used in the computation.  Whenever possible, the behavior
of last few (from two to five) terms of each sequence is extrapolated
to large $n$ choosing $\theta=2.5$ in the Ansatz Eq. (\ref{ansatz}).
The extrapolations are indicated by dashed lines.  A solid line
interpolates among the symbols, to profile clearly the trend of each
sequence and show when it appears to settle down in the asymptotic
behavior predicted by Eq. (\ref{espasint}).  The MRA sequences
computed for $D \lesssim 0.7 D_{tr}$, tend to flatten for sufficiently
large $n$ and thus can be convincingly extrapolated. On the contrary,
for $0.7 D_{tr} < D <D_{tr}$, the MRA sequences develop pronounced
oscillations so that a simple-minded extrapolation by
Eq. (\ref{espasint}) of their last few available terms, would lead to
irregular fluctuations in the final exponent estimates. Therefore in
this range of $D$, no extrapolations are indicated in
Fig. \ref{graf_chi2d2_sq_Zinn_S1}.

\begin{figure}[p]
\begin{center}
\leavevmode
\includegraphics[width=5.00 in]{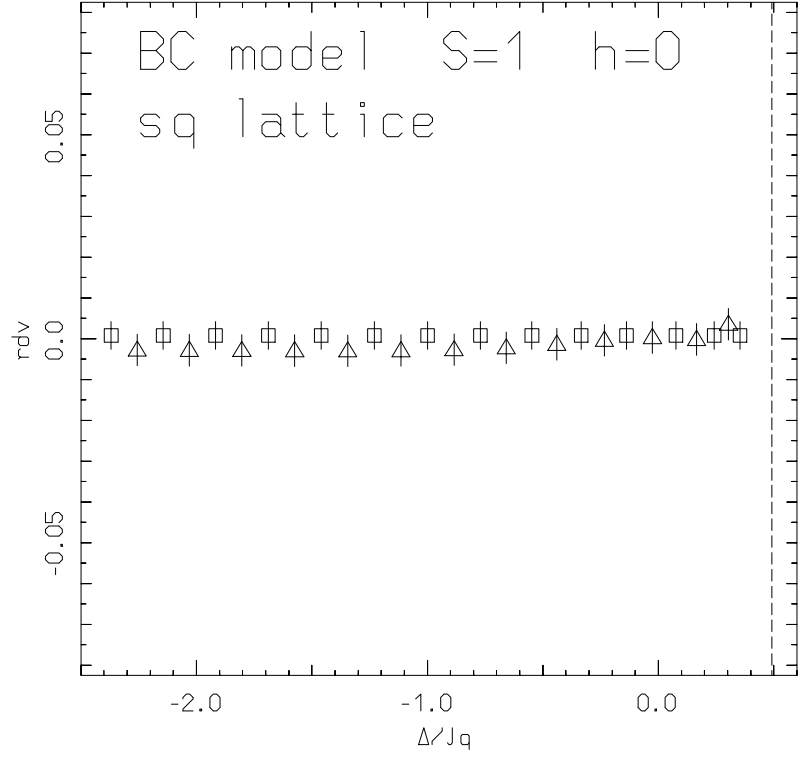}
\caption{ \label{figura_esp_sq} BC model with spin $S=1$ on the $sq$
  lattice for $h=0$.  The relative deviations $rdv$ from the Ising values
  for the exponents $\gamma^{(2;0)}(D;1)$ from $\chi_{(2;1)}(K,D;1)$ 
  (open triangles), and $\gamma^{(4;0)}(D;1)$ from $\chi_{(4;0)}(K,D;1)$
  (open squares) vs $\Delta/Jq$.
The vertical dashed line
  indicates the tricritical value\cite{quian} $\Delta_{tr}/Jq$ of the
  crystal field.}
\end{center}
\end{figure}

  In Fig. \ref{figura_esp_sq} the results of the above procedure using
 the expansions of the ordinary (higher) susceptibilities or their
 $D$-derivatives, are summarized by plotting vs $\tilde \Delta $ the
 relative deviations $rdv$ from the $2d$ Ising exponents, of the
 extrapolated values of the MRA estimator-sequences for the exponents
 $\gamma^{(2;0)}(D;1)$, and $\gamma^{(4;0)}(D;1)$. These deviations
 are defined as $rdv =
 \frac{\gamma^{(2r;0)}(D;1)}{\gamma^{(2r;0)}(0;1)}- 1$ (with $r=1,2$),
 for $D < D_{tr}$ and our estimates of them are quite small over a
 large range of values of $D$ since they exceed $1 \%$ only in
 the crossover region for $D \gtrsim 0.7 D_{tr}$. Also the exponent
 $\nu(D;1)$ can be simply obtained using the hyperscaling relation.
 The uncertainties of the exponent deviations are smaller than the
 size of symbols. The figure does not include points from MRA
 sequences in the region in which simple-minded extrapolations are not
 safe.  These results corroborate the validity of universality over a
 large part of the critical boundary.

\begin{figure}[p]
\begin{center}
\leavevmode
\includegraphics[width=5.00 in]{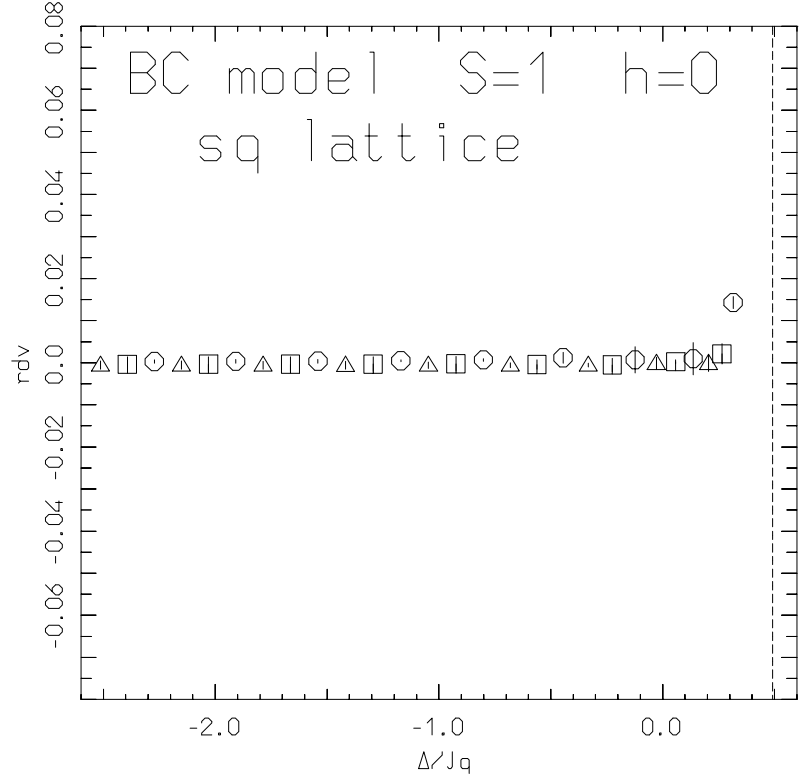}
\caption{ \label{graf_ratios_S1_sq} BC model with spin $S=1$ on the
  $sq$ lattice for $h=0$. The relative deviations $rdv$ from the $2d$
  Ising values for the universal ratios of critical amplitudes ${\cal
  I}^+_{6}$ (open circles), ${\cal I}^+_{8}$ (open triangles) and
  ${\cal J}^+_{8}$ (open squares) vs $\Delta/Jq$. The vertical dashed line
  indicates the tricritical value\cite{quian} $\Delta_{tr}/Jq$ of the
  crystal field.   }
\end{center}
\end{figure}

 The effects of the crossover at $D \gtrsim 0.4 D_{tr}$, show up also
when studying the behavior of the first few terms ${\cal I}^+_{6}$,
${\cal I}^+_{8}$ and ${\cal J}^+_{8}$ of the sets of universal ratios
of critical amplitudes defined by Eqs. (\ref{Ii}) and (\ref{Ai}). 
These quantities, computed forming first-order DAs for the appropriate
non-singular universal ratios of the higher susceptibilities, are
biased with the series estimates of the  critical boundary $T_c=T_c(D)$.
In Fig. \ref{graf_ratios_S1_sq}, we have plotted vs $\tilde \Delta$ the
relative deviations $rdv$ of these ratios from their $D=0$ (i.e. pure
Ising) values.  They remain quite small over a wide range of values of
$D$ up to $D \lesssim 0.4 D_{tr} $.  A good consistency with the
universality predictions is observed, although as $D$ approaches
$D_{tr}$, the convergence rate of the analysis methods deteriorates.

 In the Fig. \ref{graf_ampl_sus_sq_S1}, we have plotted the critical
amplitudes of $\chi_{(2;0)}(K,D;1)$ and $\chi_{(4;0)}(K,D;1)$ vs
$\tau$ for the $sq$ lattice, showing that our estimates are consistent
with the predictions of amplitude-scaling and with the location of the
TCP determined in Ref. [\onlinecite{quian}], where they should vanish.
These quantities are estimated forming DAs of the effective amplitudes
(see for example Eq. (\ref{ampef}) for the the ordinary
susceptibility) biased with the series estimates of the critical
temperatures $T_c=T_c(D)$ and the Ising values of the exponents, so
that this calculation {\it must assume } that exponent universality is
valid over all the critical boundary.

In conclusion, strong crossover effects are observed in the analysis of
the $sq$ lattice series.  Therefore, we have not indicated extrapolations in
some figures, and expect that, in this range of $D$, significantly
extended series and/or improvements of the simple numerical methods of
our analysis would be needed.

\section{The   BC model with spin $S=1$   in $3d$}

Generally, for the BC model with spin $S=1$ at zero field on
three-dimensional lattices, the HT analysis is numerically simpler
than in the $sq$ case, because the series expansions show smoother
behaviors.

\begin{figure}[p]
\begin{center}
\leavevmode
\includegraphics[width=5.00 in]{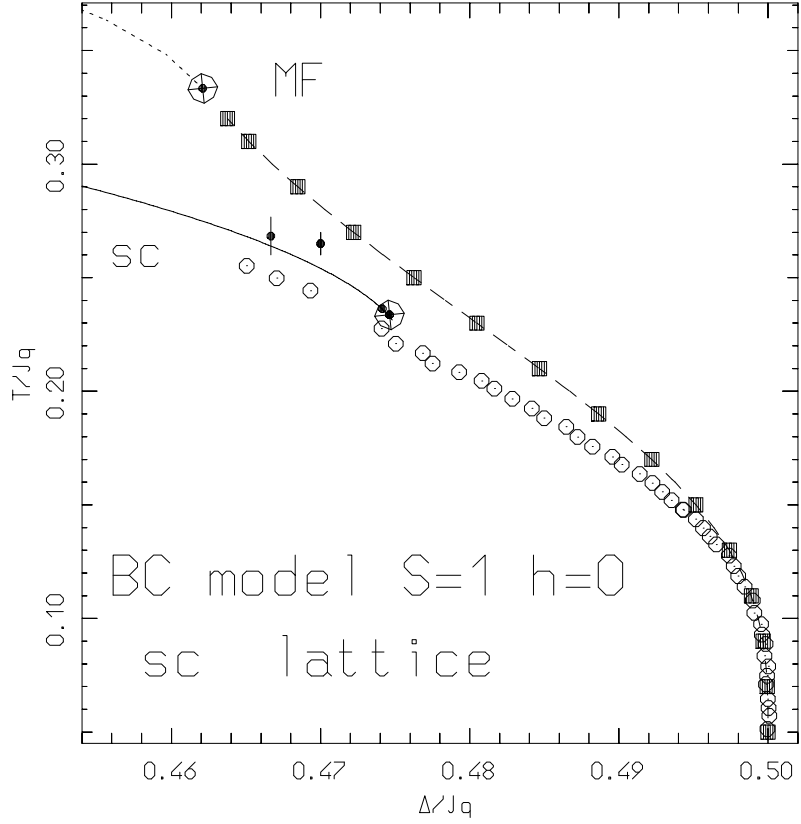}
\caption{ \label{Graf_fig3_first_sc} BC model with spin $S=1$, in the
 $h=0$ plane for the $sc$ lattice. Phase diagram in the
 anisotropy-temperature plane, by series and in the MF approximation.
 Big crossed open circles are TCPs. The upper short-dashed line is the
 MF critical phase-contour and full squares connected by long-dashed
 line are the MF first-order contour.  The solid curve is the critical
 phase-boundary from series and the TCP is taken from a
 simulation\cite{deng}.  The black dots nearby the solid curve are MC
 points\cite{Oz}.  Open circles are points of the first-order (and the
 second-order) part of the phase-boundary from intersections of the LT
 and HT expansions of the free energy. }
\end{center}
\end{figure}

 For the $sc$ lattice, Fig. \ref{Graf_fig3_first_sc} shows
 the phase diagram in the anisotropy-temperature plane.  The position
 of the TCP determined in the MC simulation of
 Ref.[\onlinecite{deserno}] is $\tilde \Delta_{tr}=0.474113(5) $ with
 $\tilde T_{tr}=0.2363(1)$.  A more recent (and probably more
 accurate) MC estimate\cite{deng}, indicated in the figure by a big
 open crossed circle, is $\tilde \Delta_{tr}= 0.474617(16)$ with
 $\tilde T_{tr}=0.23365(4)$, so that $D_{tr}=2.0313(4)$. The same
 simulation moreover indicated the value $X_{tr}=0.6485(2)$ for the
 critical concentration.

 A sample of the results from  MRA extrapolations  for the
ordinary susceptibility $\chi_{(2;0)}(K,D;0)$ (to which generous error
bars are attached) is reported in Table \ref{tab32} and compared with
those of earlier MC studies. Our extrapolations agree well with MC
data\cite{Oz,deserno} over the range of values of $\tilde \Delta$ in
which both analyses are available, except very close to the TCP.  At
$\Delta/J=2.82$, a visible difference appears between the estimate of
Ref. [\onlinecite{Oz}] and that of the series. However the latter are
likely to be more accurate up to the TCP. From the Table \ref{tab32},
an excellent agreement can be observed between the series and the
simulation results, in particular at the value\cite{min1,min2} $D=0.655$
or equivalently $\Delta/J=1.689...$, in which the corrections to
scaling are expected to be minimal, so that also the MC estimates can
be very accurate.

  A sequence of points in the first-order part of the phase-boundary
is also shown in Fig. \ref{Graf_fig3_first_sc}.  As anticipated, it is
obtained locating the intersections of the LT and HT expansions of the
free-energies in the region in which both might be approximately
valid. In this calculation, we have used the temperature-grouped $sc$
lattice LT expansion of the free-energy, that extends to order
$u^{42}$.  Both the HT and the LT expansions are resummed by the
highest available PAs. Except very close to the TCP, the resulting
estimates appear quite reasonable, but unfortunately no independent
estimates exist.  The method based on the joint use of LT and HT
expansions might be applied also at temperatures above $T_{tr}$ at
which the transition is continuous, to reproduce also the critical
phase-boundary. One should then look for tangency points of the LT and
HT free-energy curves.  While such points do exist for $\tilde T >
\tilde T_{tr}$, they yield estimates falling systematically below (in
temperature) those previously obtained from the analysis of the
susceptibility HT expansions. This fact is not surprising, at least
because as the temperatures increase, longer LT expansions would be
needed. Also in the case of the $fcc$ lattice expansions\cite{SWS},
the second-order line computed in this way suffered from the same
shortcomings.

\begin{figure}[p]
\begin{center}
\leavevmode
\includegraphics[width=5.00 in]{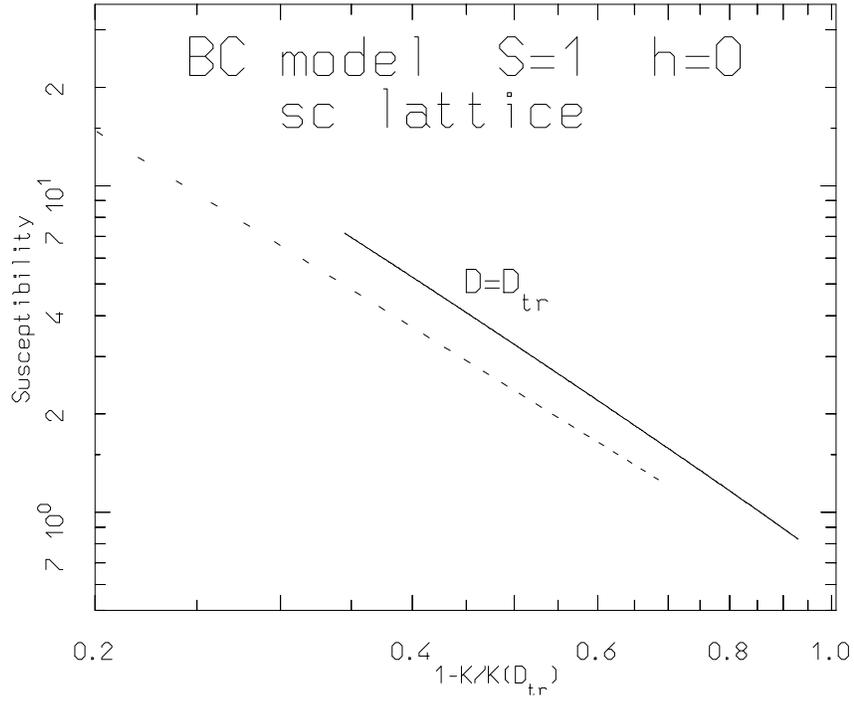}
\caption{ \label{susc_sc_tangenz} BC model with spin $S=1$ for $h=0$
on the $sc$ lattice.  The solid line is a bilog plot of the HT
expansion of the susceptibility $\chi_{(2;0)}(K,D;1)$ computed along
the tangent to the phase-contour at $D=D_{tr}$.  
The dashed line is a
simple power behavior with exponent $\gamma_{tr} = 2.$. }
\end{center}
\end{figure}

The Fig.  \ref{susc_sc_tangenz}, contains a bilogarithmic plot of the ordinary
 susceptibility evaluated along the path $D=D_{tr}-2.15(2)(1-K/K_{tr})$
 tangent to the phase-boundary at the TCP and suggesting the value $\gamma_u
 = 2.00(5)$ of the critical exponent. This implies $\phi_{u} = 2.00(5)$.
 
 In Fig. \ref{fig_TF_sc_tau}, the phase-contour is drawn
in the $(\tau, \tilde T)$ plane to show that it continues to be very nearly a
straight line down to $\tau \approx 0.1$.

 The phase diagram in the concentration-temperature plane $(X,\tilde T)$ for
the $sc$ lattice is shown in Fig. \ref{TcscvsX}. The phase-boundary is
obtained averaging the highest possible PAs of the HT expansion of
$X(K,D;1)$ evaluated along the critical contour in 
 the anisotropy-temperature plane shown in Fig.
\ref{Graf_fig3_first_sc}.  At $(D=D_{tr}$,$K=K_{tr})$ , the estimate
$X_{tr}= 0.666(3)$ is obtained for the critical concentration.  Unlike
what is observed for the $fcc$ lattice\cite{SWS}, in this case the LT
expansions are sufficiently long to show that the left-hand branch of
the first-order line reaches the TCP, with good approximation.

\begin{figure}[p]
\begin{center}
\leavevmode

\includegraphics[width=5.00 in]{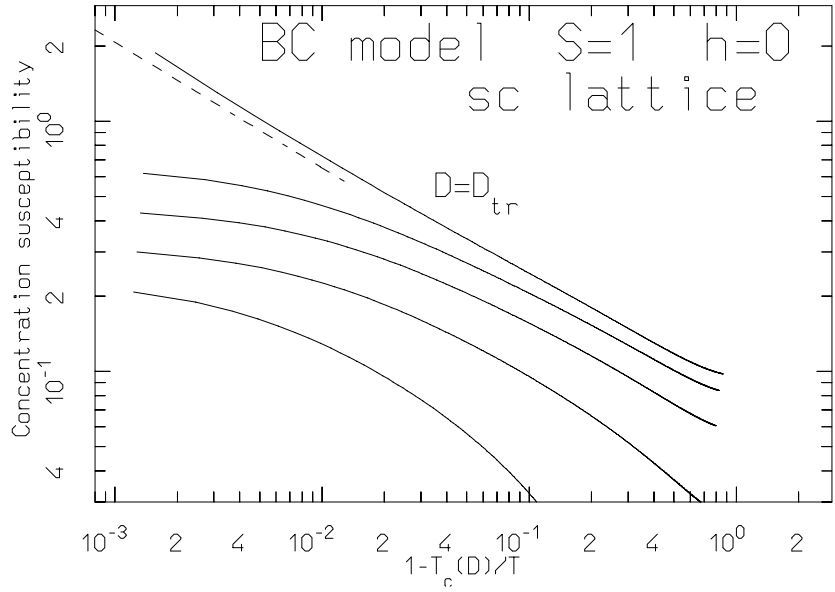}
\caption{ \label{conc_susc_sc} BC model with spin $S=1$ for $h=0$ on
the $sc$ lattice.  The uppermost solid line is a bilog plot of
the PA resummed HT expansion of the concentration susceptibility
$Y(K,D;1)$ along the path  $D=D_{tr}$ vs  $1-T(D_{tr})/T$. 
The dashed line has a simple power behavior with
exponent $\lambda = -0.5$. The lower curves show the same quantity
evaluated for $D$ taking the sequence of four values $0.74D_{tr}$
,$0.59D_{tr}$ ,$0.44D_{tr}$, $0.30D_{tr}$ and for each of them the abscissa is
 $1-T_c(D)/T$. }
\end{center}
\end{figure}

The same considerations as for the $sq$ lattice can be repeated here
for the $sc$ lattice. The behavior of $ Y(K,D;1)$ observed in our
series analysis is illustrated in Fig. \ref{conc_susc_sc} that
indicates an asymptotically finite behavior along the critical
phase-contour and a strong divergence as the TCP is approached from
above at  fixed $D= D_{tr}$.  In $3d$ the mild specific-heat-like 
divergence  expected on the critical contour is
not observed experimentally and no theoretical explanation of this
fact is known\cite{riedel2,riedel3}.  In the following Fig.
\ref{susc_sc}, this behavior is contrasted with that of the ordinary
susceptibility that diverges over all the critical boundary, albeit
showing a smaller exponent at the TCP.

\begin{table}[ht]
  \caption{ The BC model with $S=1$ on the $sc$ lattice.  
    Phase-contour $T_c=T_c(\Delta;1)$ from MC simulations and
    from the analysis of the expansions of the ordinary
    susceptibility. The estimates of the first-order part
    of the phase-contour are obtained by the LT-HT intersection method.  To
    retain the conventions of most authors, the data for the
    temperatures and the crystal fields, unlike those in
    the figures and in the text, are not normalized to the
    coordination number $q$.}
\begin{tabular}{|c| c c c c c c  c|}
 \hline $\Delta/J$ &&& $T/J$&& &&Trans.Ord.\\ 
\hline 
&This paper&Ref.[\onlinecite{Oz}]&
  Ref.[\onlinecite{deserno}]     &Ref.[\onlinecite{deng}]
  &Ref.[\onlinecite{ziere3d}] &Ref.[\onlinecite{min2}]&\\ 
\hline
0.&3.19622(2)&3.20(1)&& &&&  Second\\
  1.&2.877369(3)(1)&2.88(1)&& &&&Second \\
1.43474(1)&2.7&&& &&&  Second\\

1.5&2.670434(1)        && && &&Second\\
1.68933856...&2.57914(3)  && &&&2.5791695...&Second \\
1.83970(1)&2.5 & &&&&&Second\\

2.&2.407314(1)&2.42(1)  && &&&Second \\
2.16568(1)&2.3 & &&&&&Second\\

2.2&2.275495(1) &2.27(2)&& &&&Second \\
2.4&2.118974(1)&2.11(2)  && &&&Second \\
2.42144(1)&2.1 & &&&&&Second\\

2.523(6)& & &&&2.&&Second\\
2.52513(1)&2. & &&&&&Second\\
2.61361(1)&1.9&&&&&&Second\\
2.68752(1) &1.8&&&&&&Second \\
2.74738(1)&1.7&&&&&&Second\\
2.79370(1)&1.6&&&&&&Second \\
2.80&&1.61(5)&&&&&Second \\
2.82&&1.59(4)&&&&&Second \\
2.82693(1)&1.5&&&&&&Second \\
2.83874(1)&1.45&&&&&&Second \\
2.8446(3)& & &&&1.4182&&TCP\\
2.8448(3)&1.421(3)&&1.4182(55)& &&&TCP\\
2.8477(1)&1.403(2)&&&1.4019(3)&&& TCP \\
2.8502 &0.221(1)&&&&&&First\\
2.890&0.201(1)&&&&&& First\\
2.961&0.152(1)&&&&&&First\\
2.994&0.108(1)&&&&&&First\\
2.998&0.0835(2)&&&&&&First\\
\hline 
\end{tabular} 
\label{tab32}
\end{table}

\begin{figure}[p]
\begin{center}
\leavevmode
\includegraphics[width=5.00 in]{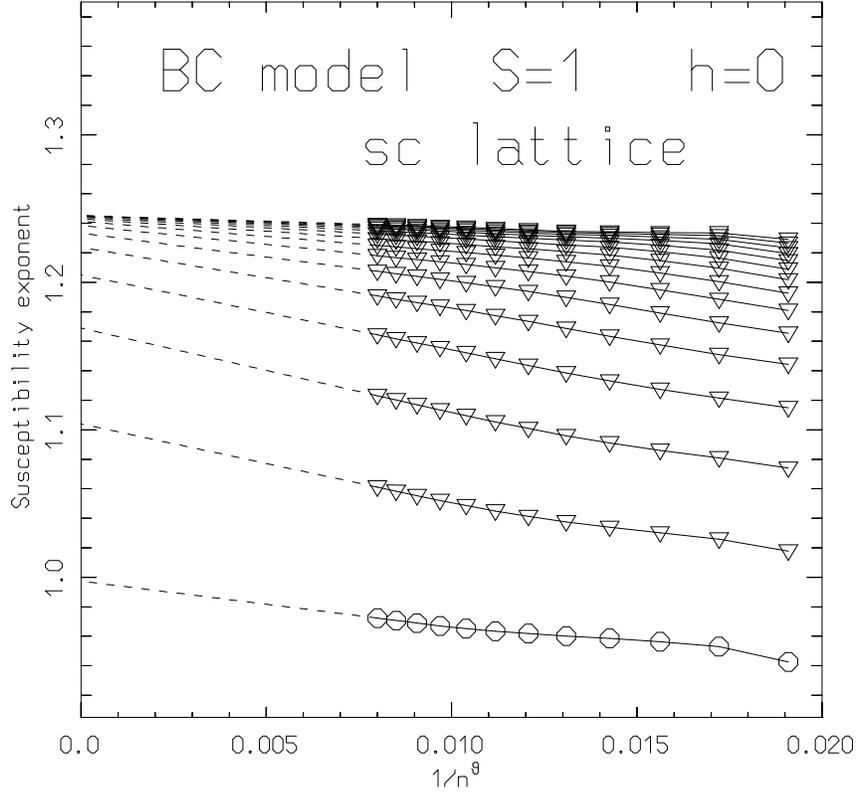}
\caption{\label{graf_esp_gam+phi_sc} BC model with $S=1$ on the $sc$
 lattice, for $h=0$.  MRA estimator-sequences for the exponent
 $\gamma^{(2;0)}(D_{tr};1)$ from the mixed susceptibility
 $\chi_{(2;1)}(K,D;1)$, vs $1/n^{\theta}$ with $\theta=1.5$, for
 several fixed values of $D$.  The lowermost sequence (open circles)
 computed for $D \approx D_{tr}$, can be extrapolated to a value
 compatible with a (logarithmically corrected) MF approximation value
 $\gamma^{(2;0)}(D_{tr};1)=1$. The other sequences (open triangles)
 are evaluated for values of $D$ smaller by $10\%$, $20\%$, etc. }
\end{center}
\end{figure}

When computing the MRA sequences for the critical exponents
$\gamma^{(2;0)}(D;1)$ from the expansion of the ordinary
susceptibility $\chi_{(2,0)}(K,D;1)$, crossover effects begin to
appear only at values of $D$ significantly closer to $D_{tr}$ than in
the $sq$ lattice analysis.  Correspondingly our simple extrapolation
procedure becomes less reliable and its results begin to deviate from
the Ising value of the exponent. This effect however might partly be
due to some residual upward concavity in the curves to be
extrapolated, which we have not attempted to account for.  The ensuing
indication is that the extrapolation of longer expansions is likely to
lead to a closer agreement with the Ising values.  The
Fig. \ref{graf_esp_gam+phi_sc} shows the smoother MRA
estimator-sequences of the exponent $\gamma^{(2;0)}(D;1)$ computed
from the mixed susceptibility $\chi_{(2,1)}(K,D;1)$.  The lowest curve
shown in this figure shows the sequence evaluated at $D \approx
D_{tr}=2.8446$. Its extrapolated value is close to the expected MF
tricritical exponent, $\gamma^{(2;0)}(D_{tr};1) = 1$, within an
uncertainty that might be ascribed to the presence of logarithmic
corrections (unaccounted for by this approach).

\begin{figure}[p]
\begin{center}
\leavevmode
\includegraphics[width=5.00 in]{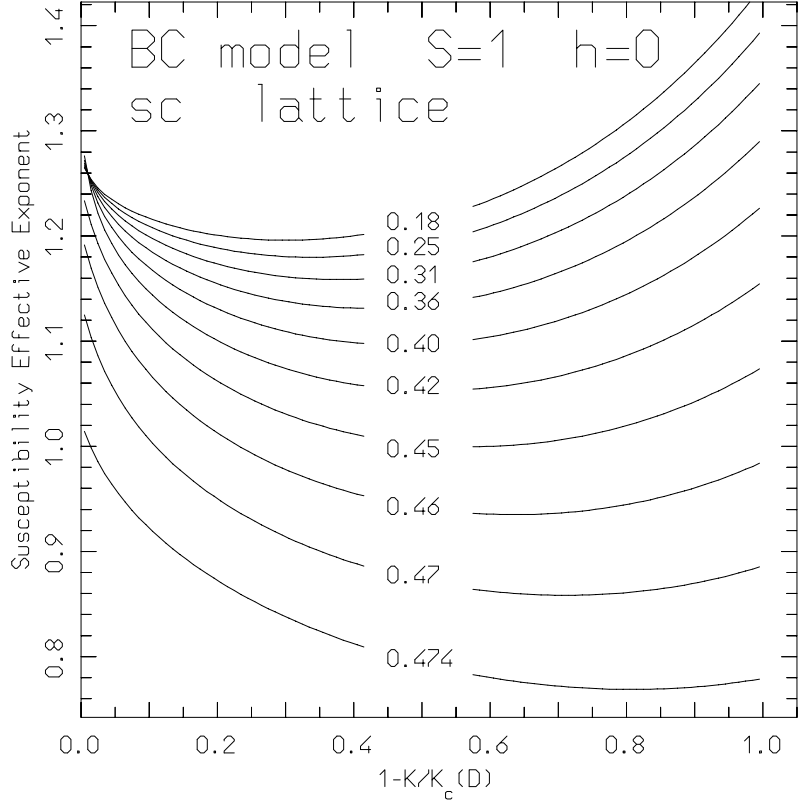}
\caption{\label{graf_eff_esp_gamsc} BC model with spin $S=1$ on the
 $sc$ lattice, for $h=0$.  The effective exponents of the
 susceptibility $\chi_{(2,0)}(K,D;1)$ vs the deviation $1-K/K_c(D)$ from the
 corresponding critical temperatures, for the values of $\Delta/Jq$
 indicated on the curves.  The lowermost curve, computed for
 $\Delta_t/Jq \approx 0.4743$, can be extrapolated to a value of
 $\gamma^{(2)}_t(D_t;1) \approx 1.02$ compatible with the
 (logarithmically corrected) MF approximation value
 $\gamma^{(2)}(D_t;1)=1$.}
\end{center}
\end{figure}

For several values of $\tilde \Delta$ indicated on
the curves, the Fig. \ref{graf_eff_esp_gamsc} shows the behavior of
high-order PAs of the effective exponent $\gamma^{(2;0)}(D;1)$ vs the 
deviation $1-K/K_c(\Delta)$ from the corresponding critical temperatures.
 This computation uses  the expansion of $\chi_{(2,0)}(K,D;1)$.

\begin{figure}[p]
\begin{center}
\leavevmode
\includegraphics[width=5.00 in]{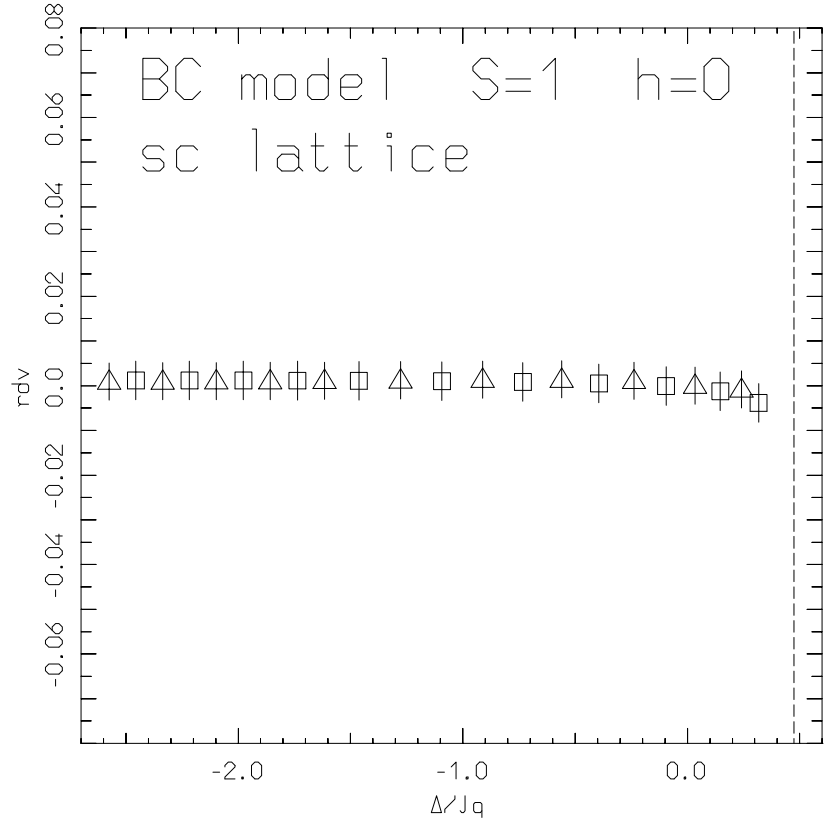}
\caption{ \label{figura_esp_sc} BC model with spin $S=1$ on the $sc$
 lattice, for $h=0$.  The relative deviations $rdv$ from the Ising values
 of the extrapolated MRA estimator-sequences for the  exponents
 $\gamma^{(2;0)}(D;1)$(open triangles) of $\chi_{(2,0)}(K,D;1)$, 
and $\gamma^{(4;0)}(D;1)$  (open squares) of $\chi_{(4,0)}(K,D;1)$.
 The vertical dashed line is the tricritical value $\Delta_{tr}/Jq$
  of the crystal field.}
\end{center}
\end{figure}

  Finally, still for  the $sc$ lattice, the
Fig. \ref{figura_esp_sc} summarizes all these results showing the
relative deviations $rdv$ from the Ising values of the extrapolated
estimator-sequences (defined as for the $sq$ lattice) for the
exponents $\gamma^{(2;0)}(D;1)$ and  $\gamma^{(4;0)}(D;1)$
determined over a large interval of values of $D$ nearly reaching the
TCP.  The  two exponents have been evaluated by extrapolating the
MRA estimator-sequences. Assuming the validity of the
hyperscaling relation
$\gamma^{(4;0)}(D;1)=2\gamma^{(2;0)}(D;1)+d\nu(D;1)$ also the universality of 
$\nu(D;1)$ can be checked.  It might be 
conventionally assumed that the slowdown of convergence rate and thus
the crossover region begins where the deviation of the estimated
exponents from the Ising values\cite{bcisiesse} exceeds $0.5 \%$,
namely at the value $D \approx 0.9 D_{tr}$.
\begin{table}[ht]
  \caption{ The BC model  with $S=1$  on the $bcc$ lattice. 
Phase-contour $T_c=T_c(\Delta;1)$ the analysis of the expansions of the
    ordinary susceptibility (or of its $D$-derivatives). Our estimates
    of the first-order part of the phase-contour are obtained by the
    LT-HT intersection method.    Temperatures
    and  crystal fields are not normalized to the
    coordination number $q$, unlike elsewhere  in the
    text.No independent estimates exist.}
\begin{tabular}{|c| c  c|}
\hline
 $\Delta/J$ & $T/J$&Trans.Ord.\\
\hline 
0.0&4.4512(1) & Second  \\
  1.36015(1)  & 4.&Second \\
  1.83802(1)  & 3.8&Second \\
  2.25133(1)  & 3.6&Second \\
  2.60624(1)  & 3.4&Second \\
  2.76339(1)  & 3.3&Second \\
  2.90771(1)  & 3.2&Second \\
  3.15972(1)  & 3.&Second \\
  3.36555(1)  & 2.8&Second \\
  3.52797(1)  & 2.6&Second \\
  3.59367(1)  & 2.5&Second \\
  3.64940(1)  &2.4&Second \\
  3.69547(1)   &2.3&Second \\
  3.73215(1)   &2.2&Second \\
  3.75975(1)   &2.1&Second \\
  3.778476(1)   &2.&Second \\
   &&TCP \\
3.795&1.897(1)&First\\
3.865& 1.610(1)&First\\
3.957& 1.199(1)&First\\
3.980 & 0.995(1)&First\\
3.997& 0.6892(1)&First\\
\hline 
\end{tabular} 
\label{tab612}
\end{table}

 The Fig. \ref{graf_ampl_sus_sc_S1} 
     plots  vs $\tau$ the critical amplitudes $A_{(2,0)}$ (full
 triangles) of $\chi_{(2;0)}(K,D;1)$ and $-A_{(4,0)}$ (full circles)
 of $\chi_{(4;0)}(K,D;1)$ 
 and shows that they vanish as the TCP is approached.
 
\begin{figure}[p]
\begin{center}
\leavevmode
\includegraphics[width=5.00 in]{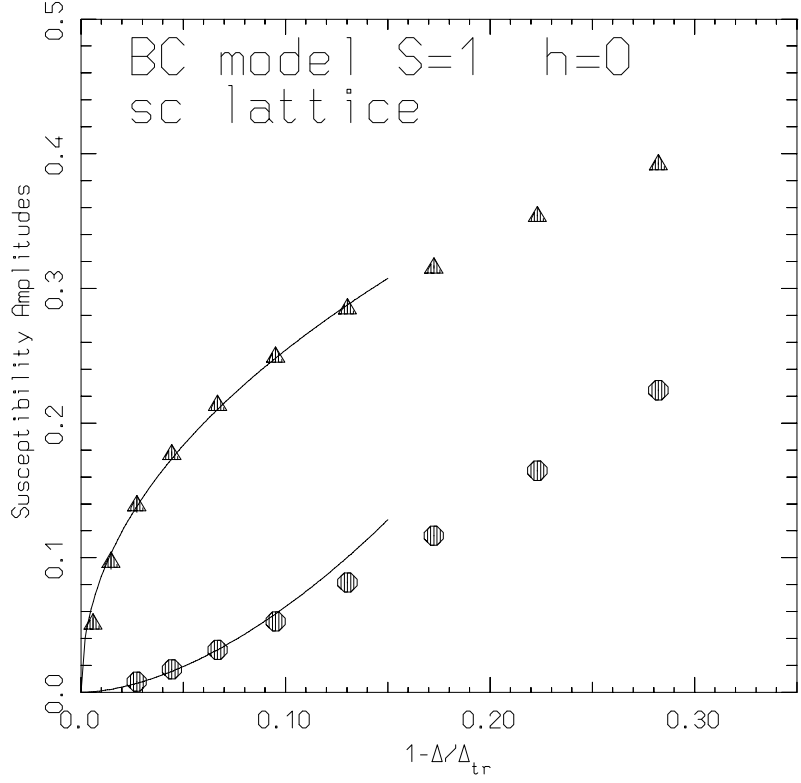}
\caption{ \label{graf_ampl_sus_sc_S1_vs_Delta_rid} BC model with spin
  $S=1$ on the $sc$ lattice, for $h=0$.  The critical amplitudes
  $A_{2,0}$ of $\chi_{(2;0)}(K,D;1)$ (full triangles), and $-A_{4,0}$
  of $\chi_{(4;0)}(K,D;1)$ (full circles)  vs 
  $1-\Delta/\Delta_{tr}$. The solid curves are the
  behaviors of the amplitudes near the TCP 
 predicted by the amplitude-scaling property with
  $\phi_u=2$. }
\end{center}
\end{figure}

 In Fig. \ref{graf_ampl_sus_sc_S1_vs_Delta_rid} the solid lines are
 from a fit comparing the predictions of the amplitude-scaling
 property, with the behaviors the critical amplitudes $A_{(2,0)}$
 (full triangles) of $\chi_{(2;0)}(K,D;1)$ and $-A_{(4,0)}$ (full
 circles) of $\chi_{(4;0)}(K,D;1)$ vs $1-\Delta/\Delta_{tr}$ under the
 assumption that $\phi=2$ and the position of the TCP is well determined in
 Ref. [\onlinecite{deng}].

\begin{figure}[p]
\begin{center}
\leavevmode
\includegraphics[width=5.00 in]{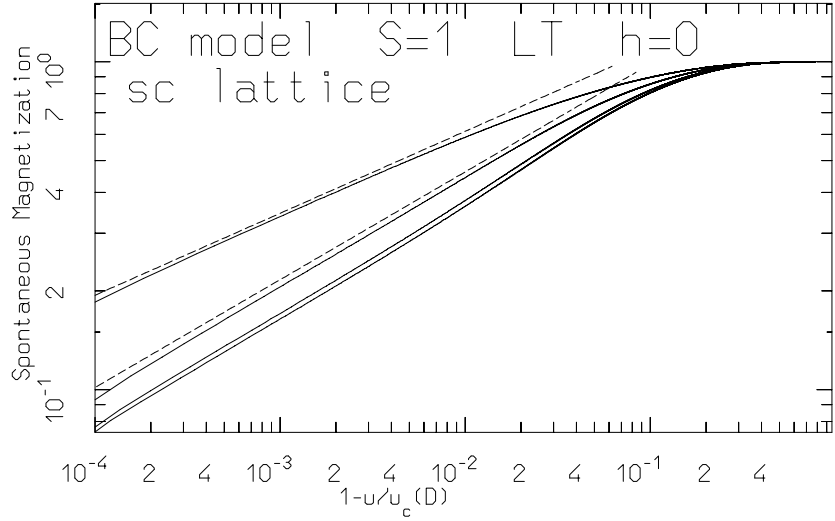}
\caption{ \label{graf_sm_sc_u} BC model with $S=1$ on the $sc$
  lattice, for $h=0$.  A bilog plot of the spontaneous magnetization
  vs $1-u/u_c(D)$.  The uppermost curve is computed for
  $D=D_{tr} \approx 2.031$, while the others for the values
  $D=D_{tr}-0.35$, $D=D_{tr}-0.70$, etc. The various curves are obtained by
  a [5/6] PA of the LT expansion of the spontaneous magnetization.
  The uppermost dashed line indicates an asymptotic value
  characterized by the exponent $\beta=0.25(1)$, as expected at the TCP. The
  lowermost dashed line indicates pure power behavior with
 exponent $\beta=0.33(1)$, shared
  by the curves associated to the  values of $D << D_{tr}$.  }
\end{center}
\end{figure}

\begin{figure}[p]
\begin{center}
\leavevmode
\includegraphics[width=5.00 in]{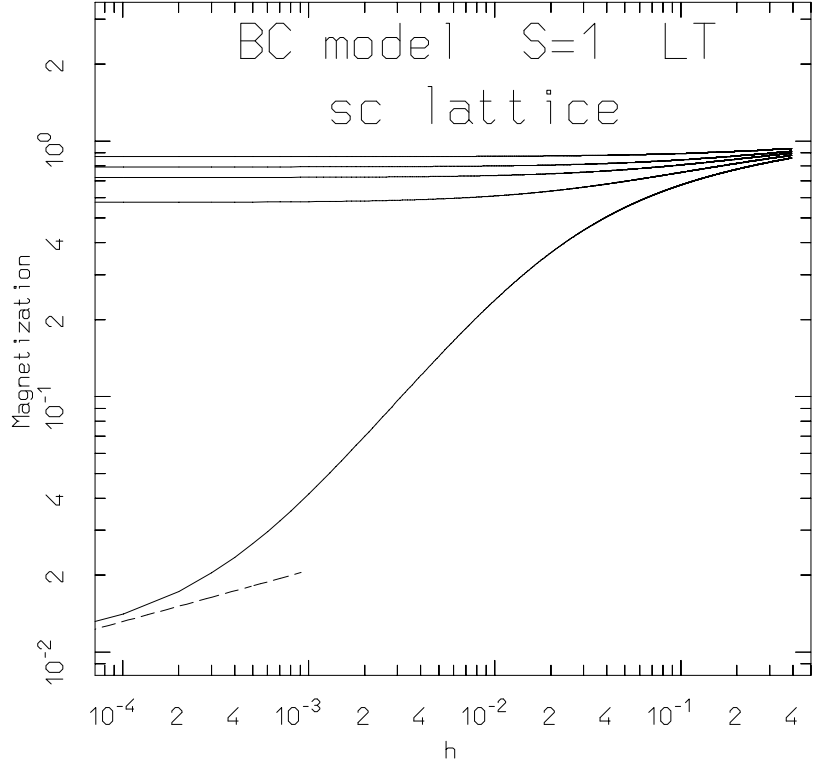}
\caption{ \label{graf_magsp_h_D1354_sc_bil} BC model with $S=1$ on the
  $sc$ lattice. A bilog plot of the magnetization at fixed $u(D)$, vs
  $h$.  The  asymptotic behavior of the lowermost curve 
  characterized by an exponent $\delta_{tr} \approx 5.$ 
is computed along the tricritical isotherm
  i.e.  for $u=u_c(D_{tr})$. The other curves correspond to a sequence of 
  decreasing values of $u(D)$.}
\end{center}
\end{figure}

The LT expansions with $H \neq 0$ for the three-dimensional lattices
 are used to study the expansion of the magnetization in powers of
 $\mu=\exp(-H)$. The (field-grouped) magnetization is resummed by a
 simple [5/6] PA that uses all available coefficients.  In particular,
 Fig. \ref{graf_sm_sc_u} shows the spontaneous magnetization vs $u$
 for several equally spaced values of $ -1.44 \leq D \leq D_{tr}
 \approx 2.0313$. In Fig. \ref{graf_magsp_h_D1354_sc_bil} the same PA
 is used to represent the behavior of the magnetization vs $h$, at
 fixed $D=D_{tr} \approx 2.0313$ and at a few other values of $D$ such
 that $0 \leq u(D) \leq u_c(D_{tr})$.  The asymptotic behaviors of the
 curves shown in these figures are consistent with the expected values
 of the exponents $\beta_{tr}$ and $\delta_{tr}$. Longer LT expansions
 would be needed to obtain also reliable estimates of the
 uncertainties.

\begin{figure}[p]
\begin{center}
\leavevmode
\includegraphics[width=5.00 in]{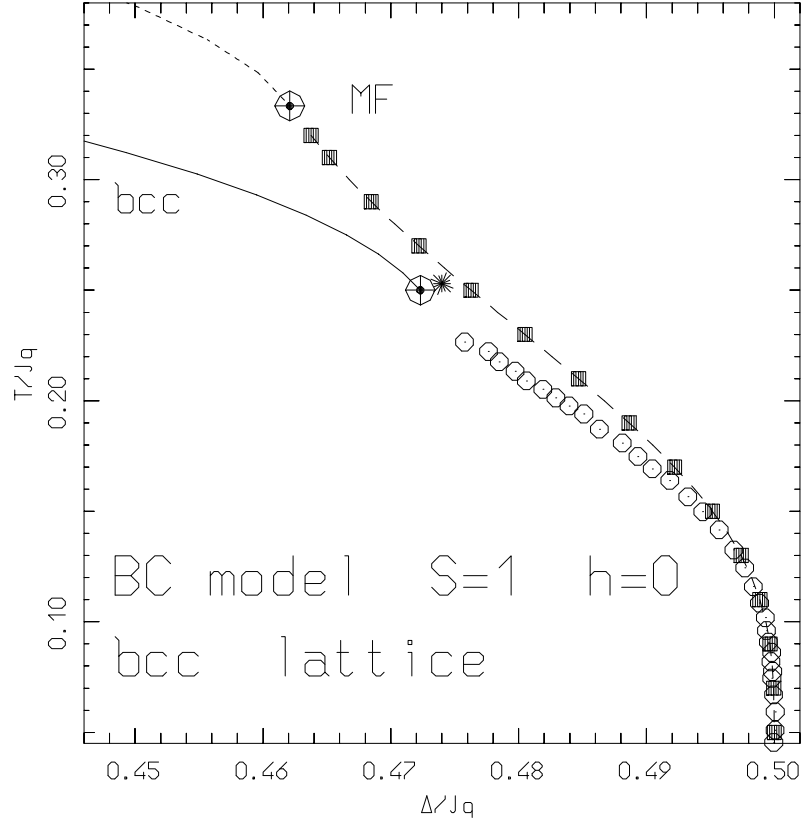}
\caption{ \label{Graf_fig3_first_bcc} BC model with spin $S=1$, on the
 $bcc$ lattice for $h=0$. Phase diagram in the $(\Delta/Jq,T/Jq)$
  plane, computed by series and in the MF
 approximation. Big crossed open circles are TCPs, the star is taken
 from Ref. [\onlinecite{grollau}].  The upper short-dashed line is the
 MF critical border. Full squares connected by a long-dashed line are
 the MF first-order contour.  The solid curve is the series result for
 the critical phase-contour.  Open circles are points of the first-order part
 of the phase-contour.}
\end{center}
\end{figure}

Generally, completely similar results are obtained from the analysis
 of the $bcc$ lattice, so that for brevity most of the corresponding
 figures are omitted.  For this lattice (see the Table
 \ref{tab612} and the Fig. \ref{Graf_fig3_first_bcc}), the series
 analyses show a faster convergence and a narrower crossover region
 than in the $sc$ case.  In spite of these more favorable features, we
 have continued to attach quite generous error bars also to the
 estimates for this lattice.

A single estimate\cite{grollau} of the TCP position for the $bcc$
lattice: $\tilde \Delta_{tr}=3.792$, $\tilde T_{tr}=2.024$ (with
undetermined uncertainties) can be found in the literature.  Its
location is indicated by a star in the
Fig. \ref{Graf_fig3_first_bcc}. The results of the series analysis
suggest that slightly smaller values for both coordinates might be
preferred. We conjecture that $3.7785 \leq \tilde \Delta_{tr} \leq
3.795$ and $2. \geq \tilde T_{tr} \geq 1.90(3)$.
 
 The first-order part of the phase-boundary for the $bcc$ system,
obtained  as in the $sc$ lattice analysis, is also shown in this figure.  
We have used the temperature-grouped LT expansion of the
free-energy that extends to order $u^{56}$.  Quite reasonable estimates
are obtained even very near the expected TCP. Also for the $bcc$
lattice, as for the $sc$ lattice, no independent results for this
region of the phase-boundary are available for comparison.  

Results similar to those of the Fig. \ref{figura_esp_sc} are obtained
also in the case of the $bcc$ lattice.  For $D \lesssim .4 D_{tr}$, the
estimated limits of the MRA sequences for the exponents show relative
deviations $ < 10^{-3}$ from the Ising values.

Moreover, in the case of the $bcc$ lattice, properties similar to
those of the $sc$ lattice are shown in Fig. \ref{TcbccvsX} for the
phase diagram in the concentration-temperature plane.  The
series estimate $X_{tr}= 0.69(1)$ of the critical concentration, is
obtained for the $bcc$ lattice. For brevity, no figure is devoted to
the concentration susceptibility $ Y(K,D;1)$ because its behavior is
completely similar to that for the $sc$ lattice.

 The Fig. \ref{graf_ampl_sus_bcc_S1} shows the dependence on $\tau$ of
    the critical amplitudes $A_{(2,0)}$ (full triangles) of
    $\chi_{(2;0)}(K,D;1)$ and $-A_{(4,0)}$ (full circles) of
    $\chi_{(4;0)}(K,D;1)$,
    for $\tau$ in a small vicinity of the TCP of the $bcc$ lattice.

\begin{figure}[p]
\begin{center}
\leavevmode
\includegraphics[width=5.00 in]{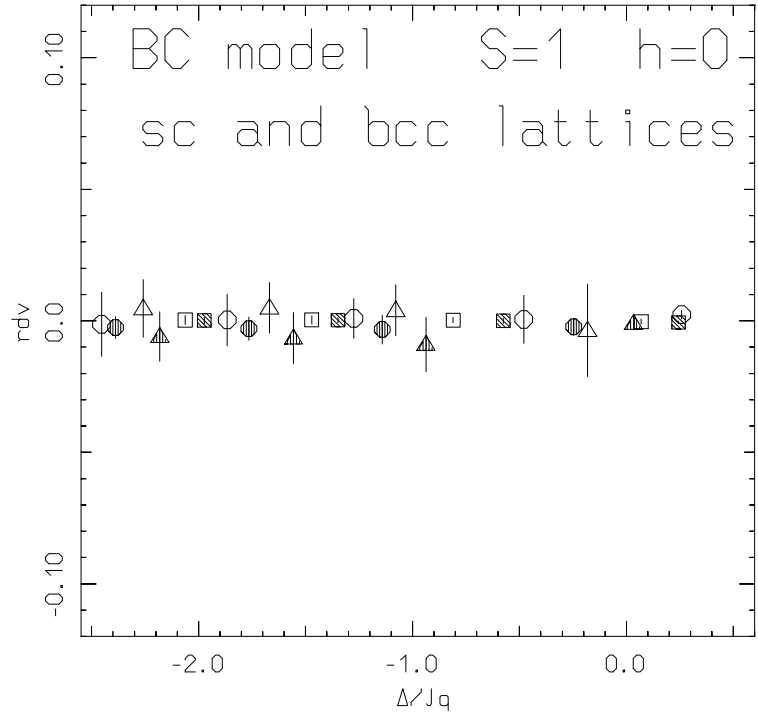}
\caption{ \label{graf_ratios_sc_bcc_S1} BC model with spin $S=1$ on
the $sc$ and the $bcc$ lattices, for $h=0$. 
 The relative deviations $rdv$ from the Ising values for the
estimates of the universal ratios of critical amplitudes ${\cal
I}^+_{6}$, ${\cal I}^+_{8}$ and ${\cal J}^+_8$ vs $\Delta/Jq$.
These quantities are  open circles, open triangles and
open squares respectively, in the case of the $sc$ lattice. The
analogous full symbols shoe the same ratios for the $bcc$
lattice. }
\end{center}
\end{figure}

The estimates of the universal ratios of critical amplitudes ${\cal
  I}^+_{6}$, ${\cal I}^+_{8}$ and ${\cal J}^+_{8}$, are shown in the
  Fig. \ref{graf_ratios_sc_bcc_S1}, for both the $sc$ and the $bcc$
  lattices. For these ratios, it is convenient to plot vs $\tilde
  \Delta$ the relative deviations $rdv$ from the Ising values.  They
  are quite small up to $\tilde \Delta \lesssim 0.4$, whereas for
  larger values of $\Delta$, it is the convergence rate of our methods
  that slows down.

\section{The   BC model with spin $S=3/2$  on the $sq$ lattice}
The phase diagram in the $(\tilde T,\tilde \Delta)$ plane for the spin $S=3/2$
 system on the $sq$ lattice (together with those for the $sc$ and the
 $bcc$ lattices) is shown in Fig.  \ref{figura_fase_32_Ing}.  The
 phase-boundary in the MF approximation, which is given by the curve
 highest in temperature is also reported for comparison.

   Also for this value of the spin, very accurate determinations of
the phase-boundary and exponents up to $\tilde \Delta \lesssim 0.45$ and
for $\tilde \Delta \lesssim 0.55$ are obtained by forming MRAs of the HT
expansion of $\chi_{(2;0)}(K,D;3/2)$. In the range $0.45 \lesssim
\tilde \Delta \lesssim 0.55$, the MRA sequences oscillate and thus their
extrapolations are not straightforward. In Sect. IV D, this fact was
related to the presence of nearby singularities accompanied by the
weakening of the critical amplitudes for $\tilde \Delta \approx 1/2$.
 It is suggested by the MF approximation and subsequently confirmed by
a MC simulation\cite{xavier}, that a critical end-point of the
first-order transition in the LT region of the phase diagram lies very
close to the $sq$ critical border, at $\tilde \Delta \approx 0.492$ with
$S^2 \tilde T \approx 0.09$.  It is then likely that the asymptotic behavior 
of the HT expansion coefficients of the susceptibility  is
sensitive also to this nearby singularity and that  in a vicinity of this
value of $\tilde \Delta$, it is more convenient to form the MRAs of
$\chi_{(2;1)}(K,D;3/2)$ instead of  those of $\chi_{(2;0)}(K,D;3/2)$
to determine the phase-boundary.

The numerical results for the phase-boundary can be used, together
 with the known Ising exponents, to bias the determination of the
 critical amplitudes of $\chi_{(2;0)}(K,D;3/2)$ and
 $-\chi_{(4;0)}(K,D;3/2)$, that are plotted for graphical convenience
  vs $(1+exp(2d))^{-1}$ in Fig. \ref{graf_ampl_sus_sq_S1}, together
 with the analogous quantities for the spin $S=1$ system to contrast
 the respective behaviors. After decreasing to a very small minimum at $\tilde
 \Delta \approx 1/2$, the amplitudes for $S=3/2$ system sharply
 rise. This behavior is not surprising because for large values of
 $\tilde \Delta$, the spin $S=3/2$ model reduces to a spin $S=1/2$
 model, for which these amplitudes are sizable.

\begin{figure}[p]
\begin{center}
\leavevmode
\includegraphics[width=5.00 in]{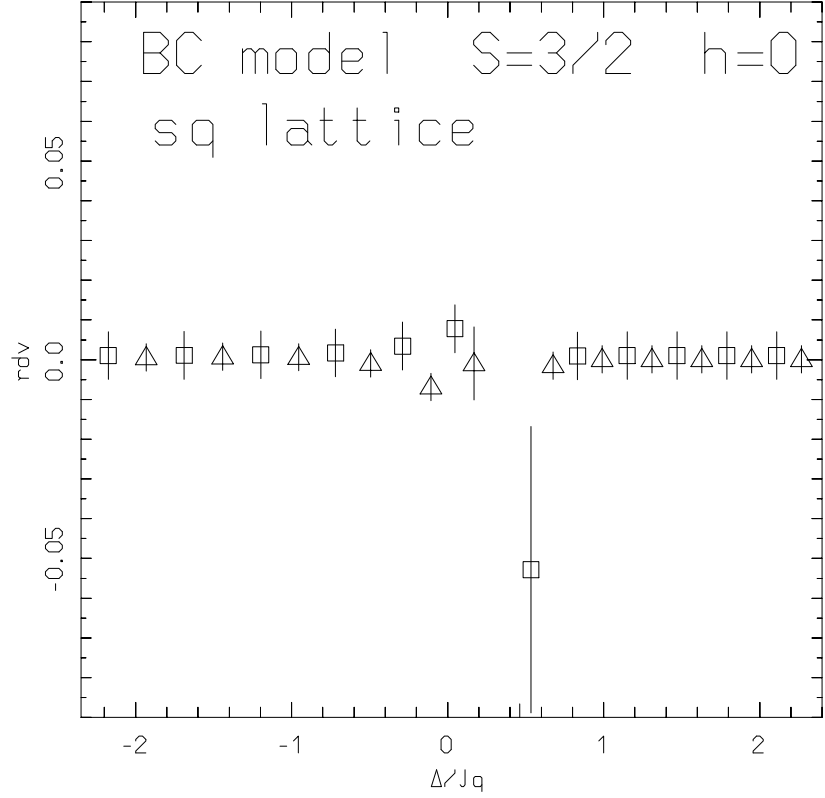}
\caption{ \label{figura_esp_sq_32} BC model with spin $S=3/2$ on the
  $sq$ lattice for $h=0$. The relative deviation $rdv$ of the extrapolated 
  MRA estimator-sequences from the Ising values for the exponents
  $\gamma^{(2;0)}(D;3/2)$(open triangles) of $\chi_{(2;0)}(K,D;3/2)$
and  $\gamma^{(4;0)}(D;3/2)$(open squares) of $\chi_{(4;0)}(K,D;3/2)$ vs 
 $\Delta/Jq$. }
\end{center}
\end{figure}

  As shown in Fig. \ref{figura_esp_sq_32}, only for values of $\tilde
\Delta \approx 0.5$, the estimates of the exponent
$\gamma^{(2;0)}(D;3/2)$ along the phase-boundary, obtained
extrapolating the MRA sequences with the Ansatz Eq. (\ref{ansatzesp}),
deviate up to a few percent from the expected Ising value.

  The fluctuations in a small interval around $\tilde \Delta=1/2$, are
not so strong as to indicate a violation of universality, but only a
slower convergence rate of our numerical procedures.

\begin{figure}[p]
\begin{center}
\leavevmode
\includegraphics[width=5.00 in]{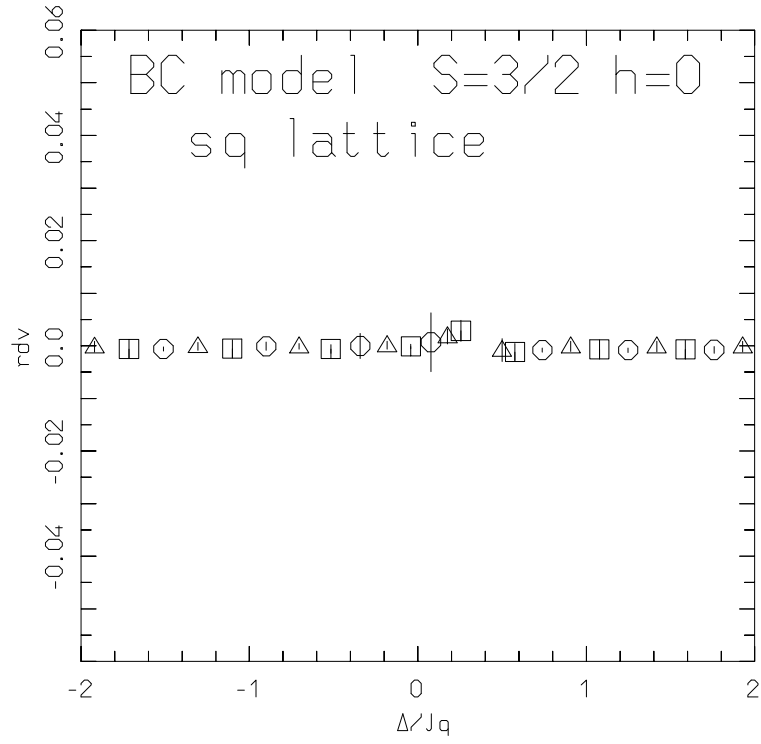}
\caption{ \label{graf_ratios_S32_sq} BC model with spin $S=3/2$ on the $sq$
  lattice for $h=0$.  The relative deviations $rdv$ of the universal
  ratios of critical amplitudes ${\cal I}^+_{6}$ (open circles),
  ${\cal I}^+_{8}$ (open triangles) and ${\cal J}^+_{8}$ (open
  squares) from the corresponding $2d$ Ising values  vs
  $\Delta/Jq$   in the range  $-2<\Delta/Jq<2$. }
\end{center}
\end{figure}

In Fig. \ref{graf_ratios_S32_sq}, the relative deviations $rdv$ from
the Ising values for the universal ratios of critical amplitudes
${\cal I}^+_{6}$, ${\cal I}^+_{8}$ and ${\cal J}^+_{8}$, are plotted
vs $\tilde \Delta$ in the interval $-2. \leq \tilde \Delta \leq 2$ to
show that they remain Ising-like over a large interval of values of
$\tilde \Delta$ and display no serious anomalies that might signal the
presence of a TCP.  At larger values of $\tilde \Delta$, we observe a
slowdown of the convergence rate of the approximations rather than a
failure the universality properties.

  In conclusion, for this lattice, no indications appear of a TCP with its
  ensuing first-order line, so that the behavior of the model is
 Ising-like over a large range of values of $D$ and the qualitative
 predictions of the MF approximation for the half-odd-spin BC models
 are fully confirmed.

\section{The   BC model  with spin $S=3/2$  in $3d$}
 The phase-boundaries of the spin $S=3/2$ systems for both the $sc$
and the $bcc$ lattices are  drawn in the same Fig.
\ref{figura_fase_32_Ing}.  For the former system, a critical-end-point
at\cite{grollau32} $(\tilde \Delta \approx 0.491, S^2 \tilde T \approx 0. 095)$
 or at\cite{ilko} $(\tilde \Delta \approx 0.4922, S^2 \tilde T \approx 0.103$,
  not far from the  phase-contour (but less close to it than  in the
$sq$ case) is likely to influence the convergence of the simplest MRAs
formed with the expansions of the ordinary susceptibility
$\chi_{(2;0)}(K,D;3/2)$. A similar problem might occur for the $bcc$
lattice. As observed for the $sq$ lattice, in a neighborhood of
$\tilde \Delta=1/2$, a faster convergence in the determination of the
critical boundary can be achieved using the HT expansions of the
$D$-derivative of the susceptibilities.

\begin{figure}[p]
\begin{center}
\leavevmode
\includegraphics[width=5.00 in]{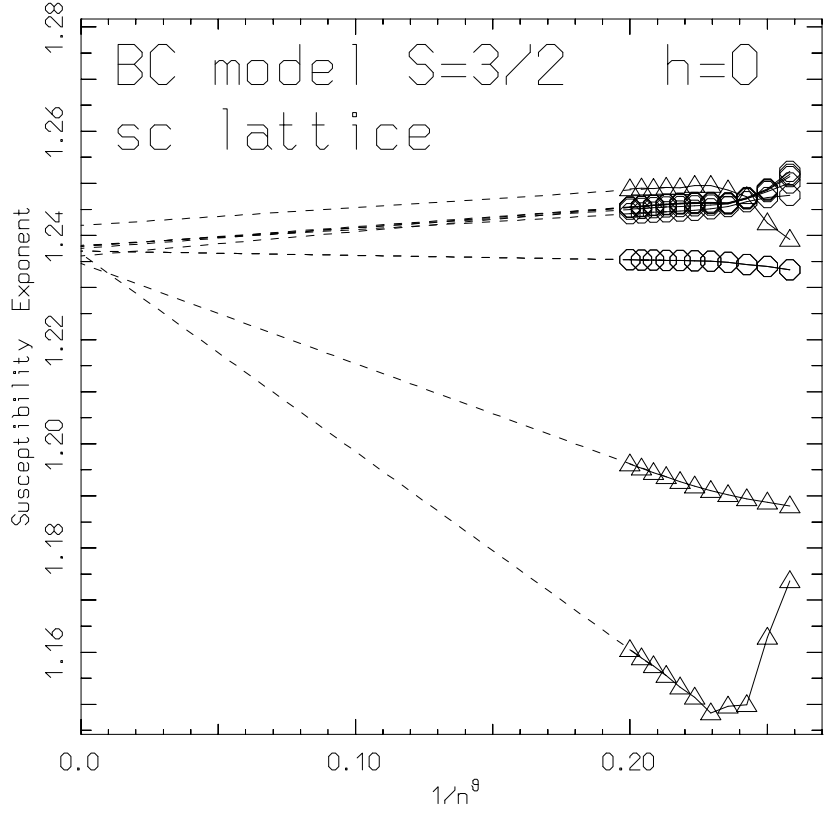}
\caption{ \label{graf_sc_Z_S32} BC model with spin $S=3/2$ on the $sc$
lattice, for $h=0$.  The MRA estimator-sequences for the 
exponent $\gamma^{(2;0)}(D;3/2)$ vs $ 1/n^{\theta}$, with $n$ the
number of terms included in the series and $\theta= 0.5$, for several
values of $-1.5 \lesssim \Delta/Jq \lesssim 0.85$. The terms of the
sequences are in general indicated by open circles, except those with
$ 0.44 \lesssim \Delta/Jq \lesssim 0.57$ indicated by open triangles.
To profile the behavior of each sequence, the symbols of the
successive terms are connected by segments. The dashed lines are the
sequence extrapolations to large $n$.}
\end{center}
\end{figure}

In Fig. \ref{graf_sc_Z_S32}, the sequences of MRA estimators of the
critical exponent $\gamma^{(2;0)}(D;3/2)$ are plotted vs
$1/n^{\theta}$ with $\theta=0.5$, for the $sc$ lattice.  The diagram
obtained for the $bcc$ lattice is completely similar and therefore it
needs no separate illustration.  The MRA sequences  are computed for
several values of $-1.5 \lesssim \tilde \Delta \lesssim 0.85$, 
using Eq. (\ref{zinnesp}).  As in the analogous
figures for the other BC systems considered so far, a solid line
interpolates among the terms of each sequence, while the last few
points of the sequence are extrapolated to large expansion order by a
fit of Eq. (\ref{espasint}).
 It is clear that in a wide range of values of $D$, no
significant anomalies are observed all along the phase-boundary and
the exponent estimates remain Ising-like within $\approx 10^{-3}$.
The terms of the MRA sequences for  $\tilde \Delta \approx 0.443$ and  
$\tilde \Delta \approx 0.507$,  are indicated by triangles.

For the $bcc$ lattice, the next Fig.  \ref{graf_bcc_espef_S32} is
 devoted to the effective exponent of the susceptibility. The curves
 are computed for various values of $D$ by forming the highest-order,
 defect-free, diagonal or near-diagonal PAs of the HT expansions of
 the effective exponents and each one is plotted vs the corresponding
 deviation $1-K/K_c(D)$ from the critical temperature.  The curves
 show that the effective critical exponents are Ising-like in a
 vicinity of $K_c(D)$ that becomes very narrow as $\tilde \Delta = 1/2$ is
 approached from below, while it expands for smaller or larger values of
 $\tilde \Delta$. The figure obtained for the $sc$ lattice, is
 completely similar and therefore is omitted.

 It is also useless to include for the $sc$ and the $bcc$ lattices,
figures summarizing the relative deviations $rdv$ from the Ising values
of the extrapolated MRA estimator-sequences for the exponents
$\gamma^{(2;0)}(D;3/2)$ of $\chi_{(2;0)}(K,D;3/2)$, and
$\gamma^{(4;0)}(D;3/2)$ of $\chi_{(4;0)}(K,D;3/2)$.
over a large range of values of $D$. It is enough to remark that the
 BC system is confirmed to remain Ising-like along the phase-boundary
 up to large values of $D$, and the few-percent fluctuations of the
 exponent estimates, observed in a neighborhood of $\tilde \Delta=1/2$ are
 certainly due only to a slowdown of the convergence rate of the
 approximations and should not be taken as indications of anomalies.

Similarly the figures showing the relative deviations from the 
Ising values for the universal ratios 
of critical amplitudes ${\cal I}^+_{6}$, ${\cal I}^+_{8}$ and ${\cal
J}^+_{8}$ on the $sc$ and the $bcc$ lattices 
 can be omitted. Again no such anomalies are observed in the estimates
 that might suggest the presence of a TCP.  As for the
 exponents, the small fluctuations in the estimates are likely to be
 due only to a local slowdown of the approximations convergence rate.

 In the Figs. \ref{graf_ampl_sus_sc_S1} and
\ref{graf_ampl_sus_bcc_S1}, the critical amplitudes of
$\chi_{(2;0)}(K,D;3/2)$ and $\chi_{(4;0)}(K,D;3/2)$ are plotted vs 
$(1+exp(2d))^{-1}$ for graphical convenience,
 for the $sc$ and of the $bcc$ lattices, to
emphasize the qualitative difference of their behavior from that of
the spin $S=1$ case. Just as observed for the $sq$ lattice, after
a fast decrease as $\tilde \Delta \to 1/2$, the amplitudes rise sharply
with $\tilde \Delta$ as they should since for large values of this field,
the spin $S=3/2$ BC model tends to a spin $S=1/2$ model.

  In conclusion, also in $3d$ the behavior of the spin $S=3/2$ model
appears qualitatively different from that of the spin $S=1$ model. As
$D$ varies, the exponent estimates remain Ising-like and thus
consistent with the expected universality properties and with a
previous MC simulation\cite{lara} confirming the absence of a TCP
followed by a first-order line.  The validity of the structural
prediction of the MF approximation, at least for the model with the
lowest non-trivial half-odd spin value is thereby confirmed.

\section{ Summary and Conclusions }

We have derived an extensive body of HT and LT expansions for many
thermodynamical observables of the BC model with spin $S=1$ and
$S=3/2$ on the $sq$, the $sc$ and the $bcc$ lattices in presence of a
magnetic field.  Our aim was to understand the potential of an
approach to this model by long series expansions  and possibly to
demonstrate how it can be employed to map carefully a very wide region
of the phase diagram, testing at the same time the universality
properties  for the critical exponents and for
appropriate ratios of the critical amplitudes. The expansions of the
moments of the correlation-function remain to be computed, so that
only assuming the validity of hyperscaling, we had access to the
correlation-length exponents, but of course not to the corresponding
critical amplitudes.

 Our analyses succeed in mapping the phase diagrams both in the
 anisotropy-temperature and in the concentration-temperature planes as
 well as in the verification, over large regions of the parameter
 space, of several expected universality properties. Moreover, although
 presented here only for the particular cases of spin $S=1$ and
 $S=3/2$ (the latter not studied in detail so far by series methods),
 they confirm the validity of the general indication emerging from the
 MF approximation, that a TCP can occur only in BC models with integer
 values of the spin.

  Unlike what is observed for the $3d$  lattices, 
 for the system with spin $S=1$ on the $sq$ lattice we have
 pointed out that the crossover region is very wide and the shape of
 the critical phase-contour in the concentration-temperature plane
 deviates markedly from the MF behavior. These remarks might be
 confirmed by simulations, that however have not been reported so far.
 Moreover, as in $3d$, the concentration
 susceptibility evaluated along the critical contour is finite and rises very
 steeply as $\Delta \to \Delta_{tr}$.  

  At HT, mostly in the case of the $sq$ lattice with spin $S=1$, the
series-analysis approach meets with convergence problems in the
tricritical region.  As a result, we have not been able to improve the
accuracy of the present determinations of the TCP parameters obtained
by simulation or transfer-matrix methods, but we could only test their
consistency with our analysis. We have conjectured that this
difficulty might reflect the clustering of unphysical singularities
nearby the series convergence border, accompanied by the vanishing of
the critical amplitudes as the TCP is approached, (as it is prescribed
by amplitude-scaling).  In spite of the absence of a TCP, a similar
problem arises also in the case of spin $S=3/2$, in which a drastic increase
of the complexity of the corrections to scaling takes place when the
phase-contour approaches the critical end-point of the first-order
transition line in the ordered phase.  However, for both values of the spin,
these convergence problems of our approximations can be alleviated by
the simple prescription of studying the $D$- or $K$-derivatives of the
susceptibility instead of the susceptibility 
 to determine the phase-boundary, so that finally
the range of validity of the standard single-variable methods of
series analysis can be extended up to a small distance from the point
$\tilde \Delta \approx 1/2$.  Thus it seems that presently in this
region both the transfer-matrix and the simulation methods might be
more promising than the simplest series methods, because they do not
need to keep a strict control of the leading singularities to
determine the critical parameters.

 For spin $S=1$, at LT the only known method for mapping out the
first-order part of the phase-boundary, based on the intersection of
the LT and HT approximations of the free-energy in their expected
common region of approximate validity, has given access to this part
of the phase-contour also for the $sc$ and the $bcc$ lattices, so far
not obtained by other numerical methods.  The procedure however may loose
precision in a small vicinity of the TCP.

 In the future, it might be appropriate to return on this subject, not
 only because the BC models with spin $S \ge 2$ are still unexplored
 by series methods, but also because several features of the behavior
 of the systems considered here need further illustration and moreover
 the simplest single-variable approximation methods employed in this
 study might perhaps have missed part of the information content of
 the expansions.  In spite of all the limitations of our analysis, we
 are confident that the series data we have derived, might remain as a
 necessary tool for further extensions and for alternative analyses.

\appendix
\section{  Mean Field description of the phase structure}
The simplest MF-like  description\cite{landau,aharony} of a typical 
 tricritical behavior is obtained starting with a Landau-Ginzburg Hamiltonian
 for a scalar field $\phi(x)$ of the form
\begin{equation}
 {\cal H}_{LG}(\phi)=\frac{1}{2}(\partial_k\phi)^2+V(\phi)
\label{Hlandginz}
\end{equation}
 with potential 
\begin{equation}
V(\phi)= -H\phi+ \frac{A_2}{2}\phi^2 +\frac{A_4}{4}\phi^4+\frac{A_6}{6}\phi^6
\label{landginz}
\end{equation}
The coefficients $A_{2i}$ are functions of the parameters that
characterize the system, for example the temperature $T$ etc. In the
$H=0$ plane, a MF-like approximation is obtained restricting to
$x$-independent configurations of the field $\phi$ (i.e. neglecting
its spatial fluctuations) and assuming $A_6 > 0$ to ensure
thermodynamical stability. One can then conclude\cite{aharony} that a
TCP occurs for values of the parameters such that $A_2=A_4=0$. It
separates a second-order line described by the equation $A_2=0$ ($A_4
> 0$), from a first-order line described by the equation $A_2= 3
A_4^2/16 A_6$, (with $A_4 < 0$).

A more realistic spin-dependent description of the essential features
 of the phase diagram of the BC model is
 formulated\cite{Blume,Capel,BEG} turning to the standard MF
 approximation whereby the Hamiltonian of Eq. (\ref{hamBC}) 
with spin $S$ is replaced by the
 solvable trial Hamiltonian
\begin{equation}
{\cal H}_{0}(\eta, \Delta, T; \{s_i\}) =
 -\frac{\eta}{S} \sum_i s_i + \frac{\Delta}{S^2} \sum_i s^2_i
\end{equation}
 of non-interacting spins in external fields $\eta$  and $\Delta$.

If $<>_0$ indicates the ensemble average with respect to ${\cal H}_0$,
 the specific magnetization  $m(\eta, \Delta, T) \equiv <s_0>_0$ is
\begin{equation}
m(\eta, \Delta, T) = \frac{S}{\beta}\frac{\partial {\rm ln}
z_0(\eta,\Delta,T)}{\partial \eta}
\label{phi2e}
\end{equation}
with
\begin{equation}
z_0(\eta, \Delta, T) = \sum_{s=-S}^S\exp(\beta \eta \frac{s}{S}- D
\frac{s^2}{S^2})
\label{eqz}
\end{equation}
and $D = \beta \Delta$.

The convexity inequality\cite{bogo1,bogo2}
\begin{equation}
f(H, \Delta, T) \le \phi(H, \eta, \Delta, T) \equiv 
-\frac{1}{\beta}{\rm ln}(z_0) + \frac{1}{N}< {\cal H} - {\cal H}_0 >_0
 \end{equation}
is used to optimize the choice of the effective magnetic field $\eta$.
We have
\begin{equation}
\phi(H, \eta, \Delta, T) =  -\frac{1}{\beta}{\rm ln}(z_0) - 
\frac{1}{2}Jq\frac{m^2}{S^2} + (\eta + H)\frac{m}{S}
\label{phi2}
\end{equation}
with $\phi$  interpreted as MF free-energy.
The extremality  condition 
\begin{equation}
\frac{\partial \phi}{\partial \eta} = 0
\label{phieta}
\end{equation}
is taken as equilibrium condition of the MF theory.

Since from Eq. (\ref{phi2})
\begin{equation}
\frac{\partial \phi}{\partial \eta} = 
-\frac{1}{\beta}\frac{\partial {\rm ln}z_0}{\partial \eta} + \frac{m}{S} +
\frac{\partial m}{\partial \eta} (-\frac{qJ}{S^2} + \frac{\eta + H}{S})
\end{equation}
 from Eqs.(\ref{phi2e}) and (\ref{phieta}), it follows
\begin{equation}
\eta = \frac{Jqm}{S} - H
\label{etamh}
\end{equation}
The dependence on $\eta$ can be eliminated computing $\eta(m, \Delta,
 T)$ from Eq. (\ref{phi2e}) and substituting this expression in $\phi$ to
 form the MF Helmholtz free-energy $f$
\begin{equation}
f(H, \Delta, T) \equiv \phi(H, \eta(m, \Delta, T), \Delta, T)
\label{feqphy}
\end{equation}

which does not depend on $m$, since
\begin{equation}
\frac{\partial \phi(H, \eta(m, \Delta, T), \Delta, T)}{\partial m} = 
\frac{\partial \phi(H, \eta, \Delta, T)}{\partial \eta} \frac{\partial \eta(m,\Delta,T)}{\partial m} = 0
\label{fm}
\end{equation}
due to Eq. (\ref{phieta}).

By Eq. (\ref{phi2}) 
\begin{equation}
f(H, \Delta, T) = G(m, \Delta, T) + \frac{m}{S} H
\label{legendre}
\end{equation}
and
\begin{equation}
G(m, \Delta, T) = -\frac{1}{\beta}{\rm ln}z_0(\eta(m,\Delta,T), \Delta, T)
    - \frac{1}{2}Jq\frac{m^2}{S^2} + \frac{m}{S}\eta(m, \Delta, T)
\label{eqG}
\end{equation}
 denotes the Gibbs free-energy, the appropriate potential to study
 equilibrium at constant $H$.  Eq. (\ref{legendre}) is the Legendre
 transformation relating the potentials $f(H, \Delta, T)$ and $G(m,
 \Delta, T)$, that implies
\begin{equation}
\frac{\partial G(m, \Delta, T)}{\partial m} = -\frac{H}{S}
\end{equation}
Eq. (\ref{phi2e}) for $\eta(m, \Delta, T)$ cannot be solved exactly,
but for small $m$, we are allowed to expand $\eta(m, \Delta, T)$ and
$G(m,\Delta, T)$ in powers of $m$.

 Defining \cite{plascak}
$\alpha_k = \frac{\partial^k {\rm ln}(z_0)}{\partial \eta^k}|_{\eta=0}$,
and inverting Eq. (\ref{phi2e})
\begin{equation}
\eta(m, \Delta, T) = \frac{\beta}{\alpha_2}\frac{m}{S} - 
\frac{\beta^3\alpha_4}{6 \alpha_2^4}\frac{m^3}{S^3} + 
 \beta^5(\frac{\alpha_4^2}{12 \alpha_2^7} - \frac{\alpha_6}{120\alpha_2^6}) 
\frac{m^5}{S^5} + O(m^7)
\label{etam}
\end{equation}
The expansion of the Gibbs free-energy for small $m$ is
\begin{equation}
G = A_0 + \sum_{i=1} \frac{A_{2i}}{2i} (\frac{m}{S})^{2i}.
\label{Gfree}
\end{equation}
Using Eq. (\ref{eqG}), it follows
\begin{equation}
A_0 = -\frac{\alpha_0}{\beta},\qquad
A_2 = \frac{\beta}{\alpha_2} -Jq,\qquad
A_4 = -\frac{\beta^3}{6}\frac{\alpha_4}{\alpha_2^4},\qquad
A_6 = \frac{\beta^5}{120\alpha_2^7}(10\alpha_4^2 - \alpha_2\alpha_6)
\label{Akgen}
\end{equation}

The critical phase-boundary is obtained for $A_2 = 0$
 with $A_4 > 0$ and so  is described by the equation
\begin{equation}
\frac{\beta}{\alpha_2} = qJ
\label{2ndcr}
\end{equation} 
which is exact because along the critical line  $m=0$.

In the $H=0$ plane, the first-order part of the phase-boundary can
 also be studied in this expansion, but only nearby its end-point,
 where $m$ is small. A first-order transition occurs where
 $G(0,\Delta,T) = G(m,\Delta,T)$ with $G'(m,\Delta,T) = 0$.  For small
 $m$ with positive $A_6$, the $O(m^8)$ term can be neglected in
 Eq. (\ref{Gfree}), so that the first-order transition is obtained for
\begin{equation}
A_2 \approx \frac{3}{16} \frac{A_4^2}{A_6}, \qquad
\frac{m^2}{S^2} \approx -4 \frac{A_2}{A_4}
\label{firstcr}
\end{equation}
These equations are valid only for small $m$. As $T$ is lowered from
 the end-point of the first-order line, $m$ increases and
 the approximation becomes invalid.

Let us now discuss the critical lines that border the wings.
Inserting Eq. (\ref{etamh}) in Eq. (\ref{phi2})
\begin{equation}
f(m,H,\Delta,T) \equiv -\frac{1}{\beta}{\rm ln}(z_0(\eta=\frac{Jqm}{S} - H)) + 
\frac{1}{2S^2}Jqm^2
\label{gpot}
\end{equation}
For $f' \equiv \frac{\partial{f}}{\partial m}=0$ this is the MF
  free-energy.  The critical lines that separate two phases are
  characterized by the vanishing of the first three derivatives of the
  free-energy with respect to $m$. Let us argue why it is so. At a
  minimum of the potential $f'=0$, $f'' \ge 0$. Assume now that there
  are two such minima $m_A$ and $m_B$.  Since $\int_{m_A}^{m_B} f''(m)
  dm = 0$, $f''$ must take both positive and negative values. Then
  $f'' > 0$ for $m_A \le m < m_a$ and $f'' < 0$ for $m_b \le m < m_B$,
  where $m_a$ and $m_b$ are two points at which $f'' = 0$.  This
  implies that there exists a point $m_* \in [m_a, m_b]$ at which
  $f''' = 0$.  If the values $m_A$ and $m_B$ occur in two different
  phases and varying $T$ they merge at the point
  $m_A=m_a=m_*=m_b=m_B$, then $f'=f''=f'''=0$.

This is the situation for the points on the critical lines of the
wings that separate two phases.
Solving the two equations $f'' = 0$ and $f'''=0$  in the variables
$\eta=\frac{Jqm}{S} - H$ and $K$  and then using $f'=0$,
we can write $m$ and $H$ in terms of $K$. This can be done analytically
for spin $S=1$ and numerically for spin $S=3/2$.

The previous remarks apply also to the whole critical phase-contour in the
$H=0$ plane. In this case the equations
$f'=f'''=0$ are trivially satisfied for $m=H=0$ and the equation
$f''=0$ must lead to the same result as using $A_2 = 0$ in the approach
outlined above that uses the expansion of the Gibbs free-energy.

\subsection{The spin S=1 model}
For this value of the spin, Eq. (\ref{eqz}) takes the form
\begin{equation}
z_0 = 1 + 2\exp(-D)\cosh(\beta \eta)
\label{eqz1}
\end{equation}
so that expanding
${\rm ln}z_0(\eta) = {\rm ln}z_0(0) + {\rm ln}(1 + 
\frac{1}{\delta}(\cosh(\beta \eta)-1))$
\begin{equation}
\alpha_2 = \frac{\beta^2}{\delta},\qquad
\alpha_4 = \beta^4(\frac{1}{\delta} - \frac{3}{\delta^2}),\qquad
\alpha_6 = \beta^6(\frac{1}{\delta} - \frac{15}{\delta^2} +
\frac{30}{\delta^3})
\label{alphas1}
\end{equation}
with $\delta \equiv 1/\tau= 1 + \frac{1}{2}\exp(D)$.

Using\cite{BEG}  Eqs.(\ref{Akgen}) and (\ref{alphas1})
\begin{equation}
A_2 = \frac{\delta}{\beta} - Jq,\qquad
A_4 = \frac{\delta^2}{2\beta}(1 - \frac{\delta}{3}),\qquad
A_6 = \frac{\delta^3}{2\beta}(1 - \frac{3}{4}\delta + \frac{3}{20}\delta^2).
\label{Aks1}
\end{equation}
Since $A_6$ is strictly
positive, the Landau-Ginzburg potential truncated at the sixth order
is stable so that for small $m$, $G(m)$ can be approximated by
\begin{equation}
G(m) \approx A_0 + \frac{A_2}{2}m^2 + \frac{A_4}{4}m^4 + \frac{A_6}{6}m^6
\label{eqG6}
\end{equation}
The MF critical phase-contour is given by $A_2=0$ with $A_4 > 0$.  In
terms of the variable $\tilde T \equiv \frac{T}{Jq}$ and $\tilde
\Delta = \frac{\Delta}{qJ}$, this line is described by $\tilde \Delta
= {\tilde T}{\rm ln}[2(1/{\tilde T} - 1)]$, with $\frac{1}{3} < \tilde
T < 1$ and terminates at a TCP of coordinates $\tilde T= \frac{1}{3}$,
$\tilde \Delta = \frac{2}{3}{\rm ln}(2)$, at which $A_2=A_4=0$.  The
critical boundary (dashed line) and the position of the TCP (big
crossed circle) are shown in Figs.  \ref{Graf_fig3_first_sq},
\ref{Graf_fig3_first_sc} and \ref{Graf_fig3_first_bcc}.  Using
Eq. (\ref{firstcr}), the first-order line in a vicinity of the
TCP is $\tilde T \approx \tau +
\frac{5}{16}(1-\frac{1}{3\tau})^2 + O((\tau-\frac{1}{3})^3)$, while the
critical phase-contour is given by $\tilde T = \tau$.  Therefore at the
TCP, the two lines share the same slope, but show a
different curvature.

Let us now set $\tilde f \equiv f/{Jq}$ and consider the other
critical lines.  From Eq. (\ref{gpot}) 
\begin{equation}
\tilde f = -\tilde T{\rm ln}(1 + 2e^{-D}\cosh(\beta Jqm - \beta H))
+ \frac{1}{2}m^2
\label{phi1}
\end{equation}
 we have
\begin{equation}
    \frac{\partial{\tilde f}}{\partial m} 
= m - \frac{2 \sinh(z)}{e^D + 2\cosh(z)} \qquad \qquad
\frac{\partial^2{\tilde f}}{\partial m^2} =
    1 -Kq\frac{4 + 2e^D \cosh(z)}{(e^D + 2\cosh(z))^2}
\label{phim2}
\end{equation}
\begin{equation}
\frac{\partial^3{\tilde f}}{\partial m^3} =
    -2(\beta qJ)^2 \sinh(z) \frac{e^{2D} 
- 2e^D \cosh(z) - 8}{(e^D + 2 \cosh(z))^3}
\label{phim3}
\end{equation}
here  $z = m/\tilde T - \beta H$.

We are now ready to discuss the pair of first-order phase-transition
 surfaces, the wings, that extend in the full $H, D, T$ space
 symmetrically with respect to the $H=0$ plane.  They separate three
 phases, defined by the minima $m_0, m_\pm$ of the potential, that
 correspond to the solutions of the Eq. $f' = 0$ with $f'' > 0$.
 In the $H=0$ plane, the two wings join  into a first-order transition line.
As explained previously,
 for $T > 0$  the wings  are bordered by 
critical lines along which the first three derivatives of $f$
vanish\cite{lawrie},
therefore for $z \ne 0$ we obtain
\begin{equation}
\cosh(z) = \frac{Kq - 2}{\sqrt{4-Kq}},\qquad e^D = \frac{4}{\sqrt{4-Kq}},
\qquad m_{*}^2 = \frac{Kq-3}{Kq}
\label{crm}
\end{equation}
It follows that\cite{BEG}
\begin{equation}
\tilde \Delta 
= \frac{\tilde T}{2}{\rm ln}\Big(\frac{4\tilde T}{4\tilde T - 1}\Big)
 \qquad \qquad  \frac{H}{Jq} = \pm\sqrt{1 - 3\tilde T} 
\mp {\tilde T}{\rm ln}\Big{(}
\frac{1 - 2{\tilde T} + \sqrt{1 - 3{\tilde T}}}
{\sqrt{{\tilde T}(4{\tilde T} - 1)}}\Big{)}
\label{crH}
\end{equation}
for $\frac{1}{4} \le \tilde T \le \frac{1}{3}$.

The corresponding value of the magnetization is
\begin{equation}
m_* = \pm\sqrt{1-3{\tilde T}}
\label{crmm}
\end{equation}

These  critical lines terminate at the tricritical point.

The critical phase-boundary in the $H=0$ plane 
corresponds to $m=0$, so that Eq. (\ref{phim2}) gives
\begin{equation}
\Delta = T{\rm ln}(\frac{2}{\tilde T} - 2)
\end{equation}
as already obtained  from the equation $A_2 = 0$.

\subsection{Critical exponents in MF}
Let us now review the computation of the tricritical exponents in MF in
the ordered phase\cite{lawrie}.  Near the critical phase boundary, $m$
is small and we are allowed to expand the free-energy in powers of
$m$, as in Eq.(\ref{Gfree}).  To account for the tricritical point it
is sufficient to truncate the expansion at the sixth-order in $m$
\begin{equation}
f = A_0 + \frac{A_2}{2}m^2 + \frac{A_4}{4}m^4 + \frac{A_6}{6}m^6
\label{Gfree1}
\end{equation}
At equilibrium
\begin{equation}
H = -(A_2 + A_4 m^2 + A_6 m^4)m
\label{G1h}
\end{equation}
In the $H = 0$ plane, the minimum can be either $m=0$ or
\begin{equation}
m^2 = \frac{-A_4 + (A_4^2 - 4 A_2 A_6)^\frac{1}{2}}{2A_6}
\label{minm}
\end{equation}
which is a solution of
\begin{equation}
A_2 + A_4 m^2 + A_6 m^4 = 0
\label{G1e4}
\end{equation}
In the tricritical region, the scaling fields  are $g = A_2$ and
$t = A_4$, while
in the critical region,  just beneath the critical phase-border
$A_2 = 0$, $A_4 > 0$, the scaling field is $\dot{t} = A_2$.

At a minimum $m \neq 0$, 
using Eqs. (\ref{Gfree1}), (\ref{minm}) and (\ref{G1e4})
\begin{equation}
f = A_0 + \frac{1}{24 A_6^2}\Big(A_4^3 - 6A_2A_4A_6 - 
(A_4^2-4A_2A_6)^\frac{3}{2}\Big)
\label{fmin}
\end{equation}
The magnetic susceptibility at $H=0$ is computed from Eq. (\ref{G1h})
\begin{equation}
\chi = -\frac{1}{\beta}(\frac{\partial m}{\partial H})|_{H=0} = 
\frac{1}{\beta}(A_2 + 3 A_4 m^2 + 5 A_6 m^4)^{-1}
\end{equation}
For $m \neq 0$, we use Eq.(\ref{G1e4}) to get
\begin{equation}
\chi = -\frac{1}{\beta}(4A_2 + 2A_4 m^2)^{-1} =
\frac{-1}{4 A_2\beta}\Big(1 + \frac{A_4}{\sqrt{A_4^2 - 4A_2 A_6}}\Big)
\label{chio}
\end{equation}
Two functions related to the concentration are
\begin{equation}
\chi_{(0;g)} = \frac{\partial f}{\partial g} =
\frac{1}{4A_6}\Big(-A_4 + (A_4^2 - 4 A_6 A_2)^\frac{1}{2}\Big)
\label{chi0g}
\end{equation}
\begin{equation}
\chi_{(0;gg)} = \frac{\partial \chi_{(0;g)}}{\partial g} =
-\frac{1}{2}(A_4^2 - 4 A_6 A_2)^{-\frac{1}{2}}
\label{chi0gg}
\end{equation}
Consider now the ordered ($m \neq 0$) region    with $A_4 \neq 0$.
In this region $f-A_0$, $m$ and $\chi$ can be written as
a power of $|t|$ times a scaling function depending on the variable
 $x = \frac{g}{|t|^{\phi_u}}$, with $\phi_u=2$.

From Eq.(\ref{minm})
\begin{equation}
|m| = |t|^{\beta_u} {\cal M}^{(\pm)}(x) \qquad  with  \qquad  
{\cal M}^{(\pm)}(x) = \frac{1}{\sqrt{2A_6}}\Big(\mp1 
+ \sqrt{1-4A_6x}\Big)^\frac{1}{2}
\end{equation}
so that $\beta_u = \frac{1}{2}$.
The $\pm$ superscript  corresponds to the sign of $A_4=t$.
From Eq.(\ref{fmin})
\begin{equation}
f = A_0 + |t|^{2-\alpha_u} {\cal G}^{(\pm)}(x), \quad with \quad 
{\cal G}^{(\pm)}(x) = 
\frac{1}{24 A_6^2}[\pm(1-6A_6 x) - (1-4 A_6 x)^\frac{3}{2}]
\label{fmin1}
\end{equation}
so that $\alpha_u = -1$.
From Eq.(\ref{chio})
\begin{equation}
\chi = |t|^{-\gamma_u} X^{(\pm)}(x), \qquad with\qquad 
X^{(\pm)}(x) = \frac{-1}{4\beta x}\Big(1 \pm \frac{1}{\sqrt{1-4A_6x}}\Big)
\label{chio1}
\end{equation}
so that $\gamma_u = 2$.
From Eq.(\ref{chi0g})
\begin{equation}
\chi_{(0;g)} = |t|^{\beta_{2u}}X_{(0;1)}^{(\pm)}(x), \qquad with \qquad
X_{(0;g)}^{(\pm)}(x) = \frac{1}{4A_6}(\mp1 + (1-4A_6x)^\frac{1}{2})
\end{equation}
so that $\beta_{2u} = 1$.
From Eq.(\ref{chi0gg})
\begin{equation}
\chi_{(0;gg)} = |t|^{-\gamma_{(0,2)u}}X_{(0;gg)}^{(\pm)}(x), \qquad with \qquad
X_{(0;gg)}^{(\pm)}(x) = \frac{-1}{2 (1-4A_6x)^\frac{1}{2}}
\end{equation}
so that $\gamma_{(0,2)u} = 1$.

$f-A_0$, $m$, ..., $\chi_{(0;gg)}$ scale exactly under $t \to \lambda
t$, $g \to \lambda^2 g$. This reflects the fact that Eq.(\ref{Gfree1})
has degree six in $m$. If in the expansion of Eq.(\ref{Gfree}), higher
powers of $m$ were kept, corrections to the tricritical scaling would
appear.

In MF $<s_0^2>$ is given by
\begin{equation}
\chi_{(0;1)} = \beta \frac{\partial f}{\partial D} =
\beta \frac{\partial g}{\partial D} \frac{\partial f}{\partial g} +
\beta \frac{\partial t}{\partial D} \frac{\partial f}{\partial t}
\label{chi01}
\end{equation}
Near the tricritical point the first term behaves as $|t|^{2-\alpha_u-\phi_u}$,
and the second term  as $|t|^{2-\alpha_u-1}$. Therefore using Eq.(\ref{Aks1})
\begin{equation}
\chi_{(0;1)} \approx (\delta - 1) \chi_{(0;g)}
\end{equation}
Similarly the tricritical behavior of $\chi_{(0;2)}$ 
is the same as that of $\chi_{(0;gg)}$.

Let us now consider the MF computation of the  exponents 
along the critical phase-contour.
The scaling functions are
$A_2 = g = \dot{t} \to 0$ and $A_4 = t$ constant.
Set $x = \frac{1}{A_4^2} \dot{t} \to 0$.
Along this path the thermodynamical quantities differ from the corresponding
scaling functions by a multiplicative constant (a power of $|A_4|$),
so that the critical exponents are determined directly using the scaling
functions.
\begin{equation}
{\cal G}^{(+)}(x) = -\frac{x^2}{4} - \frac{A_6}{6} x^3 + O(x^4)
\end{equation}
At the leading order,
\begin{equation}
{\cal G}^{(+)}(x) = \dot{G} \dot{t}^{2-\alpha},  \qquad with \qquad \dot{G} = -\frac{1}{4A_4^4}
\end{equation}
so that $\alpha = 0$ is the  critical exponent of the specific heat.
\begin{equation}
{\cal M}^{(+)}(x) = (-x)^\frac{1}{2}(1 + \frac{A_6}{2}x + O(x^2))
\end{equation}
At the leading order,
\begin{equation}
{\cal M}^{(+)}(x) = (-x)^\frac{1}{2} = A_4^{-1} |\dot{t}|^\beta
\end{equation}
so that $\beta = \frac{1}{2}$ is the  critical exponent of the magnetization.
\begin{equation}
X^{(+)}(x) = -\frac{1}{2\beta x}(1 + A_6x + O(x^2))
\label{chiX1}
\end{equation}
and
\begin{equation}
X^{(+)}(x) = \dot{X}(\dot{t}) |\dot{t}|^{-\gamma}, \qquad with \qquad
\dot{X}(\dot{t}) 
= \frac{A_4^2}{2\beta}(1 + \frac{A_6}{A_4^2}\dot{t} + O(\dot{t^2}))
\end{equation}
so that $\gamma = 1$ is the  critical exponent of the susceptibility.

There are corrections to the ordinary scaling $\dot{t} \to \lambda \dot{t}$.
The term in $A_6$ is the first correction to the ordinary scaling, that 
 would be exact if $A_6=0$.

$X_{(0;g)}^{(\pm)}(x)$ and $X_{(0;gg)}^{(\pm)}(x)$ are regular, 
so that the concentration and its susceptibility
 diverge only at the TCP.

Consider now the region $A_2 \neq 0$ and use the tricritical
scaling fields $g = A_2$ and $t=A_4$
At $m \neq 0$,
\begin{equation}
f - A_0 = |g|^{2-\alpha_t} {\cal G}_t^{(\pm)}(y), \qquad with \qquad
{\cal G}_t^{(\pm)}(y) = \frac{1}{24 A_6^2}\Big(y^3 \mp 6y A_6 -
(y^2 \mp 4 A_6)^\frac{3}{2}\Big)
\end{equation}
so that $\alpha_t = \frac{1}{2}$ and $y=\frac{t}{|g|^{\phi_t}}$,
$\phi_t = \frac{1}{2}$.
\begin{equation}
|m| = |g|^{\beta_t} {\cal M}_t^{(\pm)}(y), \qquad with \qquad
{\cal M}_t^{(\pm)}(y) = \frac{1}{\sqrt{2A_6}}\Big(-y + (y^2\mp 4 A_6)^\frac{1}{2}\Big)^\frac{1}{2}
\label{mtric}
\end{equation}
so that $\beta_t = \frac{1}{4}$.
\begin{equation}
\chi = |g|^{-\gamma_t} X_t^{(\pm)}(y), \qquad with \qquad
X_t^{(\pm)}(y) = \mp \frac{1}{4\beta}\Big(1 + \frac{y}{\sqrt{y^2\mp4 A_6}}\Big)
\end{equation}
so that $\gamma_t = 1$.
\begin{equation}
\chi_{(0;g)} = |g|^{\beta_{2t}} X_{(0;g)t}^{(\pm)}, \qquad with \qquad
X_{(0;g)t}^{(\pm)} = \frac{1}{4A_6}\Big(-y + (y^2 \mp 4A_6)^\frac{1}{2}\Big) 
\end{equation}
so that $\beta_{2t} = \frac{1}{2}$
\begin{equation}
\chi_{(0;gg)} = |g|^{-\gamma_{(0;2)t}} X_{(0;gg)t}^{(\pm)}, \qquad with \qquad
X_{(0;gg)t}^{(\pm)} = -\frac{1}{2}(y^2 \mp 4 A_6)^{-\frac{1}{2}}
\end{equation}
so that  $\gamma_{(0;2)t} = \frac{1}{2}$.

To examine the  critical behavior along the critical phase-boundary for
$A_2=g=\dot{t} \to 0$ with $A_4=t > 0$ constant, we should expand at
$y \to \infty$.

If $t > 0$ and $g < 0$ one has $y \to +\infty$,
so $X_t^{(-)} = \frac{1}{2\beta} + O(y^{-2})$, and
$\chi = |\dot{t}|^{-\gamma}X_t^{(-)}$ with the susceptibility exponent
$\gamma = 1$, as  already found using $X^{(+)}(x)$ and similarly for the
other ordinary critical exponents.

While along a trajectory with $y = \frac{t}{|g|^\frac{1}{2}}$ constant
the tricritical scaling is exact, along the trajectory with $A_4 > 0$ constant,
there is a crossover from the region with $y$ small, in which the tricritical
scaling is valid, to the region with $y$ large, in which  ordinary
scaling is valid. Consider for instance the magnetization scaling
function in Eq.(\ref{mtric}).
For $y$ small 
\begin{equation} 
{\cal M}_t^{(-)}(y) \approx A_6^\frac{-1}{4}, \qquad with \qquad
|m| \approx |g|^{\frac{1}{4}} A_6^\frac{-1}{4}
\end{equation}
and the behavior is tricritical, with tricritical exponent $\beta_t = 1/4$.

For $y$ large, 
\begin{equation}
{\cal M}_t^{(-)}(y) \approx \frac{|g|^\frac{1}{4}}{A_4^{\frac{1}{2}}}, \qquad
|m| \approx |g|^\frac{1}{2} A_4^{-\frac{1}{2}} = 
|\dot{t}|^\frac{1}{2} A_4^{-\frac{1}{2}}
\end{equation}
so that ordinary critical behavior appears, with critical exponent
$\beta = 1/2$.

\subsection{The  spin S=3/2 model }

In this case Eq. (\ref{eqz}) is
\begin{equation}
z_0 = 2e^{-\frac{D}{9}}({\cal C}_1 + d_1 {\cal C}_3)
\label{eqz32}
\end{equation}
where we have defined $d_1 = exp(-\frac{8}{9}D)$,
${\cal C}_n = \cosh(n \frac{\beta}{3}(\frac{2}{3}Jqm - H))$, 
${\cal S}_n = \sinh(n \frac{\beta}{3}(\frac{2}{3}Jqm - H))$ and
$b = \frac{4}{9}\beta J q$.
The first Taylor coefficients of ${\rm ln}(z_0(\eta))$ are
\begin{equation}
\alpha_2 = \frac{\beta^2}{9}\frac{9d_1 + 1}{1+d_1} \qquad \qquad
\alpha_4 = -\frac{2\beta^4}{81} \frac{81d_1^2 - 14d_1 + 1}{(1+d_1)^4}
\end{equation}
From Eq. (\ref{Akgen}), we can determine the first coefficients of the
Landau-Ginzburg expansion of the potential
\begin{equation}
A_2 =  \frac{9(d_1+1)}{\beta(9d_1+1)} - qJ \qquad
A_4 = \frac{27}{\beta}(1+d_1)^2\frac{81d_1^2-14d_1+1}{(9d_1+1)^4}
\end{equation}
$A_4$ is always positive.
The MF critical phase-boundary  is obtained from the Eq.  $A_2 = 0$:
\begin{equation}
\tilde \Delta = \frac{9 \tilde T}{8}{\rm ln}
    \Big(\frac{ 1 - \tilde T}{  \tilde T - \frac{1}{9}}\Big)
\label{Delta32}
\end{equation}
Equivalently, in terms of 
$\tilde \tau ={\rm C}(D, h=0; 3/2) = 1/(1 + exp(8D/9))$
the phase-contour is
\begin{equation}
\tilde T=\frac{8}{9} \tilde \tau + \frac{1}{9}
\end{equation}
 since the coefficient $A_4$ is always positive, and no
 tricritical point\cite{plascak} can exist.

Let us now consider the critical phase boundaries for generic $H$.
From Eq. (\ref{gpot})
\begin{equation}
f = -\frac{1}{\beta}{\rm ln}(2e^{-\frac{D}{9}}({\cal C}_1 + d_1 {\cal C}_3)) + 
\frac{2}{9} Jqm^2
\label{fmfs32}
\end{equation}

\begin{equation}
f' = \frac{4Jq}{9}(m - \frac{1}{2}\frac{{\cal S}_1 + 3 d_1 {\cal S}_3}{{\cal C}_1+d_1 {\cal C}_3})
\label{fmfs32d1} \qquad \qquad
f'' = \frac{4Jq}{9}(1 - \frac{b\xi}{4({\cal C}_1+d_1 {\cal C}_3)^2})
\label{fmfs32d2}
\end{equation}
where
$\xi \equiv 1 + 9d_1^2 + 6d_1 {\cal C}_2 + 4d_1 {\cal C}_1 {\cal C}_3$.
\begin{equation}
f''' = -\frac{Jqb^2}{18({\cal C}_1+d_1 {\cal C}_3)^3}
[(12d_1{\cal S}_2 + 4d_1 {\cal S}_1 {\cal C}_3 + 12 d_1 {\cal C}_1 {\cal S}_3)({\cal C}_1+d_1 {\cal C}_3) - 
2\xi ({\cal S}_1 + 3d_1 {\cal S}_3)]
\label{fmfs32d3}
\end{equation}
The critical points occur at $f'=f'' = f''' = 0$.

For $\frac{2}{3}Jqm - H = 0$ one has ${\cal S}_i=0$
and ${\cal C}_i=1$, so that $f'''=0$ is trivially satisfied and $f'$ gives
$m=0=H$; the equation $f''=0$ gives again Eq. (\ref{Delta32}).

Use Eq. (\ref{fmfs32d2}) to eliminate $\xi$ from Eq. (\ref{fmfs32d3}),
obtaining an equation quadratic in $d_1$.  Eq. (\ref{fmfs32d2}) has
the form $N_2 d_1^2 + N_1 d_1 + N_0 = 0$ where $N_2 = \frac{9}{4}b -
{\cal C}_3^2$, $N_1 = \frac{3}{2}b {\cal C}_2 + b {\cal C}_1 {\cal
C}_3 - 2 {\cal C}_1 {\cal C}_3$, $N_0 = \frac{b}{4} - {\cal C}_1^2$.
From these two quadratic equations in $d_1$
\begin{equation}
d_1 = \frac{N_2 {\cal S}_1{\cal C}_1 - 3 N_0 {\cal C}_3{\cal S}_3}{3
N_1 {\cal S}_3{\cal C}_3 - N_2(3{\cal S}_3{\cal C}_1 -
\frac{3}{2}b{\cal S}_2 - \frac{b}{2}{\cal S}_1{\cal C}_3
-\frac{3}{2}b{\cal C}_1{\cal S}_3 + {\cal S}_1 {\cal C}_3)}
\label{fmfs32a}
\end{equation}
 Thus we observe two first-order phase-transition surfaces. They are
formed by two wings bordered by two critical lines ( for $H > 0$ and
for $H < 0$), along which the first three derivatives of the free
energy vanish. For $H=0$, the critical lines terminate with a critical
end-point at $\tilde T_{cep}=0.12941913384882...$, $\tilde
\Delta_{cep}=0.4875062362287...$, with $m_{cep}=\pm0.854170713633042$,
lying on the first-order transition surface with magnetization $\pm
m$, the minima of the double well. At the critical end-point, these
minima are flat, since $\frac{\partial^2{\tilde f}}{\partial m^2} =
0$.

\section{ Phenomenological Scaling}
The scaling-laws approach to the crossover behavior, introduced in
Ref.[\onlinecite{RW,pfeuty,riede1,riede2,wegne,hank}] in the context
of the transition from weakly anisotropic to fully isotropic
exchange-interactions in spin systems and illustrated in various other
contexts can also describe the transition from Ising-like to
tricritical behavior of the BC model.

First we should replace the thermodynamic fields by scaling fields
appropriate to the TCP. For this purpose let us first define the
reduced deviation from the tricritical temperature $t=T/T_{tr}-1$, and
observe that for small $t > 0$, the equation of the critical line
 (see Fig. \ref{fig_diag_tricr}) can be approximated as
\begin{equation}
\Delta_2(t) = \Delta_{tr} - at + b_2 t^{\psi_2} +...
\label{Dla}
\end{equation}
 with $\psi_2 >1$, while similarly 
 for the first-order line  (small $t < 0$), we have
\begin{equation}
\Delta_1(t) = \Delta_{tr} - at + b_1 |t|^{\psi_1} +...
\label{Dta}
\end{equation}
with $\psi_1 > 1 $. The same finite (nonuniversal) slope of
the critical line is generally assumed
\begin{equation}
a = -T_{tr}(\frac{d \Delta_2(T)}{{d T}})|_{T_{tr}} = 
    -T_{tr}(\frac{d \Delta_1(T)}{{d T}})|_{T_{tr}}
\end{equation}
on both sides of the TCP (as in some physical systems and for models
in the MF approximation).  It is also assumed that the exponents
$\psi_1$ and $\psi_2$ that characterize the lowest order correction to
the tangent approximation, are equal $\psi_1 =\psi_2= \phi_u$.

 The variable 
\begin{equation}
g \equiv \Delta - \Delta_{tr} + a t
\label{gat}
\end{equation}
will be used as a scaling field.  Along the locus $g=0$, the TCP is
approached along the tangent to the phase-contour as $|t| \to 0$. If
$t=0$, the TCP is approached at an angle with respect to the tangent,
as  $g \to 0$.
   The  fields $t$, $g$ and the magnetic field $h$ are usually 
  adopted as tricritical scaling fields.

In terms of these fields, a ``tricritical scaling'' hypothesis for the
singular part $f_s$ of the free-energy can now be formulated as
follows
\begin{equation}
f_s(g, t, h) \approx |t|^{2-\alpha_u}
W^{(\pm)}_{u}(g/|t|^{\phi_u},h/|t|^{\hat \Delta_u})
\label{scalingp}
\end{equation}
for $t, g \to 0$, with arbitrary ratio $g/t$. 
(The phase-contour lies  in the  $H=0$ plane.)
 Thus, if the TCP is approached along the line $g=0$, it is
 appropriate to refer to Eq. (\ref{scalingp}) in which the singularity
 is described explicitly by a power of $t$, for example $f_s \approx
 |t|^{2-\alpha_u} W_{u}(0,0)$. This is a natural parametrization of
 the TCP scaling property.  The exponents that characterize the TCP,
 are the specific-heat exponent $\alpha_u$, the tricritical gap
 exponent $\hat \Delta_u=\beta_u+\gamma_u^{(2;0)}$, and $\phi_u$ an
 additional ``crossover exponent''.  (Please notice also that we have
 changed into $\hat \Delta_u$ the symbol usually denoting the gap
 exponent to avoid confusion with the crystal field). In the MF
 approximation $\phi_u=2$.  The scaling function $W_{u}(x, y)$ is
 assumed to be analytic at $x=0$ and thus at $g=0$ for fixed $t \neq
 0$.

Defining $ W_{tr}(X, Y) \equiv |X|^{2-\alpha_u}
W_{u}^{(\pm)}(|X|^{-\phi_u}, Y |X|^{-\hat \Delta_u})$,
Eq. (\ref{scalingp}) can be written equivalently as
\begin{equation}
f_s(g, t, h) 
\approx |g|^{(2-\alpha_{tr})}  W_{tr}(t/|g|^{\phi_{tr}},
 h/|g|^{\hat \Delta_{tr}})
\label{scalingt}
\end{equation}
with the new set of exponents 
$\alpha_{tr}, \beta_{tr},..,\phi_{tr}$  defined by 
\begin{equation}
2 - \alpha_{tr} = \frac{2-\alpha_u}{\phi_u}, \qquad
\beta_{tr}=\frac{\beta_u}{\phi_u}, \qquad
\gamma_{tr}=\frac{\gamma_u}{\phi_u}, \qquad
\hat \Delta_{tr}=\frac{\hat \Delta_u}{\phi_u}, \qquad
\phi_{tr} = \frac{1}{\phi_u}, \qquad
\label{crexpt}
\end{equation}
The expected values of these new exponents can be read in Table II.
Also the scaling function $ W_{tr}(X, Y)$ is assumed to be
analytic at $X=0$.

 If the TCP is approached along a line $t = cg$, crossing the critical
 line at a finite angle, so that we have $t/|g|^{\phi_{tr}} \to 0$,
 because $\phi_{tr} < 1$, it is appropriate to refer to
 Eq. (\ref{scalingt}) in which the singularity appears explicitly as a
 power of $g$, for example $f_s \approx |g|^{2-\alpha_{tr}}
 W_{tr}(0,0)$.

 These remarks suggest a simple method to estimate the crossover
 exponent: we should simply compare the usual exponents
 computed along a path tangent to the phase-boundary at the TCP with those
 computed along a path forming an angle with it (as are naturally obtained
 when studying series at fixed $D=D_{tr}$).

 The scaling function $W_{u}$ of Eq. (\ref{scalingp}),  can describe
 also the Ising-like ordinary scaling behavior that is observed along
 the critical line, provided that an appropriate singularity appears
 in this quantity. Let us set $t_c(g) \equiv T_c(g)/T_{tr}-1$, so that
 $\dot t \equiv t - t_c(g)$ can be taken as a distance from the
 critical line.  Since as $\dot t \to 0$ the scaling variable
 $x=g/t^{\phi_u} \to \dot x =g/t_c(g)^{\phi_u}$, we have precisely to
 assume that at $x=\dot x$ the scaling function $W^{(\pm)}_{u}(x,y)$
 has a singularity of the form $(1-x/\dot x)^{2-\dot \alpha}$, so that
 $f_s \approx |\dot t|^{2- \alpha}$ as $\dot t \to 0$ with $g \neq
 0$.  This assumption is actually realized\cite{pfeuty} in the MF
 approximation.

 It can  be observed that for fixed small $g \neq 0 $, as
$T$ approaches the critical temperature $T_c(g)$, the system will
first behave as if $T \to T_{tr}$ and   only when $T- T_c(g)$ is very
small, past a ``crossover temperature'' $T^{\rm x}(g)$, i.e. for
$T^{\rm x}(g) >T> T_c(g) $, it  will develop the full
Ising-like critical behavior  expected away from the TCP.
This can be observed also in MF, see the end of Appendix A.

For a (mixed) susceptibility, the critical scaling reads
\begin{equation}
\chi_{(r;p)}(K, D) \approx A_{(r;p)}(K, D)|\dot{t}|^{-\gamma^{(r;p)}}
\label{crscal}
\end{equation}
The tricritical scaling is
\begin{equation}
\chi_{(r;p)}(K, D) \approx A_{(r;p)tr}\Big(\frac{t}{|g|^{\phi_{tr}}}\Big)|g|^{-\gamma_{tr}^{(r;p)}}
\label{trscal}
\end{equation}
Approaching the TCP along the scaling path $g = x
|t|^\frac{1}{\phi_{tr}}$, with the constant $x$ chosen small enough to
be close to the critical line, Eqs. (\ref{crscal}) and (\ref{trscal})
are both valid. Therefore in this region $g \approx \dot{t}$
\begin{equation}
A_{(r;p)}(K; D) \approx  |x|^{\gamma^{(r;p)} - \gamma_{tr}^{(r;p)}}
A_{(r;p)tr}^{(\pm)}(|x|^{-\phi_{tr}}) |t|^{( \gamma^{(r;p)} - \gamma_{tr}^{(r;p)})/\phi_{tr}}.
\label{scal1c}
\end{equation}
Thus, if the estimates of the amplitudes of observables like
the susceptibilities, were  
sufficiently accurate, the study of their $g \to 0$ behavior might help
to spot the TCP and to determine the crossover exponent. In
particular, in the case of the ordinary susceptibility, 
as $t \to 0$ along the line $g = x |t|^\frac{1}{\phi_{tr}}$, with $|x| << 1$,
the critical amplitude will vanish as
\begin{equation}
A_{(2;0)}(K, D) \approx  |t|^{( \gamma - \gamma_{tr})/\phi_{tr}}.
\label{ampscaling}
\end{equation}

\section{ Tools for the series analysis}
To make the paper reasonably selfcontained, it is useful to sketch
the standard numerical approximation techniques of  series
analyses.  More detailed discussions can be found in
Refs.[\onlinecite{zinnmra,guttda,bp1,bp2,bp3,bcisiesse}].
\subsection {Coefficient-ratio based methods}
  To determine the location of the critical points and their critical
  exponents using the HT expansions, it is often convenient to resort to the
  unbiased {\it modified-ratio-approximants}(MRAs) (in the loose
  lattice version), which is a smoother and faster converging
  improvement\cite{zinnmra,guttda,bcisiesse} of the traditional
  methods\cite{guttda} of extrapolation of the series-coefficient
  ratio-sequences.

  To illustrate this prescription by an example, we can refer to the HT
  expansion of the magnetic susceptibility $\chi_{(2;0)}(K,D;S)=
  \sum_r c_r(D;S)K^r$. For each fixed  $D$, a value of the  inverse 
 critical temperature can be obtained by forming the
  sequence of estimators $(K_{c}(D;S))_n$ of $K_{c}(D;S)$
  defined\cite{zinnmra,guttda} by
\begin{equation}
(K_{c}(D;S))_n=(\frac{ c_{n-2}c_{n-3}} {c_{n}c_{n-1}})^{1/4} 
exp[\frac{ s_{n}+s_{n-2}} {2s_{n}(s_{n}-s_{n-2})}]
\label{zinnbc}
\end{equation}
with 
\begin{equation}
s_{n}=\Big( {\rm ln}( \frac {c_{n-2}^2} {c_{n}c_{n-4}})^{-1}+
 {\rm ln}( \frac {c_{n-3}^2} {c_{n-1}c_{n-5}})^{-1}\Big)/2.
\end{equation}
 $c_n \equiv c_n(D;S)$ being an abridged notation for the $n$th
expansion coefficient of the susceptibility.  Provided that the leading
 correction to scaling dominates over the
subleading ones, this prescription
Eq. (\ref{zinnbc}) has the important advantage of providing at the
same time information on $K_{c}(D;S)$ and on the leading
correction-to-scaling amplitude $ a_{(2;0)}(D;S)$, defined by
Eq. (\ref{chi2nas}). Otherwise, it yields some ``effective value'' for this
 amplitude.  If the critical singularity is the nearest one to
the origin of the complex $K$ plane, the MRA estimator-sequence has
the asymptotic behavior\cite{bcisiesse} for large order $n$
 \begin{equation}
(K_{c}(D;S))_n=K_{c}(D;S)\big(1-
\frac{1}{2}\frac{ C(\gamma^{(2;0)})\theta^2(1-\theta) a_{(2;0)}(D;S)}
 {n^{1+\theta}}+o(1/n^{1+\theta})\big)
\label{kasint}
\end{equation}
where $C(\gamma^{(2;0)})$ is a known\cite{bcisiesse} positive function
of the exponent $\gamma^{(2;0)}$ of $\chi_{(2;0)}$, $a_{(2;0)}$ is
defined in Eq. (\ref{chi2nas}) and  $\theta \approx 0.52$, in the $3d$ Ising
universality class, is  the exponent of the leading correction to scaling.

A prescription\cite{zinnmra,guttda,bcisiesse} of a similar kind
provides a sequence of estimators $(\gamma^{(2;0)}(D;S))_n$ 
 for the critical exponent
 \begin{equation}
(\gamma^{(2;0)}(D;S))_n=1+\frac {2(s_n+s_{n-2})}{(s_n - s_{n-2})^2}
\label{zinnesp}
\end{equation}
In this case, the asymptotic behavior of the sequence for large order $n$ is  
 \begin{equation}
(\gamma^{(2;0)}(D;S))_n=\gamma^{(2;0)}(D;S)- 
\frac{C(\gamma^{(2;0)})\theta(1-\theta^2)a_{(2;0)}(D;S)}{n^{\theta}} +O(1/n)
\label{espasint}
\end{equation}
In general the set of the corrections to scaling rules the convergence
properties of any extrapolation method in the critical region and the
MRA method of analysis can account explicitly for the leading
terms. Thus, we expect that if the HT series are sufficiently long and
regular, the MRA estimator-sequences have settled into their
asymptotic regimes described by Eq. (\ref{kasint}), (\ref{espasint}),
and only the leading correction-to-scaling have non-negligible
amplitudes, then it is reasonable to determine $(K_{c}(D;S))$ by
fitting the simple extrapolation Ansatz
 \begin{equation}
 (K_{c}(D;S))_n=b_1(D;S)-b_2(D;S)/n^{1+\theta}.
\label{ansatz}
\end{equation}
  to the last few terms
of the estimator-sequence $(K_{c}(D;S))_n$.
 We can thus assume that $K_{c}(D;S) \approx b_1(D;S)$. 
A similar  ansatz
\begin{equation}
(\gamma^{(2;0)})_n =\tilde b_1(D;S) - \tilde b_2(D;S/n^{\theta}
\label{ansatzesp}
\end{equation}
can be used with the MRA estimator-sequence for the exponent
$(\gamma^{(2;0)}(D;S))_n$, concluding that $\gamma^{(2;0)}(D;S) \approx
\tilde b_1(D;S)$.  An analogous prescription in which the HT expansion
coefficients of $\chi_{(4;0)}$ are employed, is adopted to estimate
$\gamma^{(4;0)}(D;S)$.  These methods are {\it unbiased}, i.e. no
assumption on the value of the exponent is used in Eq. (\ref{zinnbc})
 to  compute the
critical temperature and no assumption on the critical temperature 
 in Eq. (\ref{zinnesp}) to compute the exponent.
Small multiples of the uncertainties inherent in these extrapolations
can be taken as a measure of the errors of the final estimates.

Using the formulas Eqs. (\ref{ansatz}) and (\ref{ansatzesp}), we can
estimate also the coefficients $ b_2(D;S)$ and $\tilde b_2(D;S)$,
proportional to the correction-to-scaling amplitudes, and thus get some
hint of the uncertainties to be expected. Procedures of this kind were
suggested long ago in Ref. [\onlinecite{zinnmra,fisher_amp}] and were later 
pursued by several authors\cite{zinnmra,min1,min2,bcisiesse}
to achieve optimal determinations of the universal critical parameters
by studying a model with minimal leading corrections to scaling,
singled out in a family of one-parameter-dependent models known to belong to
the same universality class. In the case of the BC model, the family
parameter is $D$.

When the structure of the correction-to-scaling terms becomes more
 complex and oscillations are observed in the highest-order terms of
 the MRA estimator-sequences, as it happens in the crossover regions,
 simple (but less accurate) ratio-method estimators\cite{guttda} such
 as
 \begin{equation}
(K_{c}(D;S))_n= \frac {n-1+\gamma^{(2;0)}} {na_{n}/a_{n-1}} +o(1/n)
\label{orf}
\end{equation}
that is biased with some accurate value of $\gamma^{(2;0)}$
 and
 \begin{equation} 
(\gamma^{(2;0)}(D;S))_n=n +1 - \frac{na_{n-1}}{a_{n}K_{c}(D;S)} +o(1/n)
\label{orfe}
\end{equation}
that is biased with  some accurate value of $K_{c}(D;S)$,
 might sometimes be more robust than  the MRA approach. 

\subsection {Pad\`e and differential  approximant methods}
In many cases the MRAs, if cautiously extrapolated to large orders of
expansion, show an apparent accuracy comparable or higher than that
obtained by the {\it differential approximants} (DAs) method for which
such extrapolations are  controversial.

The DA method also used in the series analysis, is a
generalization\cite{guttda} of the well known Pad\'e approximant(PA)
method and can similarly be either biased or unbiased.  Both methods
can be employed to evaluate either the expansions of the quantities
that remain finite at the critical points or the parameters of the
singularities for quantities that diverge there.  The DA method uses
the solution, called differential approximant, of an initial value
problem for an ordinary linear inhomogeneous differential equation of
the first or higher order in the expansion variable. The coefficients
of the equation are polynomials in that variable such that the series
expansion of the solution of the equation equals, up to some
appropriate order, the series to be approximated.  Truncating the
series under investigation at various lengths or using series of a
fixed length and choosing different degrees for the polynomial
coefficients, various DAs (i.e. solutions of various differential
equations) can be formed.  Following this procedure, for each quantity
under study, a sample of estimates can be obtained from the
highest-order approximants, namely those formed using all or most
available expansion coefficients, whose average and spread can be
computed, after possibly discarding evident outliers.  If the sample
average remain essentially stable as the order of truncation of the
series increases and it can be believed that stability indicates
convergence, then this average can be taken as the best estimate of
the parameter and a (generous) multiple of the spread of the sample may be
trusted to be a reasonable measure of uncertainty. It should be
stressed that the uncertainties associated with the analysis of a
series either by the DAs and by the MRAs do not have such a precise
statistical meaning as for MC methods, but remain subjective to some
extent.  Our analyses, will be corroborated by checking the
consistency, within the numerical uncertainties, between the MRA and
DA estimates of the critical parameters, whenever both approaches are
feasible.  The critical amplitudes and their ratios have to be
determined by (biased) PAs and DAs.

\begin{table}{l}
  \caption { BC model with spin $S=1$ on a simple-cubic lattice,
    subject to a magnetic field $H$ and a crystal field $D$. The
    coefficients $L_n(u,x;1)$ of the LT series expansion in
    powers of $\mu=\exp(h)$ for the free-energy density
     Eq. (\ref{LTexpansion}) are expressed in terms of the LT variable
    $u=\exp(-\beta)$ and of the crystal-field variable 
    $x=\exp(D)$. 
}
\begin{tabular}{l}

$L_1  = +x u^6  $\\

$L_2  = -7/2x^2u^{12}+3x^2u^{11}+u^{12}  $\\

$L_3  = +64/3x^3u^{18}-36x^3u^{17}+15x^3u^{16}-7x u^{18} 
+6x u^{16}  $\\

$L_4 =-651/4x^4u^{24}+405x^4u^{23}-657/2x^4u^{22}+83x^4u^{21}
+3x^4u^{20}+64x^2u^{24}-36x^2u^{23}-72x^2u^{22}$\\$+30x^2u^{21}
+15x^2u^{20}-7/2u^{24}+3u^{20}  $\\

$L_5 =+7031/5x^5u^{30}-4608x^5u^{29}+5532x^5u^{28}
-2804x^5u^{27}+426x^5u^{26}+48x^5u^{25}-651x^3u^{30}$\\$+810x^3u^{29}
+567x^3u^{28}
-828x^3u^{27}-57x^3u^{26}+126x^3u^{25}+32x^3u^{24}
+64x u^{30}-72x u^{28}-21x u^{26}$\\$+30x u^{24}  $\\

$L_6 =-39452/3x^6u^{36}+53370x^6u^{35}-84738x^6u^{34}
+64574x^6u^{33}-44289/2x^6u^{32}+1575x^6u^{31}$\\$+496x^6u^{30}
+18x^6u^{29}+7031x^4u^{36}-13824x^4u^{35}-471x^4u^{34}
+14020x^4u^{33}-5091x^4u^{32}$\\$-2496x^4u^{31}+421x^4u^{30}
+348x^4u^{29}+63x^4u^{28}
-1953/2x^2u^{36}+405x^2u^{35}
+1620x^2u^{34}-486x^2u^{33}$\\$-423x^2u^{32}-48x^2u^{31}-360x^2u^{30}
+126x^2u^{29}+129x^2u^{28}
+12x^2u^{27}+64/3u^{36}-36u^{32}+15u^{28}$\\

$L_7 =+909434/7x^7u^{42}-628236x^7u^{41}+1240035x^7u^{40}
-1261904x^7u^{39}+674652x^7u^{38}$\\$-157380x^7u^{37} -1360x^7u^{36}
+3888x^7u^{35}+378x^7u^{34}+8x^7u^{33}-78904x^5u^{42}
+213480x^5u^{41}$\\$-103104x^5u^{40}-172268x^5u^{39}+170994x^5u^{38}
-2724x^5u^{37}-27700x^5u^{36} -2568x^5u^{35}+1779x^5u^{34}
$\\$+900x^5u^{33}+114x^5u^{32}+14062x^3u^{42}-13824x^3u^{41}
-24435x^3u^{40}+22128x^3u^{39}+10551x^3u^{38}$\\$-6018x^3u^{37}  
+817x^3u^{36} -3348x^3u^{35}-1233x^3u^{34}+750x^3u^{33}
+456x^3u^{32}+96x^3u^{31}-651x u^{42}$\\$+810x u^{40} +567x u^{38}
-808x u^{36}-117x u^{34}+186x u^{32}+12x u^{30}  $\\

$L_8 =-10690323/8x^8u^{48}+7496787x^8u^{47}
-35373351/2x^8u^{46}+22521935x^8u^{45}-65448621/4x^8u^{44}$\\$
+6392769x^8u^{43}-1895165/2x^8u^{42}-106113x^8u^{41}
+44793/2x^8u^{40}+4622x^8u^{39}+306x^8u^{38}+x^8u^{36}$\\$ 
+909434x^6u^{48}-3141180x^6u^{47}+2963502x^6u^{46}+1262118x^6u^{45}
-3515763x^6u^{44}+1363200x^6u^{43}$\\$+407264x^6u^{42}
-201462x^6u^{41}
-54135x^6u^{40}-1836x^6u^{39}+6702x^6u^{38}+1938x^6u^{37}
+219x^6u^{36}$\\$-197260x^4u^{48}+320220x^4u^{47}+261486x^4u^{46}
-571648x^4u^{45}-31320x^4u^{44}+257250x^4u^{43}
$\\$-21338x^4u^{42}
+15330x^4u^{41}-33027/2x^4u^{40}-18840x^4u^{39}-1887x^4u^{38}
+2196x^4u^{37}$\\$+1932x^4u^{36}+372x^4u^{35}+18x^4u^{34}
+14062x^2u^{48}-4608x^2u^{47}-27648x^2u^{46} +6426x^2u^{45}
$\\$+5091x^2u^{44}+2172x^2u^{43}+16912x^2u^{42}-4560x^2u^{41}
-7092x^2u^{40}-420x^2u^{39}-2232x^2u^{38}$\\$+834x^2u^{37}
+777x^2u^{36}+144x^2u^{35}+144x^2u^{34}-651/4u^{48}+405u^{44}
-657/2u^{40}+83u^{36}+3u^{32}  $\\

$L_9 =+127579807/9x^9u^{54}-90480828x^9u^{53}
+248294610x^9u^{52}-379686836x^9u^{51}+348702921x^9u^{50}$\\$
-190517760x^9u^{49}+54753064x^9u^{48}-3978300x^9u^{47}
-1368954x^9u^{46}+60804x^9u^{45}+40050x^9u^{44}$\\$+5544x^9u^{43}
+127x^9u^{42}+24x^9u^{41}-10690323x^7u^{54}+44980722x^7u^{53}
-61926729x^7u^{52}$\\$+9022472x^7u^{51}+54503559x^7u^{50}
-45973002x^7u^{49}+4829245x^7u^{48}+6738564x^7u^{47}
-896967x^7u^{46}$\\$
-490174x^7u^{45}-142017x^7u^{44}+22500x^7u^{43}
+17479x^7u^{42}+4278x^7u^{41}+384x^7u^{40}+8x^7u^{39}
$\\$+2728302x^5u^{54}-6282360x^5u^{53}-1062648x^5u^{52}
+11093040x^5u^{51}-4277985x^5u^{50}-5268780x^5u^{49}  
$\\$+2597970x^5u^{48}+381708x^5u^{47}+289224x^5u^{46}-3732x^5u^{45}
-154596x^5u^{44}-42816x^5u^{43}-16392x^5u^{42}$\\$+11136x^5u^{41}
+6456x^5u^{40}+1344x^5u^{39}+132x^5u^{38}-789040/3x^3u^{54}
+213480x^3u^{53}+600072x^3u^{52}$\\$-419688x^3u^{51}-326820x^3u^{50}
+110796x^3u^{49}-141215x^3u^{48}+187176x^3u^{47}+162291x^3u^{46}
$\\$-72954x^3u^{45}-27303x^3u^{44}-25044x^3u^{43}-6391x^3u^{42}
+4506x^3u^{41}+1188x^3u^{40}+2016x^3u^{39}$\\$+828x^3u^{38}
+72x^3u^{37}+7031x u^{54}-9216x u^{52}-10611x u^{50}+15310x u^{48}
+5058x u^{46}-8040x u^{44}$\\$-917x u^{42}+1254x u^{40}+84x u^{38}
+48x u^{36} $\\

 \end{tabular} 
\label{tab5a}
\end{table}

\scriptsize
\begin{table}{l}
  \caption {{\it (Continued from the preceding Table)} 
  BC model with spin $S=1$ on a simple-cubical lattice,
    subject to a magnetic field $H$ and a crystal field $D$. The
    coefficients $L_n(u,x;1)$ of the LT series expansion in
    powers of $\mu=\exp(h)$ for the free-energy density
    Eq. (\ref{LTexpansion})    are expressed in terms of the LT variable
    $u=\exp(-\beta)$ and of the crystal field variable 
    $x=\exp(D)$.
}
\begin{tabular}{l}

$L_{10} =-1540944687/10x^{10}u^{60}
+1102444428x^{10}u^{59}-3449297064x^{10}u^{58}+6156900766x^{10}u^{57}
$\\$-6835882485x^{10}u^{56}+23965701903/5x^{10}u^{55}-2018275270x^{10}u^{54}
+414942978x^{10}u^{53}$\\$+2839656x^{10}u^{52}-12412763x^{10}u^{51}
-614784x^{10}u^{50}+236808x^{10}u^{49}+67267x^{10}u^{48}+4131x^{10}u^{47}
$\\$+396x^{10}u^{46}+24x^{10}u^{45}+127579807x^8u^{60}-633365796x^8u^{59}
+1136891808x^8u^{58}-633220552x^8u^{57}$\\$-611949780x^8u^{56}
+1042711956x^8u^{55}-432968505x^8u^{54}-65933616x^8u^{53}
+73586991x^8u^{52}$\\$+977856x^8u^{51}-2551188x^8u^{50}
-1745292x^8u^{49}-155650x^8u^{48}+87072x^8u^{47}+45669x^8u^{46}
+8348x^8u^{45}$\\$+849x^8u^{44}+24x^8u^{43}-74832261/2x^6u^{60}
+112451805x^6u^{59}-40853013x^6u^{58}-174019476x^6u^{57}
$\\$+326069793/2x^6u^{56}+52714461x^6u^{55}-94539076x^6u^{54}
+8386548x^6u^{53}+11536857/2x^6u^{52}+4477701x^6u^{51}
$\\$+1683855x^6u^{50}-1249242x^6u^{49}-181277x^6u^{48}
-297030x^6u^{47}
-28746x^6u^{46}+40770x^6u^{45}+20220x^6u^{44}$\\$+4716x^6u^{43}
+562x^6u^{42}+24x^6u^{41}+4547170x^4u^{60}-6282360x^4u^{59}
-9825078x^4u^{58}+14528896x^4u^{57}$\\$+5447442x^4u^{56}
-8416140x^4u^{55}+626182x^4u^{54}-2440008x^4u^{53}
-945117x^4u^{52}
+2950388x^4u^{51}$\\$+456423x^4u^{50}-204420x^4u^{49}-332881x^4u^{48}
-141396x^4u^{47}+24819x^4u^{46}-16444x^4u^{45}+6561x^4u^{44}
$\\$+10980x^4u^{43}+4220x^4u^{42}+744x^4u^{41}+24x^4u^{40}
-197260x^2u^{60}+53370x^2u^{59}+426960x^2u^{58}
-80736x^2u^{57}
$\\$-18732x^2u^{56}-51564x^2u^{55}-476312x^2u^{54}+102984x^2u^{53}
+384921/2x^2u^{52}+17103x^2u^{51}
+162828x^2u^{50}$\\$-43710x^2u^{49}
-76855x^2u^{48}-4608x^2u^{47}-21024x^2u^{46}+5982x^2u^{45}
+6633x^2u^{44}+918x^2u^{43}$\\$+888x^2u^{42}
+240x^2u^{41}+396x^2u^{40}
+36x^2u^{39}+7031/5u^{60}-4608u^{56}+5532u^{52}-2804u^{48}
+426u^{44}
+48u^{40} $\\
$L_{11} =+18794572864/11x^{11}u^{66}
-13540389348x^{11}u^{65}+47569139712x^{11}u^{64}
-97076564452x^{11}u^{63}$\\$+126406988784x^{11}u^{62}
-108143883564x^{11}u^{61}+59739201959x^{11}u^{60}
+14304038720x^9u^{60}
$\\$-2477592492x^9u^{59}-19491928200x^{11}u^{59}+2620578876x^{11}u^{58}
+306005260x^{11}u^{57}-86214999x^{11}u^{56}
$\\$-12635748x^{11}u^{55}
+423644x^{11}u^{54}+602928x^{11}u^{53}
+70275x^{11}u^{52}+6656x^{11}u^{51}+660x^{11}u^{50}
+24x^{11}u^{49}$\\$-1540944687x^9u^{66}+8819555424x^9u^{65}
-19439069532x^9u^{64}+17599284188x^9u^{63}
+2383226190x^9u^{62}$\\$-18579302220x^9u^{61}
+14304038720x^9u^{60}-2477592492x^9u^{59}
-1744972041x^9u^{58}+620343152x^9u^{57}$\\$+70418718x^9u^{56}
+2196588x^9u^{55}-14062078x^9u^{54}-3441096x^9u^{53}
-111972x^9u^{52}+314324x^9u^{51}$\\$+98685x^9u^{50}
+18228x^9u^{49}+1828x^9u^{48}+72x^9u^{47}
+510319228x^7u^{66}-1900097388x^7u^{65}
$\\$-108143883564x^{11}u^{61}+59739201959x^{11}u^{60}+1598116578x^7u^{64}
+2151778972x^7u^{63}-3985693602x^7u^{62}
$\\$+589317072x^7u^{61}+2069817648x^7u^{60}
-946297212x^7u^{59}-162915150x^7u^{58}+9669012x^7u^{57}
$\\$+40864122x^7u^{56}+38590740x^7u^{55}-11288488x^7u^{54}
+653808x^7u^{53}-2042730x^7u^{52}-981400x^7u^{51}
$\\$-21864x^7u^{50}+129864x^7u^{49}+63514x^7u^{48}
+14892x^7u^{47}+2220x^7u^{46}+168x^7u^{45}
-74832261x^5u^{66}$\\$+149935740x^5u^{65}+117115302x^5u^{64}
-368112532x^5u^{63}+13494099x^5u^{62}
+268457844x^5u^{61}$\\$-67818484x^5u^{60}-13818084x^5u^{59}
-8156250x^5u^{58}-48587000x^5u^{57}+18009936x^5u^{56}
+14417604x^5u^{55}$\\$+4763664x^5u^{54}-2262660x^5u^{53}
-2736036x^5u^{52}+301952x^5u^{51}-149535x^5u^{50}
-123420x^5u^{49}$\\$+30794x^5u^{48}+43992x^5u^{47}
+20628x^5u^{46}+4280x^5u^{45}+396x^5u^{44}+24x^5u^{43}
+4547170x^3u^{66}$\\$-3141180x^3u^{65}-12064851x^3u^{64}
+6959616x^3u^{63}+6816114x^3u^{62}-1299924x^3u^{61}
+6726046x^3u^{60}
$\\$-5978568x^3u^{59}-7517436x^3u^{58}
+2638500x^3u^{57}+208077x^3u^{56}+1763880x^3u^{55}
+1676511x^3u^{54}
$\\$-747330x^3u^{53}-290739x^3u^{52}
-260484x^3u^{51}-123447x^3u^{50}+49494x^3u^{49}
+16420x^3u^{48}$\\$+11232x^3u^{47}+4140x^3u^{46}
+3720x^3u^{45}+2508x^3u^{44}+504x^3u^{43}
+32x^3u^{42}-78904x u^{66}+106740x u^{64}$\\$+173112x
u^{62}-252388x u^{60}-135885x u^{58}+214710x u^{56}
+46144x u^{54}-75192x u^{52}-6654x u^{50}+7104x
u^{48}$\\$+288x u^{46}+852x u^{44} +72x u^{42} $ \\
 \end{tabular} 
\label{tab5b}
\end{table}

\begin{table}{l}
  \caption {The first nine expansion coefficients $g_n(D,h)$ of the HT
    series expansion in powers of $K$ for the free-energy density
    Eq. (\ref{HTexpansion}) of the BC model with spin $S=1$ and
    nearest-neighbor interaction on a simple-cubic lattice, in a
    reduced crystal field $D=K\Delta/J$ and a reduced magnetic field
    $h=KH/J$. The coefficients of higher order, which are too long to
    be reproduced here, can be found in
    Ref. [\onlinecite{arXiv}]. Notice that in this Table we have set
    for brevity $A \equiv A(D,h;1)$ and $B \equiv B(D,h;1) $, omitting
    the functional dependence of these quantities on $D$ and $h$,
    while $x=\exp(D)$ and $y=exp(h)$ as in the text. }
\begin{tabular}{l}
  \hline
     $ g_0 ={\rm ln}(1+y/x+1/xy)
 =-{\rm ln}(1-{\rm B})  $\\
     $  g_1 =  
     3{\rm A}^2     $\\

       $g_2 = 
     3{\rm B}^2/2+
     15{\rm B}{\rm A}^2
     -33{\rm A}^4/2 $\\

      $ g_3 = 
     {\rm A}^2/2+
     15{\rm B}{\rm A}^2+
     117{\rm B}^2{\rm A}^2/2+
     20{\rm A}^4
     -240{\rm B}{\rm A}^4+
     146{\rm A}^6 $\\

      $ g_4 = 
     {\rm B}^2/8+
     15{\rm B}^3/4
     -9{\rm B}^4/8+
     5{\rm B}{\rm A}^2+
     105{\rm B}^2{\rm A}^2+
     243{\rm B}^3{\rm A}^2+
     32{\rm A}^4+
     75{\rm B}{\rm A}^4
     -4401{\rm B}^2{\rm A}^4/2$ \\$
     -420{\rm A}^6+
     3753{\rm B}{\rm A}^6
     -6381{\rm A}^8/4 $\\

      $ g_5 = 
     {\rm A}^2/40+
     15{\rm B}{\rm A}^2/4+
     885{\rm B}^2{\rm A}^2/8+
     531{\rm B}^3{\rm A}^2+
     1035{\rm B}^4{\rm A}^2+
     10{\rm A}^4+
     573{\rm B}{\rm A}^4
     -960{\rm B}^2{\rm A}^4$ \\$
     -16623{\rm B}^3{\rm A}^4
     -805{\rm A}^6/2
     -7293{\rm B}{\rm A}^6+
     111591{\rm B}^2{\rm A}^6/2+
     7572{\rm A}^8
     -60012{\rm B}{\rm A}^8+
     98298{\rm A}^{10}/5 $\\

      $ g_6 = 
     {\rm B}^2/240+
     5{\rm B}^3/8+
     211{\rm B}^4/16
     -6{\rm B}^5+
     41{\rm B}^6/4+
     2{\rm B}{\rm A}^2/3+
     187{\rm B}^2{\rm A}^2/4+
     3339{\rm B}^3{\rm A}^2/4+ $\\$
     10737{\rm B}^4{\rm A}^2/4+
     17241{\rm B}^5{\rm A}^2/4+
     1967{\rm A}^4/120+
     2321{\rm B}{\rm A}^4/4+
     16941{\rm B}^2{\rm A}^4/4
     -71087{\rm B}^3{\rm A}^4/4$ \\$
     -899127{\rm B}^4{\rm A}^4/8+
     572{\rm A}^6
     -16118{\rm B}{\rm A}^6
     -62856{\rm B}^2{\rm A}^6+
     632166{\rm B}^3{\rm A}^6+
     1192{\rm A}^8+
     225204{\rm B}{\rm A}^8$\\$
     -1250433{\rm B}^2{\rm A}^8
     -132208{\rm A}^{10}+ 
     981858{\rm B}{\rm A}^{10}
     -261941{\rm A}^{12} $\\

     $ g_7 = 
     {\rm A}^2/1680+
     3{\rm B}{\rm A}^2/8+
     2917{\rm B}^2{\rm A}^2/80+
     3111{\rm B}^3{\rm A}^2/4+
     39189{\rm B}^4{\rm A}^2/8+
     13482{\rm B}^5{\rm A}^2$\\$+
     141453{\rm B}^6{\rm A}^2/8+
     13{\rm A}^4/6+
     1705{\rm B}{\rm A}^4/4+
     17699{\rm B}^2{\rm A}^4/2+
     64725{\rm B}^3{\rm A}^4/4
     -372027{\rm B}^4{\rm A}^4/2$\\$
     -703539{\rm B}^5{\rm A}^4+
     82283{\rm A}^6/120+
     6767{\rm B}{\rm A}^6/4
     -2176275{\rm B}^2{\rm A}^6/8
     -248049{\rm B}^3{\rm A}^6+
     6051543{\rm B}^4{\rm A}^6$\\$
     -23672{\rm A}^8+
     228360{\rm B}{\rm A}^8+
     3578184{\rm B}^2{\rm A}^8
     -19113288{\rm B}^3{\rm A}^8+
     95961{\rm A}^{10}
     -5566386{\rm B}{\rm A}^{10}$\\$ +
     26486325{\rm B}^2{\rm A}^{10}+
     2296224{\rm A}^{12}
     -16374696{\rm B}{\rm A}^{12}+
     25804572{\rm A}^{14}/7$\\

      $g_8= 
     {\rm B}^2/13440+
     3{\rm B}^3/64+
     2371{\rm B}^4/640+
     423{\rm B}^5/8
     -461{\rm B}^6/32+
     1689{\rm B}^7/16
     -45{\rm B}^8/64+
     {\rm B}{\rm A}^2/21$\\$+
     75{\rm B}^2{\rm A}^2/8+
     17319{\rm B}^3{\rm A}^2/40+
     50583{\rm B}^4{\rm A}^2/8+
     219345{\rm B}^5{\rm A}^2/8+
     258183{\rm B}^6{\rm A}^2/4+
     145161{\rm B}^7{\rm A}^2/2$\\$+
     5897{\rm A}^4/1680+
     2529{\rm B}{\rm A}^4/8+
     187383{\rm B}^2{\rm A}^4/20+
     611243{\rm B}^3{\rm A}^4/8
     -466653{\rm B}^4{\rm A}^4/16$\\$
     -3130035{\rm B}^5{\rm A}^4/2
     -8319465{\rm B}^6{\rm A}^4/2+
     639{\rm A}^6+
     217141{\rm B}{\rm A}^6/10
     -187287{\rm B}^2{\rm A}^6
     -6076527{\rm B}^3{\rm A}^6/2$\\$+
     1757556{\rm B}^4{\rm A}^6+
     51507546{\rm B}^5{\rm A}^6
     -69649{\rm A}^8/5
     -652824{\rm B}{\rm A}^8+
     28766757{\rm B}^2{\rm A}^8/4+
     78201879{\rm B}^3{\rm A}^8/2$\\$
     -478585635{\rm B}^4{\rm A}^8/2+
     562028{\rm A}^{10}
     -68391{\rm B}{\rm A}^{10}
     -124741005{\rm B}^2{\rm A}^{10}+
     512545839{\rm B}^3{\rm A}^{10}
     -3813290{\rm A}^{12}$\\$+
     125336223{\rm B}{\rm A}^{12}
     -1088156373{\rm B}^2{\rm A}^{12}/2
     -39985236{\rm A}^{14}+
     277367829{\rm B}{\rm A}^{14}
     -432195261{\rm A}^{16}/8$\\

     $ g_9 = 
     {\rm A}^2/120960+
     85{\rm B}{\rm A}^2/4032+
     5375{\rm B}^2{\rm A}^2/896+
     2497{\rm B}^3{\rm A}^2/8+
     190767{\rm B}^4{\rm A}^2/32+
     165681{\rm B}^5{\rm A}^2/4$\\$+
     4821229{\rm B}^6{\rm A}^2/32+
     1195431{\rm B}^7{\rm A}^2/4+
     2383713{\rm B}^8{\rm A}^2/8+
     205{\rm A}^4/756+
     67849{\rm B}{\rm A}^4/504+$\\$
     21511{\rm B}^2{\rm A}^4/3+
     977359{\rm B}^3{\rm A}^4/8+
     922845{\rm B}^4{\rm A}^4/2
     -2478237{\rm B}^5{\rm A}^4/2
     -11566918{\rm B}^6{\rm A}^4
     -23547948{\rm B}^7{\rm A}^4$\\$+
     2052793{\rm A}^6/3024+
     701845{\rm B}{\rm A}^6/24+
     3152285{\rm B}^2{\rm A}^6/16
     -8477493{\rm B}^3{\rm A}^6/2
     -51566121{\rm B}^4{\rm A}^6/2$\\$+
     48392016{\rm B}^5{\rm A}^6+
     803230243{\rm B}^6{\rm A}^6/2+
     31357{\rm A}^8/3
     -1034012{\rm B}{\rm A}^8
     -7019841{\rm B}^2{\rm A}^8+
     131377442{\rm B}^3{\rm A}^8$\\$+
     307273428{\rm B}^4{\rm A}^8
     -2598632472{\rm B}^5{\rm A}^8
     -2972317{\rm A}^{10}/20+
     52005649{\rm B}{\rm A}^{10}/2
     -392748291{\rm B}^2{\rm A}^{10}/4$\\$
     -1959609677{\rm B}^3{\rm A}^{10}+
     15939623007{\rm B}^4{\rm A}^{10}/2
     -30528358{\rm A}^{12}/3
     -121326545{\rm B}{\rm A}^{12}+
     3609445080{\rm B}^2{\rm A}^{12}$\\$
     -12781309029{\rm B}^3{\rm A}^{12}+
     210342571{\rm A}^{14}/2
     -2686363629{\rm B}{\rm A}^{14}+
     21942403551{\rm B}^2{\rm A}^{14}/2+
     699783404{\rm A}^{16}$\\$
     -4758350316{\rm B}{\rm A}^{16}+
      2450507294{\rm A}^{18}/3$\\
     \hline
 \end{tabular}
\label{tab1}
\end{table}

\begin{figure}[p]
\begin{center}
\leavevmode
\includegraphics[width=5.00 in]{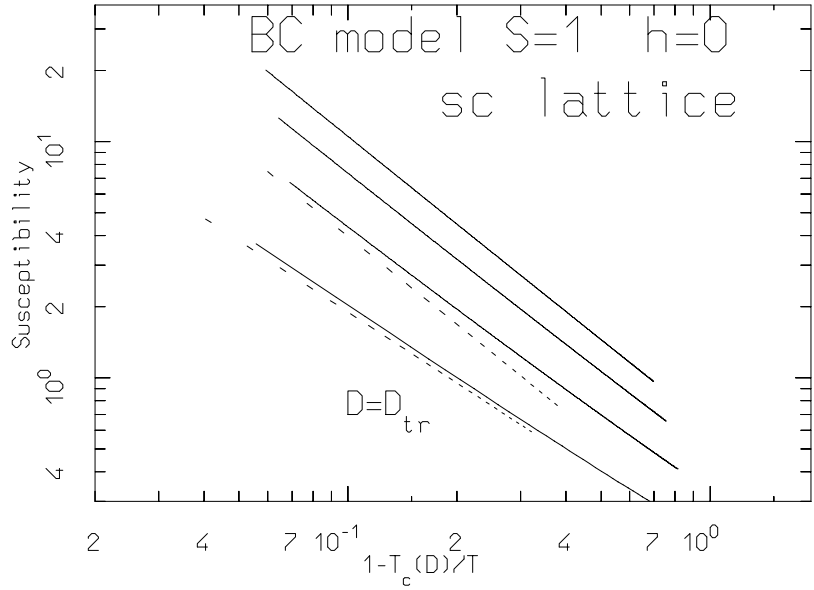}
\caption{ \label{susc_sc} BC model with spin $S=1$ for $h=0$ on the
$sc$ lattice.  The lowermost solid line is a bilog plot of the
resummed HT expansion of the ordinary susceptibility
$\chi_{(2;0)}(K,D;1)$ along the path $D=D_{tr}$ vs $1-T(D_{tr})/T$.  The
lowermost dashed line has a simple power behavior with exponent
$\gamma_{tr} = 1.$
The upper curves show the same quantity evaluated for $D$ taking the
sequence of three values $0.69D_{tr}$ ,$0.4D_{tr}$ , $0.1D_{tr}$ and
for each of them the abscissa is $1-T_c(D)/T$. The uppermost dashed line
has a simple power behavior with exponent $\gamma = 1.237$.  }
\end{center}
\end{figure}

\begin{figure}[p]
\begin{center}
\leavevmode
\includegraphics[width=5.00 in]{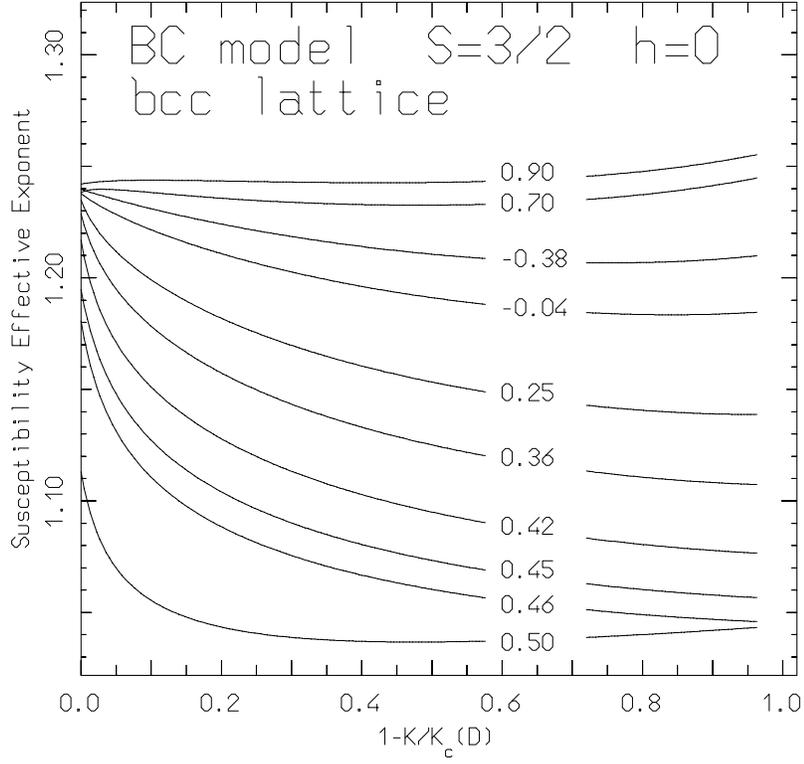}
\caption{ \label{graf_bcc_espef_S32} BC model with spin $S=3/2$ on the
$bcc$ lattice, for $h=0$.  The effective exponents of the
susceptibility $\chi_{(2,0)}(K,D;1)$, for several fixed values of
$\Delta/Jq$, vs the deviation $1-K/K_c(D)$ from the corresponding
critical temperatures.  The value of $\Delta_c/Jq$ is indicated on each
curve.  }
\end{center}
\end{figure}

\end{document}